\documentclass[12pt]{article}
\usepackage[utf8]{inputenc}
\usepackage{multicol} % http://ctan.org/pkg/multirow
\usepackage{graphicx}
\usepackage{caption}
\usepackage{subcaption}
\usepackage{amsmath,amssymb}
\usepackage[margin=1in]{geometry}
\usepackage{color}
\usepackage[colorlinks=true,linkcolor=blue,citecolor=blue]{hyperref}
\usepackage{float}
\usepackage{comment}
\usepackage[usenames,dvipsnames]{xcolor}
\usepackage[ruled,vlined]{algorithm2e}
\usepackage{authblk}
\newcommand{\specialcell}[2][c]{\begin{tabular}[#1]{@{}c@{}}#2\end{tabular}}

\title{Static and sliding contact of rough surfaces: effect of 
asperity-scale properties and long-range elastic interactions}
\author{Srivatsan Hulikal}
\author{Nadia Lapusta}
\author{Kaushik Bhattacharya\footnote{Corresponding author email: bhatta@caltech.edu}}
\affil{Department of Mechanical and Civil Engineering, California Institute of Technology, Pasadena, CA 91125}

\date{}

\begin{document}
\maketitle
% =============================================================================
% ============================ ABSTRACT =======================================
\begin{abstract}
Friction in static and sliding contact of rough surfaces is important in
numerous physical phenomena.  We seek to understand macroscopically observed  static
and sliding contact behavior as the collective response of a large number of
microscopic asperities. To that end, we build on Hulikal {\it et al.}\cite{Hulikal:1} 
and develop an efficient numerical framework that can be used
to investigate how the macroscopic response of multiple frictional contacts
depends on long-range elastic interactions, different constitutive assumptions
about the deforming contacts and their local shear resistance, and surface
roughness.  We approximate the contact between two rough surfaces as that
between a regular array of discrete deformable elements attached to a elastic
block and a rigid rough surface. The deformable elements are viscoelastic or
elasto/viscoplastic with a range of relaxation times, and the elastic
interaction between contacts is long-range.  We find that the model reproduces
main  macroscopic features of evolution of contact and friction for
a range of constitutive models of the elements, suggesting that macroscopic
frictional response is robust with respect to the microscopic behavior.
Viscoelasticity/viscoplasticity contributes to the increase of friction with contact
time and leads to a subtle history dependence. Interestingly, long-range
elastic interactions only change the results quantitatively compared to the
meanfield response.  The developed numerical framework can be used to study
how specific observed macroscopic behavior depends on the 
microscale assumptions.  For example, we find
that sustained increase in the static friction coefficient during long hold
times suggests viscoelastic response of the underlying material with multiple
relaxation time scales. We also find that the experimentally observed
proportionality of the direct effect in velocity jump experiments to the
logarithm of the velocity jump points to a complex material-dependent shear
resistance at the microscale. 
\end{abstract}
% ============================ /ABSTRACT ======================================
% =============================================================================

\section{Introduction}

Contact between surfaces plays an important role in many natural phenomena and
engineering applications. Most surfaces are rough at the microscale and thus
the real area of contact is only a fraction of the nominal area. Interactions
between surfaces such as the flow of electric current, heat, the normal and
shear forces happen across this small real area of contact
\cite{holm_R:1,bowdenFP:1}. Contact area is determined by a number of factors:
surface topography, material properties, the applied load, sliding speed etc.
Since the applied load is sustained over a small area, stresses at the contacts
can be high and time-dependent properties of the material become important. The
macroscopic friction resulting from the collective and interactive behavior of
a population of microscopic contacts shows complex time and history dependence.

In many materials, the static friction coefficient $\mu_s$ increases with the
time of contact, and a logarithmic increase is found to be a good empirical
fit:
\begin{equation}\label{eq:static_friction_evolution_experimental_dieterich}
\mu_s(t) = \mu_0 + A \log (Bt+1), 
\end{equation} 
where $t$ is the time of stationary contact, $\mu_0, A$, and $B$ are constants
dependent on the two surfaces across the interface
\cite{rabinowiczE:1,dieterich:1}. Typically, for rocks, $\mu_0$ is $0.7$-$0.8$,
$A$ is $0.01$-$0.02$ and $B$ is of the order of $1$ second$^{-1}$
\cite{dieterich:2}.  Experiments on different materials show \cite{dieterich:3}
that the kinetic friction coefficient $\mu_k$ depends not only on the current
sliding speed but also on the sliding history.  A class of empirical laws
called ``Rate-and-State" (RS) laws has been proposed to model this behavior
\cite{dieterich:4,ruina:1}.  ``Rate" refers to the relative speed across the
interface and ``state" refers to  one or more internal variables used to
represent the memory in the system. These laws, used widely in simulations of
earthquake phenomena, have been successful in reproducing many of the observed
features of earthquakes \cite{tse:1,dieterich:9,kaneko:1,barbot:1,noda:1}. One
commonly used RS formulation with a single state variable takes the form:
\begin{equation}\label{eq:RS}
\mu_k = \mu_0 + a \ln (\frac{v}{v^*}) + b \ln
(\frac{v^*\theta}{D_c}), \quad \dot{\theta} = 1 - \frac{v \theta}{D_c},
\end{equation}
where $\mu_k$ is the coefficient of friction, $v$ is the sliding velocity,
$a,b,v^*, \mu_0$, and $D_c$ are constants, and $\theta$ is an internal variable
with dimensions of time. The second equation is an evolution law for $\theta$.
At steady state, $\dot{\theta} = 0$, $\theta_{ss}(v) = D_c/v$. Using this, the
steady state friction coefficient is given by:
\begin{equation}\label{eq:RS_steady}
\mu_{ss}(v) = \mu_0 + (a-b) \ln(\frac{v}{v^*}).
\end{equation}
If $a-b>0$, the steady-state friction coefficient increases with increasing
sliding speed and if $a-b<0$, the steady-state friction coefficient decreases
with increasing sliding speed. The two cases are known as velocity
strengthening and velocity weakening respectively. The steady-state velocity
dependence has implications for sliding stability, a requirement for stick-slip
being $a-b<0$ \cite{ruina:1,ruina:2}.  For more details on experimental results
on static/sliding friction and the rate-and-state laws, see \cite{dieterich:2,
dieterich:4,ruina:1,ruina:2,dieterich:3} and references therein.

The goal of our work is to understand the microscopic origin of these
observations. As already mentioned, most surfaces have roughness features at
many length-scales and the macroscopic friction behavior is a result of the
statistical averaging of the microscopic behavior of contacts along with the
interactions between them. We would like to understand what features that are
absent at the microscopic scale emerge at the macroscopic scale as a result of
collective behavior, which factors at the microscale survive to influence the
macroscopic behavior, and which ones are lost in the averaging process.
Further, we want to understand which microscopic factors underlie a particular
aspect of macroscopic friction, for example the timescale of static friction
growth.

To bridge the micro and macro scales, the smallest relevant length-scales must
be resolved while, at the same time, the system must be large enough to be
representative of a macroscopic body. A numerical method like the Finite
Element Method, though useful to study the stress distribution at contacts,
plasticity and such, results in a large number of degrees of freedom
\cite{yan_W:1,hyun_S:1} and can be intractable during sliding of surfaces. To
overcome this, boundary-element-like methods have been proposed in the
literature and many aspects of rough surface contact have been studied using
these methods \cite{gupta1974contact,
webster1986numerical,ren1993contact,ju1996spectral,polonsky_IA:1,liu2001three,
bora_CK:1}. As far as we know, time-dependent behavior and sliding of rough
surfaces have not been studied, and this is the main focus here.  

Our prior work \cite{Hulikal:1} considered the collective behavior of an
ensemble of independent viscoelastic elements in contact with a rough rigid
surface and interacting through a mean field. We showed that the model
reproduces the qualitative features of static and sliding friction evolution
seen in experiments. We also showed that the macroscopic behavior can be
different from the microscopic one; for example, even if each contact is
velocity-strengthening, the macroscopic system can be velocity-weakening.
However, that model neglected a number of potentially important features of the
problem, such as the elastic  interactions between contacts, spatial
correlation of surface roughness, and viscoplasticity of contacts.

Here, we build on our previous work and model the contact between two rough
surfaces as that between a regular array of discrete deformable elements
attached to a elastic block and a rigid rough surface.  The deformable elements
are viscoelastic/viscoplastic with a range of relaxation times. Further, the
interactions between contacts response is taken to be long-range -- the length
change of any element depends on the force acting on all other elements --
consistent with the Boussinesq response of a semi-infinite solid.

Our numerical model combines and builds on features from previous theoretical
models that have been proposed to connect the asperity scale to the
experimentally observed features of the macroscopic frictional behavior. One
class of such models expands the classical formulation of Bowden and Tabor
\cite{bowdenFP:1} and links the macroscopic shear response to the
velocity-dependent shear resistance of the individual contacting asperities
multiplied by the evolving total contact area
\cite{brechetY:1,estrin:1,dieterich:3,berthoudP:1,baumberger:2,putelat:1}. We
use these ideas to compute the shear resistance in our models. Hence, in this
work, the evolving asperity population affects the friction only through the
evolution of the total contact area. An important difference of our model from
the previous theoretical works is that our numerical model can study the
dependence of the area evolution on the following factors: (i) different
constitutive assumptions about the deforming asperities, (ii) long-range
elastic interactions, and (iii) surface roughness. The presence of realistic
surface roughness, in particular, links our model to another class of
theoretical models \cite{archard:1, greenwood:1, majumdarA:1}, in which the
nature of the surface roughness and the collective behavior of asperities
becomes paramount in explaining the macroscale properties. The numerical
framework presented in this study can be used to explore other assumptions
about the local shear resistance, in which different asperities have different
shear strength that depends on their individual characteristics, and hence
surface roughness may play a dominating role, as discussed in the section on
conclusions.

We describe the model, as well as the numerical method used to solve the
resulting equations, in Section \ref{sec:model}.  We validate the model in
Section \ref{sec:validation}, and then study static contact in Section
\ref{sec:staticContact} and sliding contact in Section
\ref{sec:slidingContact}.   We conclude in Section \ref{sec:conclusion} with a
summary of our main findings and some discussion of open issues.

\section{Model}\label{sec:model}

%%%%%%%%%%%%%%%%%%%%%%%%%%%%%%%%%%%%%%%%%%%%%%%%%%%%%%%%%%%%%%%%%%%%%%%%%%%%%%%%
\begin{figure}
\centering
\includegraphics[scale=0.4]{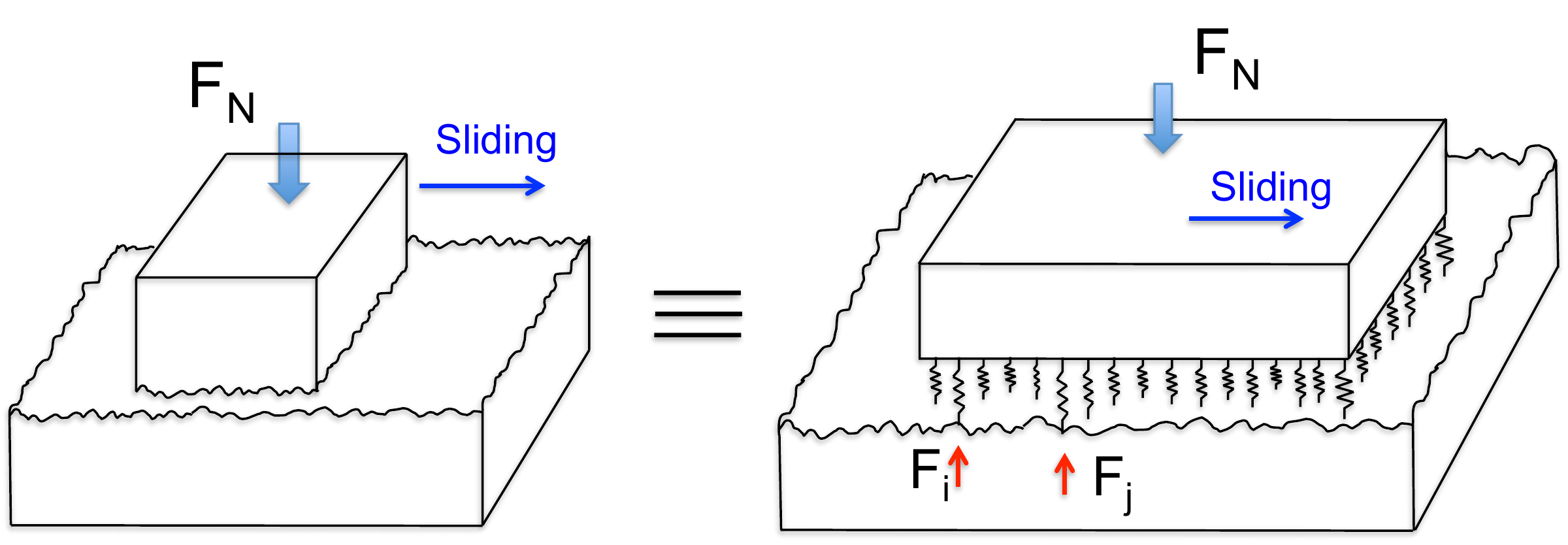}
\caption{Contact of two surfaces, one of the surfaces is approximated
by a set of discrete elements, the other surface is assumed to be 
rigid. The only degree of freedom of the elements is normal to the interface.}
\label{fig:systemDiscretization}
\end{figure}
%%%%%%%%%%%%%%%%%%%%%%%%%%%%%%%%%%%%%%%%%%%%%%%%%%%%%%%%%%%%%%%%%%%%%%%%%%%%%%%%
\subsection{Setting}

The contact between two rough surfaces is dominated by the interaction between
a small number of asperities on the two surfaces.  We model one of these
surfaces as a collection of a regular array of discrete deformable elements
attached to a rigid block while we model the other as a rigid rough surface
(Figure \ref{fig:systemDiscretization}). Contact is determined by the
interaction between a small number of discrete elements with peaks on the rigid
surface, thereby mimicking the interaction between asperities. We denote the
relaxed (or stress-free) length of the $i^\text{th}$ element by $z_i^0$ and its
reference lateral position by $(x_i^0, y_i^0)$. The height of the rough surface
at the position $(x,y)$ is $h(x,y)$.

The kinematic state of the system is described by two macroscopic variables --
the {\it dilatation} or nominal separation $d$ between the two surfaces and the
relative {\it lateral distance} $\bar{x}$ of one surface relative to the other
(or equivalently the distance of sliding) -- and $N$ microscopic variables
$u_i$ describing the {\it change in lengths of the elements} from its original
length (so that the total length of the element is $z_i^0+u_i$).  At any given
dilatation $d$ and lateral position $\bar{x}$,  the  distance between the
$i^\text{th}$ element and the rough surface is \begin{equation} \label{eq:di}
d_i = d - z_i^0 - u_i - h(x_i^0 + \bar{x},y_i^0).  \end{equation} We require
$d_i  \ge 0$ to represent the non-interpenetrability of matter.  Further, $d_i
= 0$ signifies the contact of the $i^\text{th}$ element with the rough surface.

Corresponding to the kinematic variables, we have two macroscopic forces -- the
{\it macroscopic  normal force} $F_N$ and the {\it macroscopic shear force}
$F_S$ -- as well as $N$ {\it microscopic normal forces} $F_i$ and $N$ {\it
microscopic shear strengths} $S_i$. In this work, we assume that adhesion is
negligible and hence the force on the $i^\text{th}$ element is always
compressive: $F_i \le 0$.

We postulate time-dependent constitutive relations at the microscopic scale --
one between the length change $u_i$ and force $F_i$ of each element, and
another for the microscopic strength $S_i$.  We then study the equilibrium of
the system ($F_N=\sum F_i$, $F_S = \sum S_i$) to infer the macroscopic contact
and friction laws.

%%%%%%%%%%%%%%%%%%%%%%%%%%%%%%%%%%%%%
%%%%%%%%%%%%%%%%%%%%%%%%%%%%%%%%%%%%%
%%%%%%%%%%%%%%%%%%%%%%%%%%%%%%%%%%%%%
\subsection{Rough surface} \label{sec:roughSurfaceGeneration}
%%%%%%%%%%%%%%%%%%%%%%%%%%%%%%%%%%%%%%%%%%%%%%%%%%%%%%%%%%%%%%%%%%%%%%%%%%%%%%%%
\begin{figure}[t]
\centering
\begin{subfigure}[t]{0.32\textwidth}
\includegraphics[width=\textwidth]{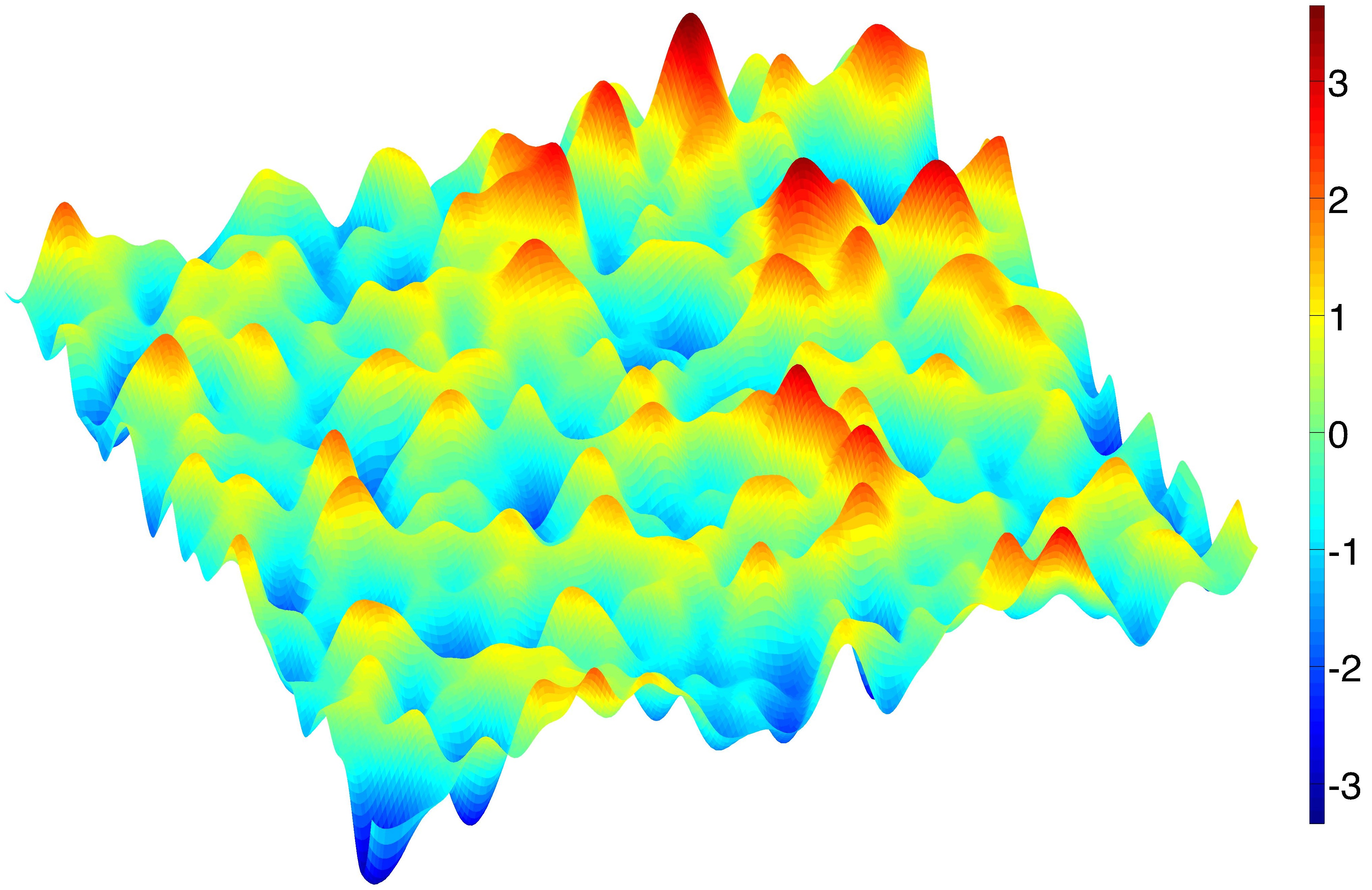}
\subcaption{}
\label{fig:gaussian256by256Surface}
\end{subfigure} 
\begin{subfigure}[t]{0.32\textwidth}
  \includegraphics[width=\textwidth]{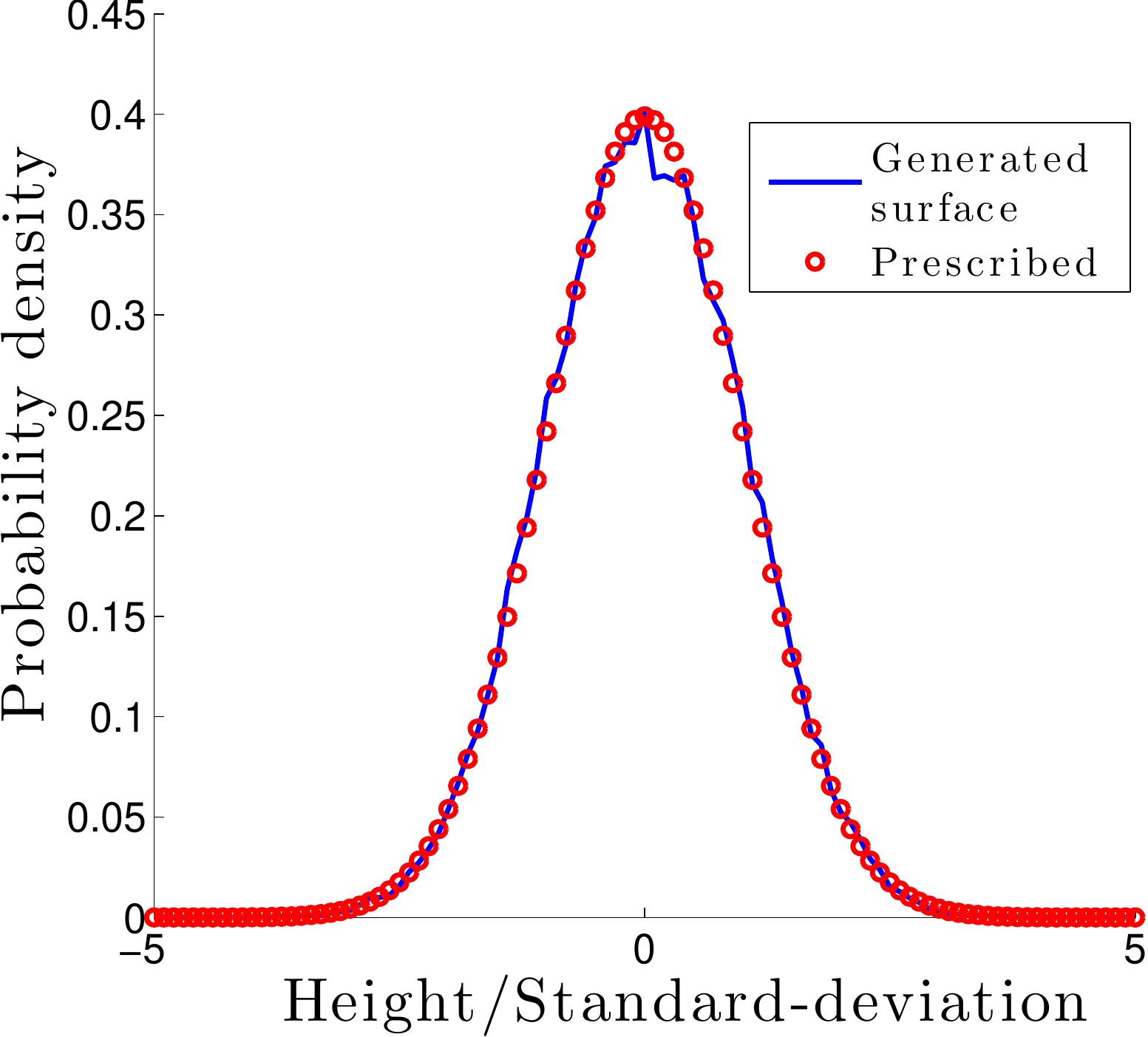}
\subcaption{}
\label{fig:pdfHeights}
\end{subfigure} 
\begin{subfigure}[t]{0.32\textwidth}
  \includegraphics[width=\textwidth]{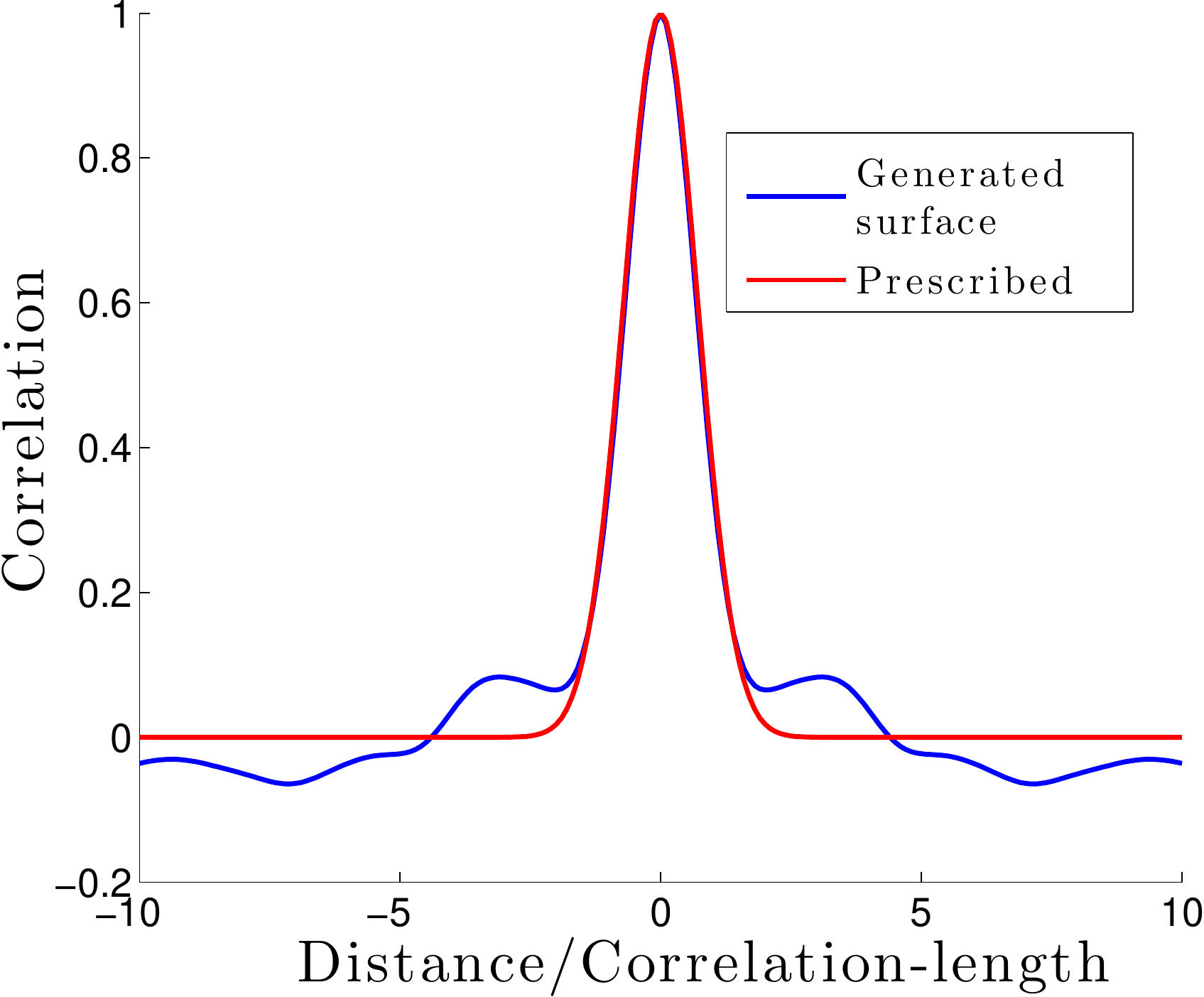}
\subcaption{}
\label{fig:gaussianCorrelation}
\end{subfigure} 
\caption{(a) Rough surface with a Gaussian distribution of heights and a
Gaussian autocorrelation (vertical features magnified). Statistical properties
of the generated surfaces: (b) probability density (Gaussian) of heights,  and
(c) Gaussian spatial correlation.}
\label{fig:surfaceStatistics}
\end{figure}
%%%%%%%%%%%%%%%%%%%%%%%%%%%%%%%%%%%%%%%%%%%%%%%%%%%%%%%%%%%%%%%%%%%%%%%%%%%%%%%%
Rough surfaces have been characterized as a stochastic process
\cite{thomas1999rough,longuethiggins:1, saylesRS:1} and this characterization
has been used extensively in exploring various aspects of contact between
surfaces \cite{greenwood:1,nayakPR:1,nayakPR:2}. The stochastic process is
specified by two functions, a probability distribution of heights which
describes features normal to the interface and an autocorrelation function
which is related to how the vertical features vary along the interface.  For
many surfaces, the probability distribution of heights is Gaussian,
$$ P(h) = \frac{1}{\sigma\sqrt{2\pi}} e^{-h^2/(2\sigma^2)}, $$
where $h$ is the height of a surface from the mean, $P(h)$ is the probability
density, and $\sigma$ is the root mean square roughness \cite{whitehouse:1}.
The autocorrelation is found to decay exponentially or as a Gaussian
\cite{whitehouse:1,Kim2001}.  In this study, we consider surfaces with a
Gaussian autocorrelation:
$$ R(\delta_x,\delta_y) = 
\langle h(x,y)h(x+\delta_x,y+\delta_y) \rangle
= \sigma^2 e^{-(\delta_x^2/\beta_x^2+\delta_y^2/\beta_y^2)},$$ 
where $R$ is the autocorrelation function, $\beta_x, \beta_y$ are the
correlation lengths along $x$ and $y$ directions, and $\langle \ \rangle$
denotes expectation with respect to the probability distribution.

Rough surfaces with these statistical properties are created by generating a
set of independent Gaussian random numbers and using a linear filter
\cite{hu_YZ:1}. The weights of the linear filter are determined from the
autocorrelation. The heights of the resulting rough surfaces are known only at
discrete locations and are interpolated using a cubic spline for intermediate
values. A typical realization of such a surface, its probability distribution
of heights and autocorrelation are shown in Figure \ref{fig:surfaceStatistics}.
The generated surface shows a good match with the prescribed properties.

Since we are interested in two rough surfaces, we need to use two stochastic
processes, one for the rigid rough surface $h(x,y)$ and one for the elements
$z_i^0$. However, in this work, we take all the elements to have the same
stress-free heights $z_i^0 = 0$. We can see from (\ref{eq:di}) that this does
not pose a loss of generality when we are in static contact ($\bar{x} = 0$)
since we may redefine $z_i^0 + h(x_i,y_i)$ to be a random variable as long as
we understand the statistics to be the joint statistics of the two surfaces.
However, this is a non-trivial assumption in the case of sliding contact,
especially in the case of viscoplastic interactions.

In what follows, we non-dimensionalize the equations.  In part, we introduce a
characteristic length $L^*$ and set $ \bar{h} = h/L^*, \bar{\sigma} =
\sigma/L^*, $ etc. Since the characteristic root-mean-square roughness
($\sigma$) of most experimental surfaces is of the order of 1 $\mu$m
\cite{whitehouse:1}, we take $L^* = 1 \ \mu$m.

%%%%%%%%%%%%%%%%%%%%%%%%%%%%%%%%%%%%%
%%%%%%%%%%%%%%%%%%%%%%%%%%%%%%%%%%%%%
%%%%%%%%%%%%%%%%%%%%%%%%%%%%%%%%%%%%%
\subsection{Constitutive relation for the deformation of the elements} \label{sec:cr}

In this work, we consider three different kinds of constitutive relations for
the deformation of the elements.  The first two are viscoelastic and the third
is viscoplastic.  We note that, in the viscoelastic models, the change in
length $u_i$ eventually (exponentially in time) returns to zero when the force
is released.

%%%%%%%%%%%%%%%%%%%%%%%%%%%%%%%%%%%%%
\subsubsection{Viscoelastic element without elastic interactions}
\label{subsubsec:local}
Here we assume that the length change $u_i$ of element $i$ depends only on its
force history and is governed by a finite number $N_T$ of relaxation times.
Therefore, given a force history $F_i(t) \le 0$,
\begin{align}\label{eq:noInteractionViscoelastic}
u_i(t) &= C^0 \left[F_i(t) +\int_0^t 
\sum_{k=1}^{N_T} A_k e^{-\lambda_k(t-\tau)}F_i(\tau) \ \mathrm{d}\tau \right]
\end{align}
where $C^0$ is the elastic compliance of the elements and $A_k$ is the
amplitude of the viscoelastic effect associated with the $k^\text{th}$
relaxation rate $\lambda_k$.  Most of our examples use a single relaxation rate
($N_T =1$).

It is convenient to non-dimensionalize the equations.  
We use  $L^*, T^*, F^*$ to nondimensionalize length, time and force and set
\begin{equation} \label{eq:non-dim}
\bar{u} = \frac{u}{L^*},
\quad\bar{t} = \frac{t}{T^*},
\quad\bar{\tau} = \frac{\tau}{T^*}, \quad\bar{F} = \frac{F}{F^*}, 
\quad \bar{A}_k = A_k T^*, \quad \bar{\lambda}_k = \lambda_k T^*, \quad
\bar{C}^0 = \frac{C^0F^*}{L^*}, \quad \dots
\end{equation}
Substituting these into (\ref{eq:noInteractionViscoelastic}), the equation remains
unchanged except all terms are non-dimensional (replace unbarred symbols with 
barred ones).

To understand this constitutive equation, assume that we have a single relaxation rate $\bar{\lambda}$.
Consider a force $\bar{F}_j(\bar{t}) = \bar{F}^0_j H(\bar{t})$, where $H(\bar{t})$ is the 
Heaviside function.  For $\bar{t} > 0$,
$$ 
\bar{u}_i(\bar{t}) = \bar{C}_0
\left[1+ \frac{\bar{A}}{\bar{\lambda}} \left(1-e^{-\bar{\lambda}\bar{t}}\right)\right] \bar{F}^0_j.$$
If $\bar{t} \ll 1/\bar{\lambda}$,
$$ \bar{u}_i(\bar{t}) \approx \bar{C}_0 \bar{F}^0_j,$$
and for $\bar{t} \gg 1/\bar{\lambda}$,
$$ \bar{u}_i(\bar{t}) \approx \bar{C}_0(1+\bar{A}/\bar{\lambda}) \bar{F}^0_j.$$
Thus, the instantaneous compliance of the system is $\bar{C}_0$,
the steady-state compliance is $(1+\bar{A}/\bar{\lambda})\bar{C}_0$,
and the length change reaches steady state at the decay rate $\bar{\lambda}$.

Since the typical timescale of static friction growth in rocks is of the order of seconds \cite{dieterich:2}, we take
$T^* = 1$ sec.  As before, we take $L^* = 1 \ \mu$m.   

%%%%%%%%%%%%%%%%%%%%%%%%%%%%%%%%%%%%%
\subsubsection{Viscoelastic element with Boussinesq interaction}
\label{subsubsecboussinesq}
Here, the length change $u_i$
of element $i$ depends {\it not} only on force history of that element, but also on the 
force history of all the other elements.  As before,
we assume that the viscoelastic response is governed by a finite number $N_T$
of relaxation times. However, the interaction between elements $i$ and $j$, for $i\ne j$, depends on the
distance $r_{ij}$ between the elements.  
Specifically, we assume that given a force history $F_i(t) \le 0$,
\begin{equation}\label{eq:boussinesqViscoelastic}
  \begin{split}
u_i(t) &= \frac{1-\nu}{2\pi G} \left[ \frac{3.8}{\Delta} \left( F_i(t) + \int_0^t
 \sum_{k=1}^{N_T} A_k  e^{-\lambda_k (t-\tau)}F_i(\tau) \mathrm{d}\tau\right) \right. \\ 
 & \phantom{asdfasdfasdf} +
\left.\sum_{j\ne i} \frac{1}{r_{ij}} \left( F_j(t) + \int_0^t
 \sum_{k=1}^{N_T} A_k  e^{-\lambda_k (t-\tau)}F_j(\tau) \mathrm{d}\tau\right) \right].
  \end{split}
\end{equation}
where $G$ is the shear modulus and $\nu$ the Poisson's ratio of the medium, 
$\Delta$ is the distance between neighboring elements
and $A_k$ is the amplitude of the viscoelastic effect
associated with the $k^\text{th}$ relaxation rate $\lambda_k$.
Most of our examples use a single relaxation rate.

The constitutive relation above is motivated by the Boussinesq solution that
describes the normal displacement $u(r)$ at position $r$ on the surface of a
homogeneous elastic half-space due to an applied normal force $F$ at the
origin (e.g. \cite{bower}):  
$$ 
u(r) = \frac{1-\nu}{2\pi G} \frac{F}{r}.
$$
This equation, the superposition principle and the viscoelastic correspondence
principle motivate the form of (\ref{eq:boussinesqViscoelastic}) when $i \ne
j$. Note that the expression becomes singular when $i=j$ since $r = 0$.  However, this
singularity is regularized if the point force is replaced by a uniform pressure
over an area. Therefore we assume that $F_i$ is uniformly distributed over a
square area of side-length $\Delta$ (the distance between the elements). The
first term (with the factor 3.8) is obtained using a solution by Love
\cite{love_AEH:1} and the viscoelastic correspondence principle.

We non-dimensionalize the equations according to (\ref{eq:non-dim}) and
$\bar{r}_{ij} = r_{ij}/L^*, \bar{\Delta} = \Delta/L^*$.  As mentioned before, we take
$L^* = 1 \ \mu$m. We assume $T^* = 1$ s and set $F^* = \frac{2\pi G {L^*}^2}{1-\nu}$
so that our constitutive relation is
\begin{equation}\label{eq:boussinesqViscoelasticNondim}
  \begin{split}
\bar{u}_i(t) &=  \frac{3.8}{\bar{\Delta}} \left( \bar{F}_i(\bar{t}) + \int_0^{\bar{t}}
 \sum_{k=1}^{N_T} \bar{A}_k  e^{-\bar{\lambda}_k (\bar{t}-\bar{\tau})} \bar{F}_i(\bar{\tau}) \mathrm{d} \bar{\tau}\right)  \\ 
 & \phantom{asdfasdfasdf} +
\sum_{j\ne i} \frac{1}{\bar{r}_{ij}} \left( \bar{F}_j(\bar{t}) + \int_0^{\bar{t}}
 \sum_{k=1}^{N_T} \bar{A}_k  e^{-\bar{\lambda}_k (\bar{t}-\bar{\tau})}\bar{F}_j(\bar{\tau}) \mathrm{d}\bar{\tau}\right) .
  \end{split}
\end{equation}

By considering a situation with one relaxation time and applying a
step load, we can see that the instantaneous and steady-state compliance of the
system are $3.8/\bar{\Delta}$ and $3.8(1+\bar{A}/\bar{\lambda})/\bar{\Delta} $
respectively when $i = j$ and $1/\bar{r}_{ij}$ and
$(1+\bar{A}/\bar{\lambda})/\bar{r}_{ij}$ respectively when $i \ne j$.  

%%%%%%%%%%%%%%%%%%%%%%%%%%%%%%%%%%%%%
\subsubsection{Elastic/viscoplastic element}
\label{subsubsec:evp}
Here we assume that the element can undergo permanent deformation through
viscoplastic creep.  We specify this constitutive relation directly in
non-dimensional quantities.  First, the total length change is split into elastic
(recoverable) and plastic (permanent) parts:
\begin{equation}
\bar{u}_i = \bar{u}_i^e + \bar{u}_i^p
\end{equation}
where $\bar{u}_i, \bar{u}_i^e, \bar{u}_i^p$ are the total, elastic, and plastic 
length changes, respectively, of the $i^\text{th}$ element.

The elastic length change is linearly related to the force, and we
take this to be nonlocal:
\begin{equation}
\bar{u}_i^e = \sum_{i,j} \bar{C}_{ij} \bar{F}_j
\end{equation}
where $\bar{C}_{ij}$ is the compliance and it is taken to be the elastic part of 
(\ref{eq:boussinesqViscoelasticNondim}).

The evolution of the plastic deformation is local, and it is given by a power-law
creep where the rate of change of the plastic length change depends on the
current force and the history of the plastic deformation
 \begin{equation}\label{eq:powerLawCreep}
\dot{\bar{u}}_i^p = \bar{A}_{cr} (|\bar{F}_i|/{\bar{F}}_i^y )^n \text{ sign}(\bar{F}_i),
\end{equation}
where ${\bar{F}}_i^y, \bar{A}_{cr}, n$ are the yield force, the creep rate, and the 
creep exponent, respectively.  $\bar{A}_{cr}$ and $n$ are material constants.
The yield force $\bar{F}_i^y$ of the $i^\text{th}$ element evolves with the deformation 
according
to the following hardening law:
\begin{equation}
\dot{{\bar{u}}}_i^{acc} = |\dot{{\bar{u}}}^p|, {\bar{F}}_i^y = {\bar{F}}_y^0(1+\bar{B} {\bar{u}}_i^{acc})^m,
\end{equation}
where ${\bar{u}}_i^{acc}, {\bar{F}}_y^0,\bar{B},m$ are the accumulated plastic length change in the 
$i^\text{th}$ element, 
the initial yield stress, the hardening rate and the hardening exponent, 
respectively.   The last three are material parameters.

%%%%%%%%%%%%%%%%%%%%%%%%%%%%%%%%%%%%%
\subsection{Calculation of macroscopic friction coefficient}

The friction coefficient is given by the ratio of the macroscopic shear strength $\bar{F}_S$
to the normal force $\bar{F}_N$:
\begin{equation} \label{eq:fric}
\mu = \frac{\bar{F}_S}{\bar{F}_N},
\end{equation}
where $\bar{F}_S = \sum_i \bar{S}_i$ and $\bar{S}_i$ is the shear force on
element $i$.  We assume that the shear force that any element in contact with
the rough surface can sustain depends on the shear strength of the
material\footnote{Note that we neglect that contribution of the viscoelastic
elements to the shear force. The total viscoelastic energy loss is of the
order of $F_N \sigma$ (for example, see Figure
\ref{fig:dilatationTimeStaticContact} where the viscoelastic energy loss is
equal to the work done by $F_N$ against the dilatation). We can estimate the
ratio of the rate of viscoelastic energy loss to the rate of total energy loss
through the nondimensional ratio ${F_N \sigma \over T_0 \langle {F_S} \rangle
v}  = { \sigma  \over T_0  \mu v} $ where $T_0$  is the typical
relaxation time, $\mu$ is the coefficient of friction and $v$ is the sliding
velocity.  For typical number, $\sigma =1\mu m$, $T_0 = 1$ sec, $\mu = O(1)$,
$v=1$mm/s, we conclude that this nondimensional ratio is $O(10^{-3})$ thereby
justifying our choice.}. Therefore, for an element in contact,
\begin{equation} \label{eq:shear}
\bar{S}_i = \begin{cases}
\bar{\tau}_S \bar{A}_i & \bar{v} = 0 \\
\displaystyle{\left(\bar{\tau}_S  + \alpha \log(\bar{v})\right) \bar{A}_i} & \text{else}\\
\end{cases}
\end{equation}
where $\bar{\tau}_S$ is the (non-dimensional) shear strength of the material,
$\bar{A}_i$ is area of element $i$, $\alpha$ is the velocity-hardening
coefficient, and $\bar{v}$ is the sliding velocity.  Importantly, we assume that
the microscopic shear force is {\it independent} of the microscopic normal
force as long as the element is in contact.
If the surfaces are in static contact, the shear strength is $\bar{\tau}_S$.  If
the surfaces are sliding at a relative speed $\bar{v}$, the asperities in
contact are sheared at a strain rate proportional to the sliding speed, and if
the shear resistance depends on the strain rate, then the local friction law
will be velocity-dependent. Taking cue from experimental results, we assume
this velocity dependence to be logarithmic. A theoretical justification for the
logarithmic dependence has been proposed by Rice et al. \cite{riceJR:1}.

Summing equation (\ref{eq:shear}) over all elements in contact,
\begin{equation} \label{eq:shear2}
\bar{F}_S = \begin{cases}
\bar{\tau}_S \bar{A} & \bar{v} = 0, \\
\displaystyle{\left(\bar{\tau}_S  + \alpha \log(\bar{v})\right) \bar{A}} & \text{else,}\\
\end{cases}
\end{equation}
where the total contact area $\bar{A}$ is given by the sum of all elements in
contact.

%%%%%%%%%%%%%%%%%%%%%%%%%%%%%%%%%%%%%
%%%%%%%%%%%%%%%%%%%%%%%%%%%%%%%%%%%%%
%%%%%%%%%%%%%%%%%%%%%%%%%%%%%%%%%%%%%
\subsection{Simulation of evolving contacts}

Now that we have the rough surfaces and the constitutive equations for the
elements, let us look at the simulation of static/sliding contact. Consider
two rough surfaces in contact under a given macroscopic normal force
$\bar{F}_N$ and sliding at a velocity $v$ relative to each other ($v=0$ for
static contact). We seek to use the model above to determine the time-evolution
of the macroscopic shear force $\bar{F}_S$ and the macroscopic dilatation
$\bar{d}$. We are also interested in the statistical features of the contact.

We assume that we know the prior history of all microscopic variables up to
some time which we set as $\bar{t}=0$.  We now solve for the microscopic
variables $\bar{u}_i(\bar{t}), \bar{F}_i(\bar{t})$ and the macroscopic variable
$\bar{d}(\bar{t})$ satisfying the appropriate constitutive relation (CR) in
Section \ref{sec:cr} subject to the constraints
\begin{eqnarray} \label{eq:constraint}
& \bar{d}_i(\bar{t}) \bar{F}_i(\bar{t}) = 0, \quad  \bar{d}_i(\bar{t}) \ge 0, \quad \bar{F}_i(\bar{t}) \le 0; \\
& \bar{F}_N(\bar{t}) = \sum_i \bar{F}_i(\bar{t})
\end{eqnarray}
where $d_i$ is given by (\ref{eq:di}) with $\bar{x}(t) = \bar{v} \bar{t}$.

We implement the model numerically by discretizing the constitutive relations
using a first-order Euler method and applying an iterative approach as
described in detail in Algorithm 1. Note that (CR) refers to the appropriate
constitutive relation of the element. 

\vspace{0.01in}

\begin{algorithm}[H] \label{alg:contact}
Given $\bar{F}_N, \bar{v}$, constitutive relation for each element denoted by (CR) below \\
Given history $\bar{u}_i(\bar{\tau}), \bar{F}_i(\bar{\tau}), \bar{d}(\bar{\tau})$ for $\bar{\tau} \le \bar{t}$\\
\Repeat ( Take a time step $\bar{t} \to \bar{t}+\Delta \bar{t}$){$\bar{t}=\bar{T}$}
{
Assume the set of elements in contact $\mathcal{I}_c$ does not change in the 
time-step $\Delta \bar{t}$.\\
Guess two dilatation rates $\dot{\bar{d_1}}$ and $\dot{\bar{d_2}}$. One of
these can be dilatation rate of the previous time-step. Compute two dilatation
guesses $\bar{d}_1(\bar{t}+\Delta \bar{t})$ and $\bar{d}_2(\bar{t}+\Delta
\bar{t})$. We will use a secant root-finding method to find the new dilatation
that conserves the total normal force. \\
\Repeat{$\sum \bar{F}_i (\bar{t}+\Delta \bar{t}) = \bar{F}_N$}
{		{
    For each rate \\
		\For{$i=1, \dots, N$}{
			\If{$i \in \mathcal{I}_c$}{set $\bar{d}_i(\bar{\tau}) = 0$ for $\bar{\tau} \in (\bar{t}, \bar{t}+\Delta \bar{t})$ and use (CR) to compute $\bar{F}_i(\bar{t}+\Delta \bar{t})$}
			\Else{ set $\bar{F}_i(\bar{\tau}) = 0$ for $\bar{\tau} \in (\bar{t}, \bar{t}+\Delta \bar{t})$ and use (CR) to compute $\bar{u}_i(\bar{t}+\Delta \bar{t})$}
		}
    Compute the total force $\sum_i\bar{F}_i(\bar{t}+\Delta \bar{t})$ at the two rates. Call these $\bar{F}_N^1,\bar{F}_N^2$. \\
    Use the secant root-finding method with the new dilatation rate guess
   $\dot{\bar{d}}_{\text{new}} = \dot{\bar{d_1}} + (\dot{\bar{d_2}}-\dot{\bar{d_1}})(\bar{F}_N-\bar{F}_N^1)/(\bar{F}_N^2-\bar{F}_N^1)$
	}
}
Use (\ref{eq:shear2}) to compute $\bar{F}_S (\bar{t}+\Delta \bar{t}) = \sum S_i (\bar{t}+\Delta \bar{t}), \ i = 1, \dots, N$\\
			{\For{$i=1, \dots, N$}{
				\If{$\bar{F}_i > 0$}{set $\bar{F}_i = 0$, remove $i$ from $\mathcal{I}_c$ (loss of contact)}
				\If{$\bar{d}_i < 0$}{set $\bar{d}_i = 0$, add $i$ from $\mathcal{I}_c$ (contact formed)}
			}}
Note $I_c, \{u_i\}, \{\bar{F}_i\}, \bar{d}, $ at $\bar{t}+\Delta \bar{t}$\\
Compute $\mu(\bar{t}+\Delta \bar{t})$ from (\ref{eq:fric})\\
}
\caption{Algorithm to simulate contact under a given global normal force.}
\end{algorithm}

\subsubsection{Computational memory and complexity considerations} In the
Boussinesq interaction case, because the deformation due to a point force
decays only as $1/r$, the compliance matrix $C_{ij}$ is dense and for a large
system, storing the matrix entries leads to large memory requirements.  We
circumvent  this challenge by computing the matrix-vector product $C_{ij}F_j$
in a matrix-free way and using an iterative solver (GMRES)
\cite{petsc-web-page} when a linear system needs to be solved. Because of the
$1/r$ decay, a brute force computation of the interactions would involve
$O(N^2)$ operations where $N$ is the number of elements, and this can be
prohibitively expensive for large systems.  Two things come to our rescue here.
First, for rough surfaces, the actual area of contact is only a small fraction
of the nominal area; so, at any instant of time, the forces are nonzero for
only a small fraction of the elements and only these elements need to be
considered in computing the displacements. Second, the $1/r$ interactions can
be computed to within prescribed error tolerance in $O(N\log(N))$ operations
using the Fast Multipole Method (FMM) \cite{greengard1987,ihler_A:1}.  In all
results presented here, we use an FMM method of order $5$ since which we found
to be sufficient by comparison with a brute force calculation.

%%%%%%%%%%%%%%%%%%%%%%%%%%%%%%%%%%%%%%%%%%%%%%%%%%%%%%%%%%%%%%%%%%%%%%%%%

\subsection{Parameters used in simulations}

Unless mentioned otherwise, the parameters used in our simulations are listed
in Table \ref{table:parameters}.
%%%%%%%%%%%%%%%%%%%%%%%%%%%%%%%%%%%%%%%%%%%%%%%%%%%%%%%%%%%%%%%%%%%%%%%%%%%%%%%%
\begin{table}
\centering
\begin{tabular}{|c|c|}
\hline
\textbf{Parameter}  & \textbf{Value}  \\ \hline
System size & 512 $\mu$m by 512 $\mu$m \\ 
No-interaction elastic compliance $\bar{C}^0$ & 16 (Section \ref{subsubsec:local})\\
Shear modulus $G$ & 30 GPa  \\ 
Poisson ratio $\nu$ & 0.25 \\ 
Rough surfaces &  Flat deformable against rough rigid \\
Probability distribution of heights & Gaussian \\ 
Rms-roughness $\sigma$ & 1 $\mu$m \\ 
Autocorrelation of heights & Gaussian \\ 
Correlation lengths $\beta_x, \beta_y$ & 10 $\mu$m \\ 
Viscoelasticity parameters & $\bar{\lambda}=1,\bar{A}=0.2$ \\ \hline
Viscoplasticity parameters & \specialcell{$n = 3,\bar{A}_{cr} = 10^{-4}$ \\ $\bar{F}_y^0 = 0.016,
\bar{B} = 0,m = 1$ (Section \ref{subsubsec:evp})} \\ \hline
Static shear strength $\tau_S$  & 2.5 GPa (Equation \ref{eq:shear})\\ 
Velocity-hardening parameter $\alpha$ & 0.0002 (Equation \ref{eq:shear})\\
\hline
\end{tabular}
\caption{Parameters used in our simulations.}
\label{table:parameters}
\end{table}
%%%%%%%%%%%%%%%%%%%%%%%%%%%%%%%%%%%%%%%%%%%%%%%%%%%%%%%%%%%%%%%%%%%%%%%%%%%%%%%%
$G, \nu$ values are typical for rocks.  The shear strength
$\tau_S$ of a contact can be a significant fraction of the shear modulus
\cite{riceJR:1,dieterich:10}, here we use  $\tau_S$ = 2.5 GPa.  $\bar{F}_y^0 =
0.016$ corresponds to a yields stress of 4 GPa.  We choose $\bar{C}^0 = 16$
since this gives us true contact areas of the order of 1\% of nominal areas,
similar to experimental observations.

The main parameters that vary in the following simulations are the material
behavior (viscoelastic parameters $\bar{\lambda},\bar{A}$ and the
viscoplasticity parameters in Table \ref{table:parameters}), the constitutive
response (with or without long-range elastic interactions), and the sliding
velocity $\bar{v}$. All
our simulations use the same rough surface.

\section{Validation using Hertzian contact}\label{sec:validation}

To validate our formulation, we simulate the Hertzian contact of a homogenous
linear-elastic sphere of radius $1$ with a rigid flat surface. The geometry of
the sphere is simulated using the undeformed lengths of the elements. The two
surfaces are initially apart, and the force and length changes of all the
elements are initialized to zero. The surfaces are then brought into contact by
decreasing the dilatation. The evolution of the length changes and forces of
the elements is computed using an algorithm similar to Algorithm 1 but with the
dilatation prescribed.
%%%%%%%%%%%%%%%%%%%%%%%%%%%%%%%%%%%%%%%%%%%%%%%%%%%%%%%%%%%%%%%%%%%%%%%%%%%%%%%%
\begin{figure}
\centering
\begin{subfigure}[t]{0.49\textwidth}
  \includegraphics[width=\textwidth]{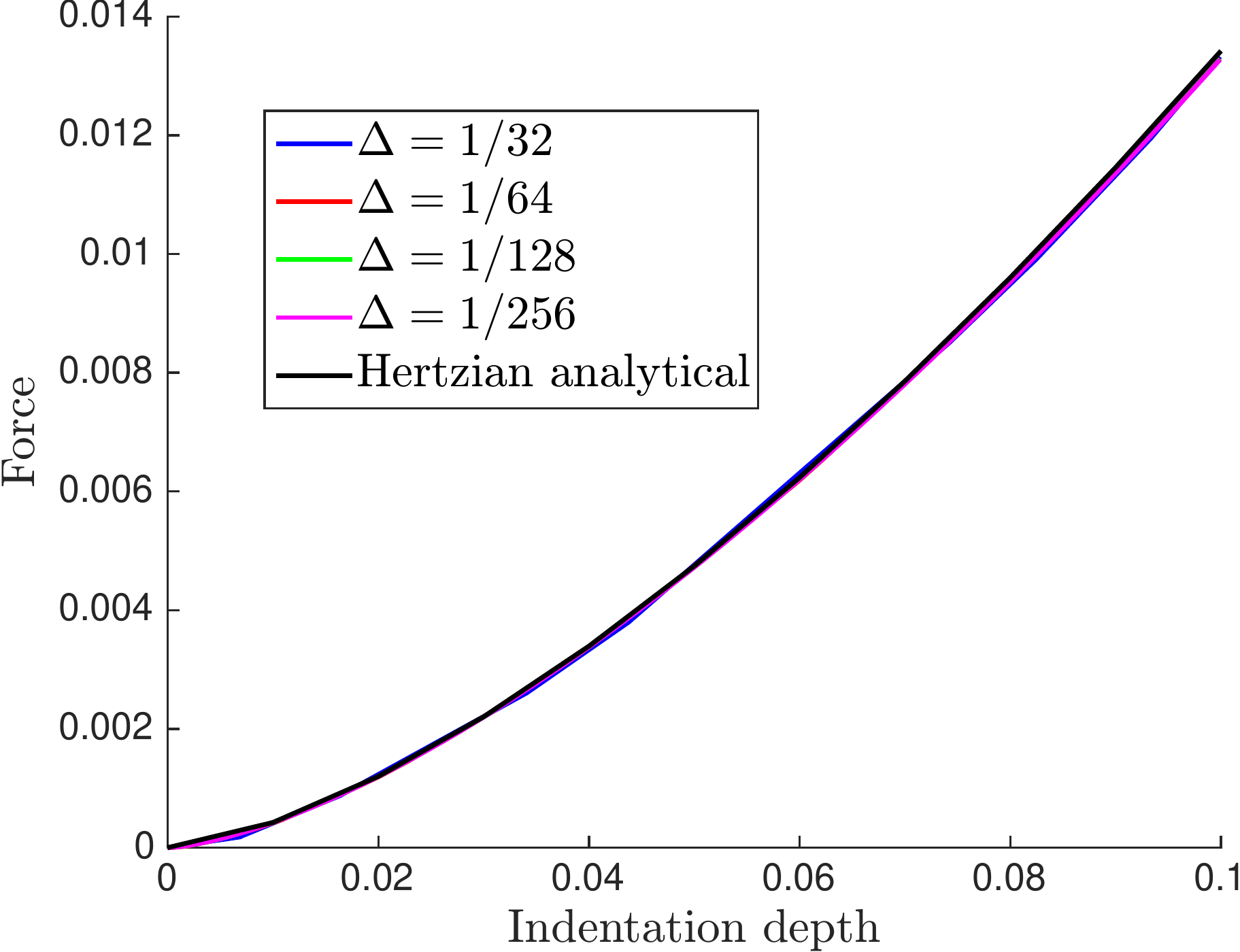}
\subcaption{}
\label{fig:forceIndentationDepthHertzianContact}
\end{subfigure} 
\begin{subfigure}[t]{0.49\textwidth}
  \includegraphics[width=\textwidth]{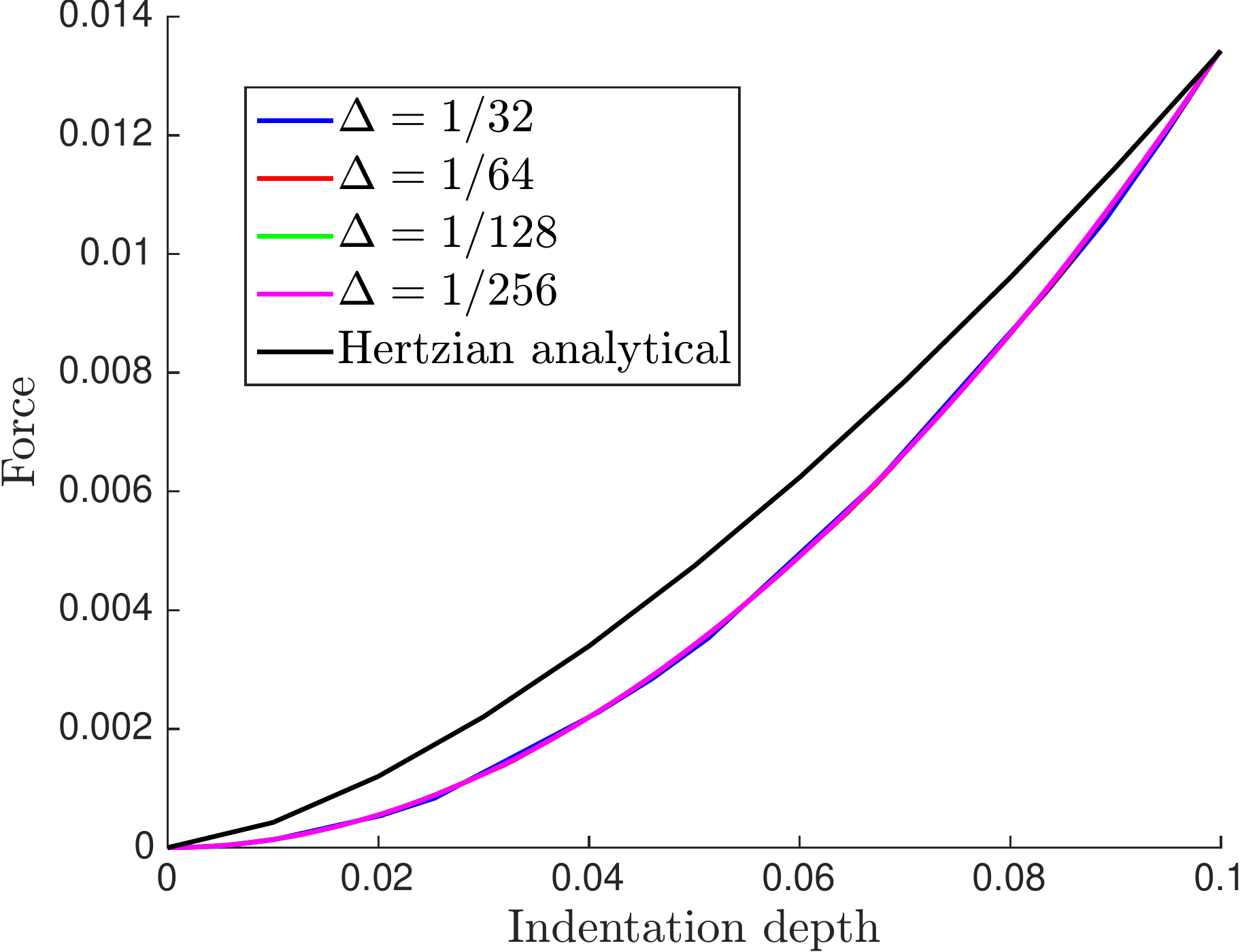}
\subcaption{}
\label{fig:forceIndentationDepthHertzianContactNoInteraction}
\end{subfigure} 
\begin{subfigure}[t]{0.49\textwidth}
  \includegraphics[width=\textwidth]{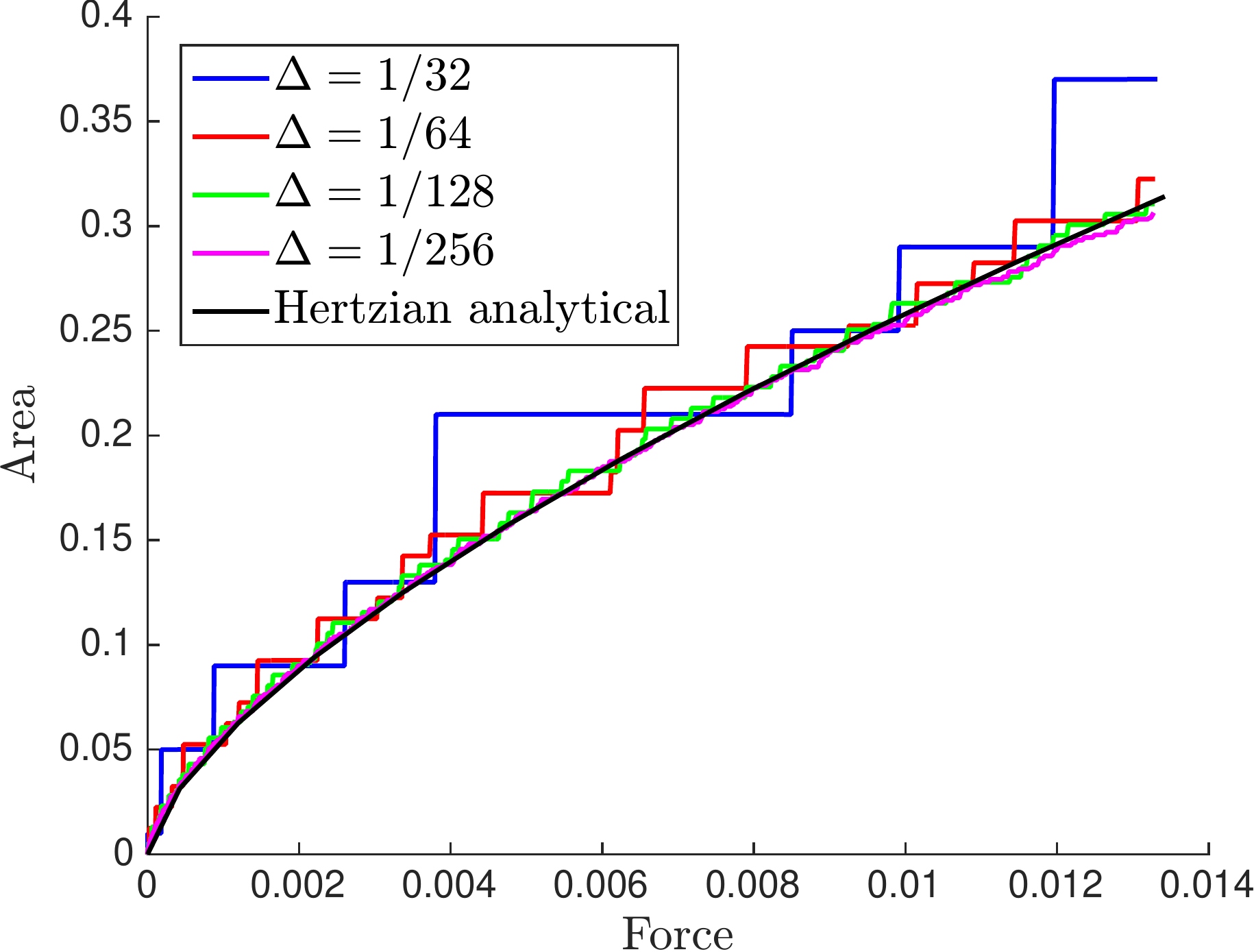}
\subcaption{}
\label{fig:areaForceHertzianContact}
\end{subfigure} 
\begin{subfigure}[t]{0.49\textwidth}
  \includegraphics[width=\textwidth]{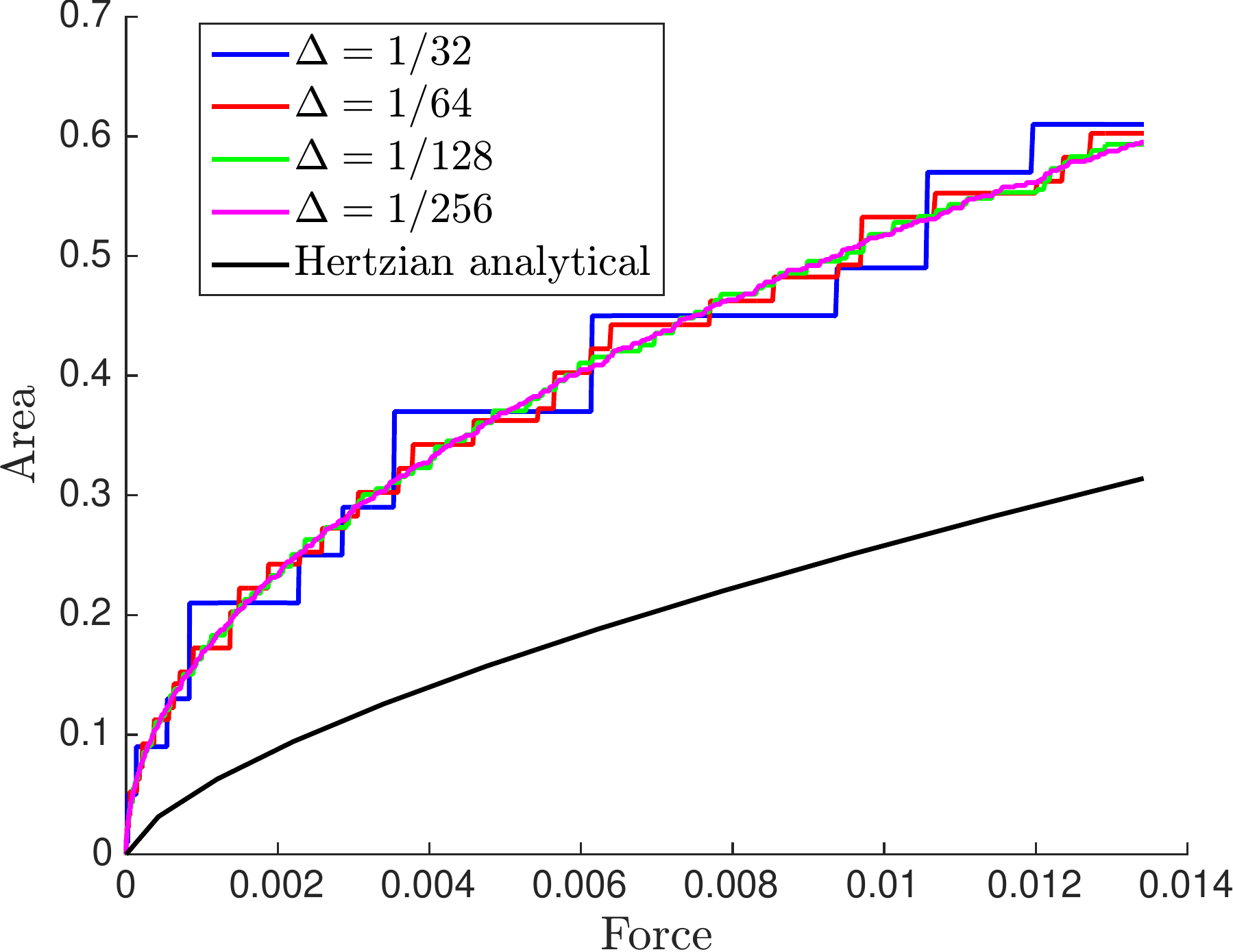}
\subcaption{}
\label{fig:areaForceHertzianContactNoInteraction}
\end{subfigure} 
\begin{subfigure}[t]{0.49\textwidth}
  \includegraphics[width=\textwidth]{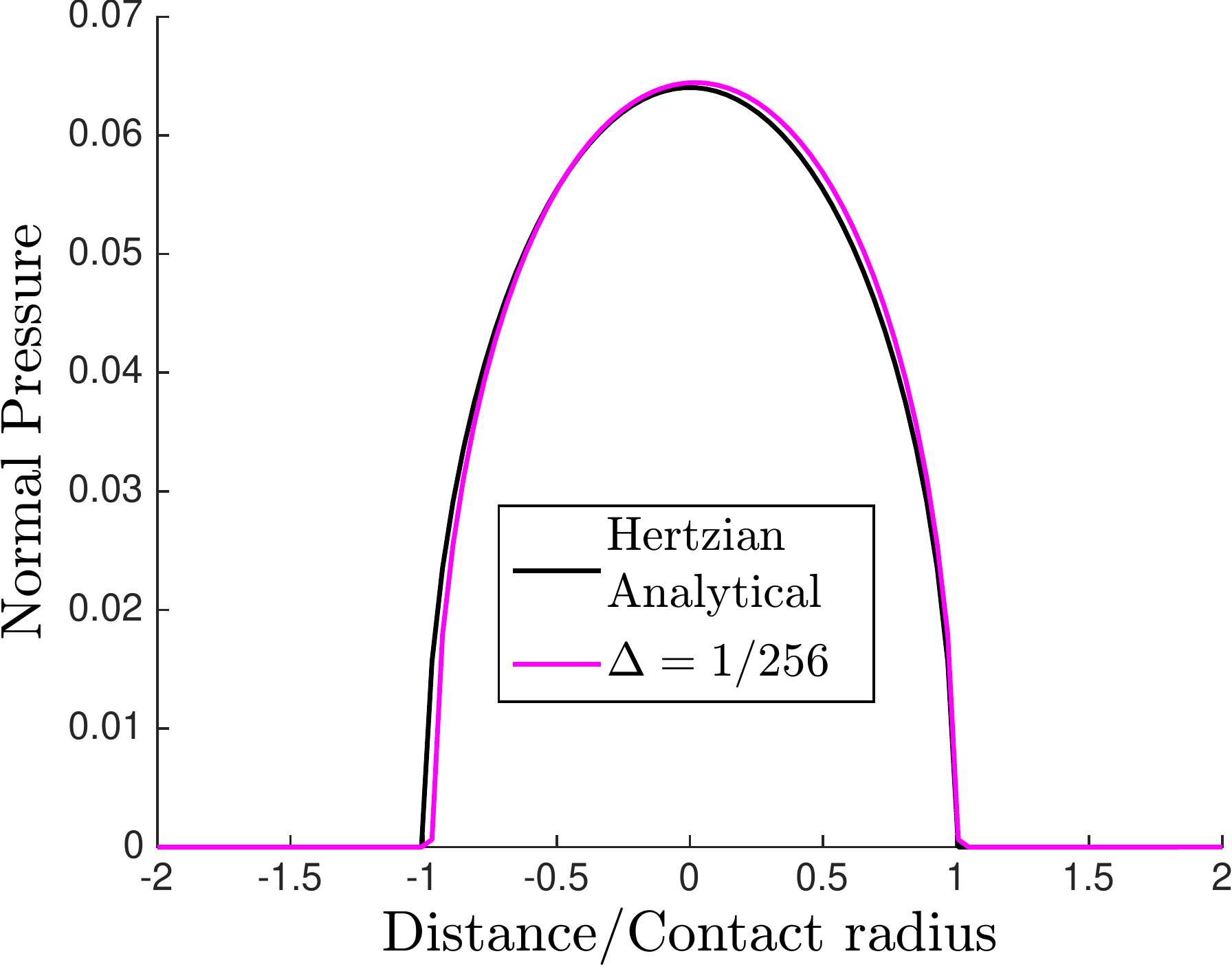}
\subcaption{}
\label{fig:pressureDistributionHertzianContact}
\end{subfigure} 
\begin{subfigure}[t]{0.49\textwidth}
  \includegraphics[width=\textwidth]{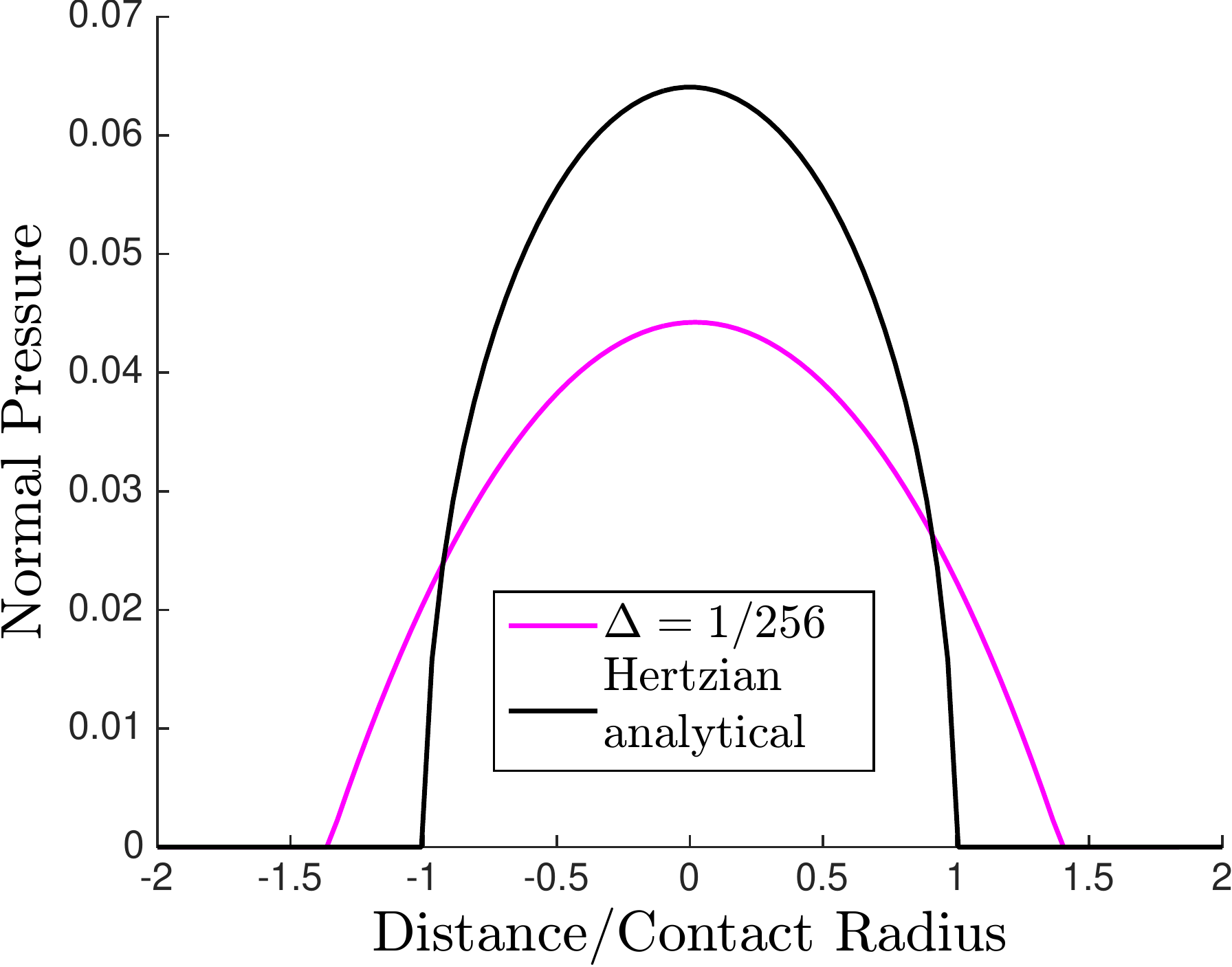}
\subcaption{}
\label{fig:pressureDistributionHertzianContactNoInteraction}
\end{subfigure} 
\caption{Comparison of analytical and numerical solutions for contact of a
linear-elastic sphere against a rigid flat: (left column) with Boussinesq
interaction, (right column) with no elastic interaction. Lines of different
colors correspond to different discretization sizes $\Delta$, the distance
between the elements.}
\label{fig:hertzianContactValidation}
\end{figure}
%%%%%%%%%%%%%%%%%%%%%%%%%%%%%%%%%%%%%%%%%%%%%%%%%%%%%%%%%%%%%%%%%%%%%%%%%%%%%%%%

In the case with Boussinesq interaction, there is an excellent match between the numerical
and analytical solutions (Figure \ref{fig:hertzianContactValidation}).
The elastic constants used in the analytical and numerical solutions are the same and
no other parameters are used in obtaining the numerical results.
For the case with no elastic interaction, $\bar{C}^0$ is chosen to make the 
force at the final indentation match the analytical solution.
However, we find that the scaling of the force and area deviates from the Hertzian solution
in this case.

\section{Static contact}\label{sec:staticContact}

We start by considering the static contact of rough surfaces. The surfaces,
initially apart, are loaded ``instantaneously" to a nominal pressure of $100$
MPa. The system is then evolved keeping the global normal force constant. The
evolution of the forces, length changes of the elements, and contact area is computed.
Let us first look at the viscoelastic case.

\subsection{Evolution of contact and force distribution}

\begin{figure}
\begin{minipage}{0.4\textwidth}
\begin{subfigure}[t]{1.0\textwidth}
\includegraphics[width=\textwidth]{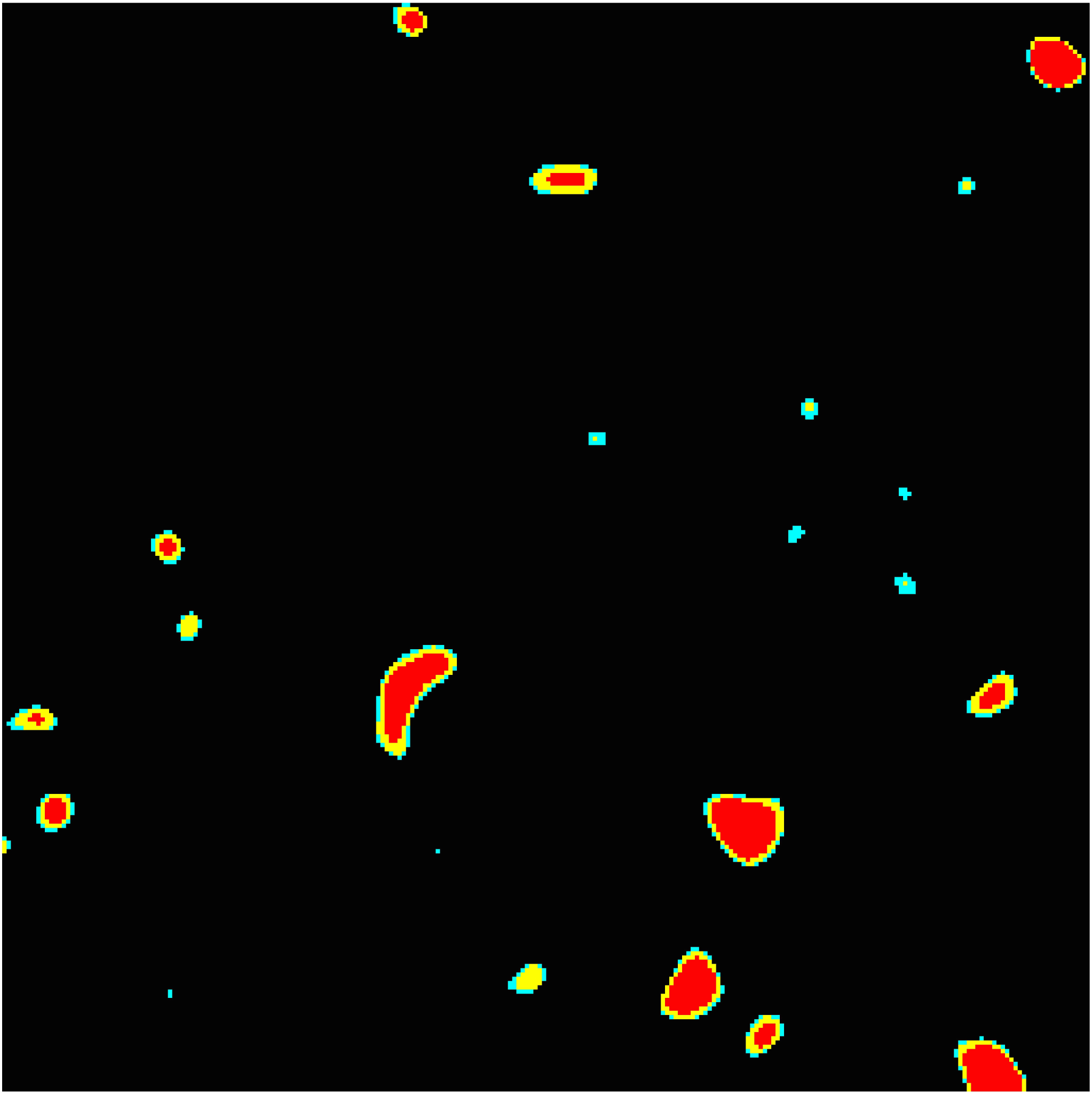}
\subcaption{}
\label{fig:contactsDieterichLikeGaussian}
\end{subfigure} 
\end{minipage}
\begin{minipage}{0.6\textwidth}
\begin{subfigure}[t]{1.0\textwidth}
\includegraphics[width=\textwidth]{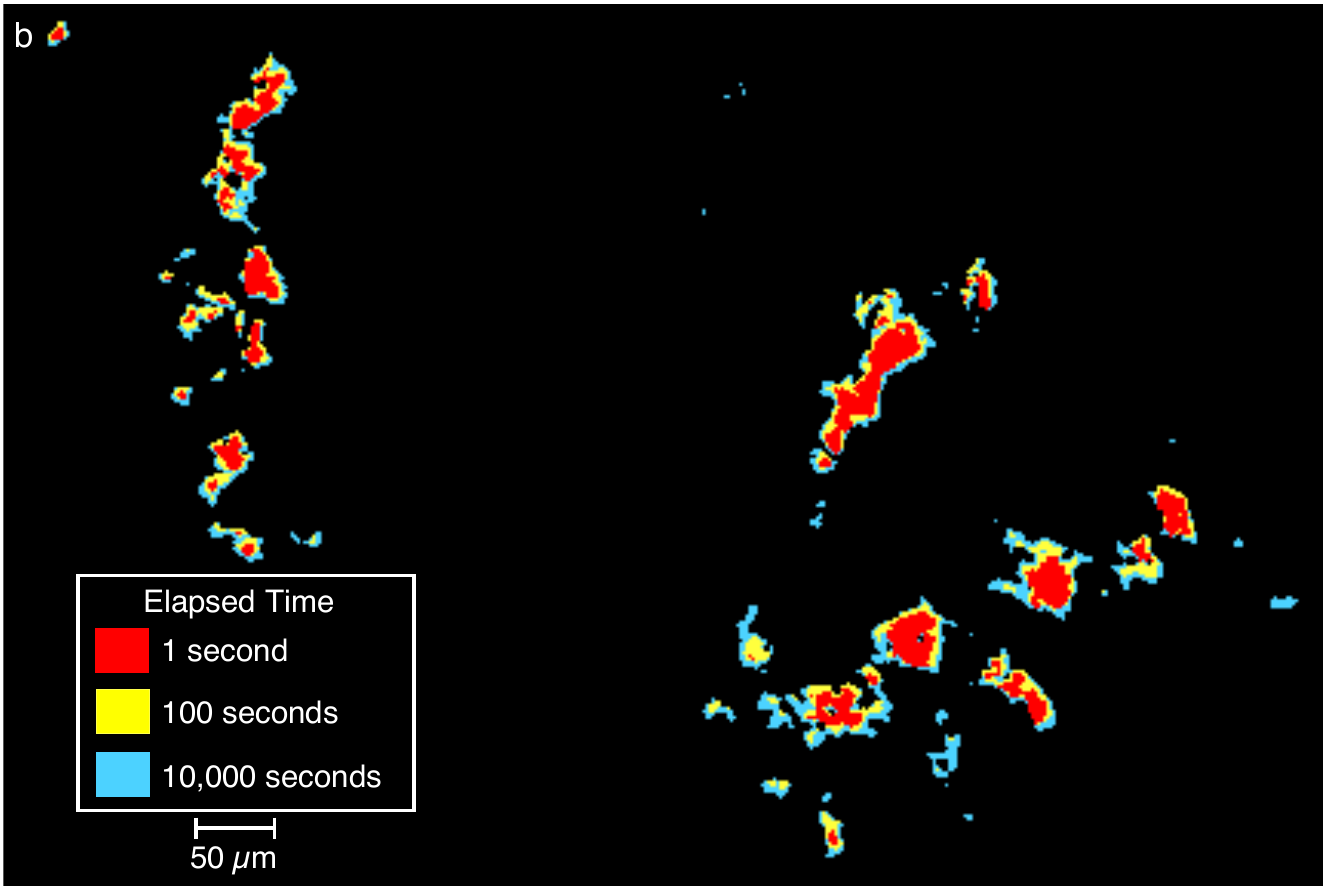}
\subcaption{}
\label{fig:contactsEvolutionStaticContactDieterichColor}
\end{subfigure} 
\end{minipage}
\begin{subfigure}[t]{0.49\textwidth}
\includegraphics[width=\textwidth]{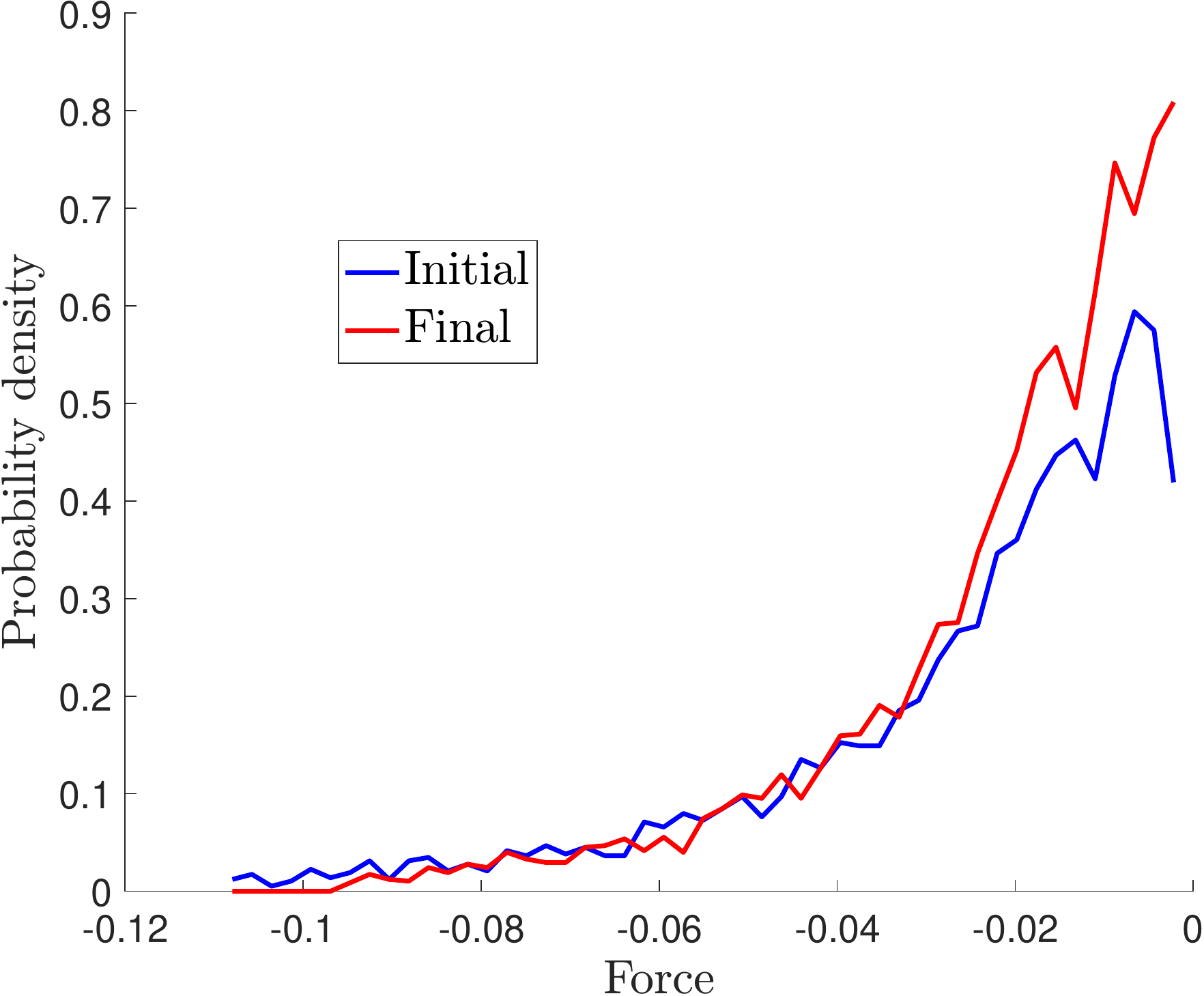}
\caption{}
\label{fig:forceDistributionInitialFinalStaticContact}
\end{subfigure} 
\begin{subfigure}[t]{0.49\textwidth}
\includegraphics[width=\textwidth]{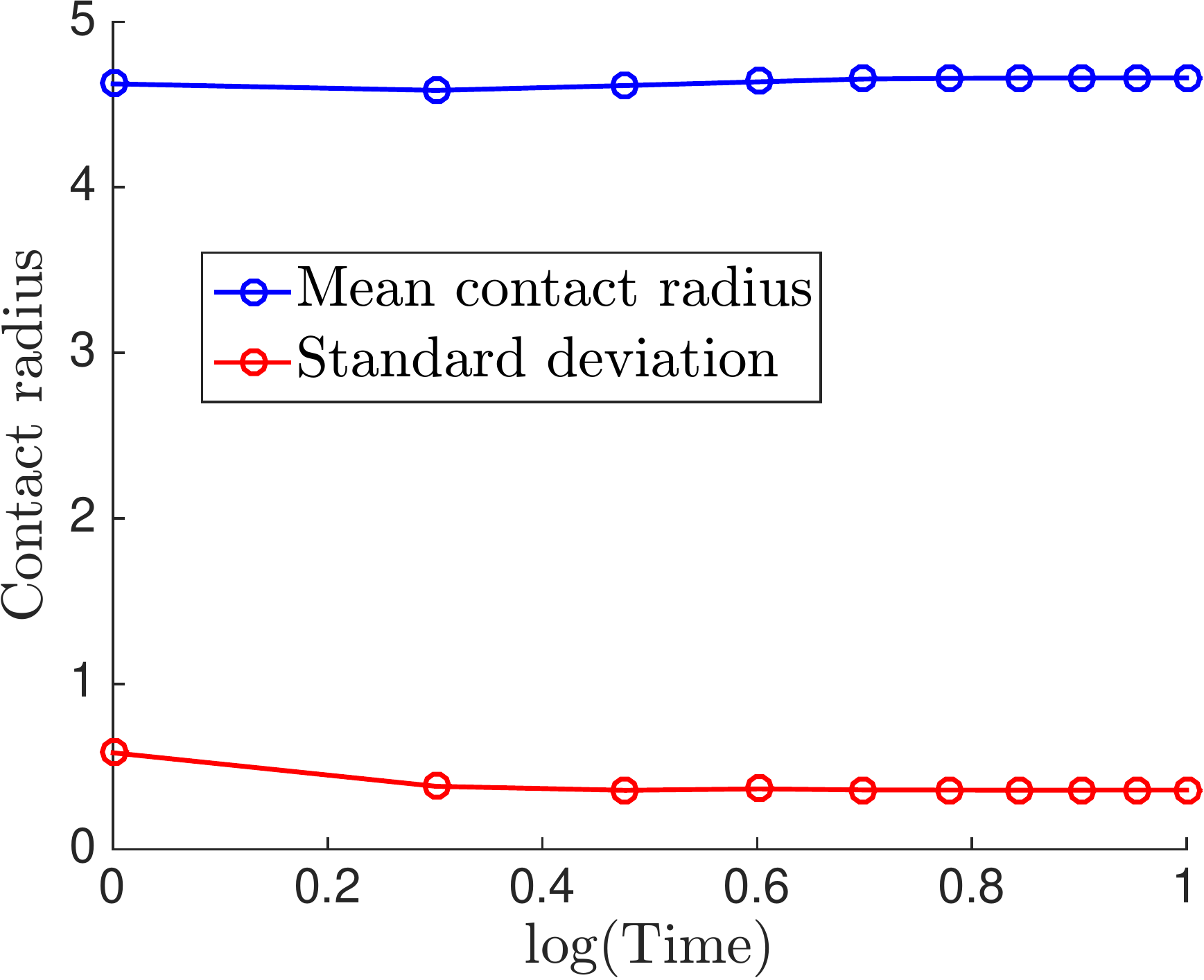}
\caption{}
\label{fig:averageContactRadiusTimeStaticContactNoInteraction}
\end{subfigure} 
\caption{Evolution of contacts, in (a) our simulations and (b) experiments
\cite{dieterich:3}, during a static contact test. In (a), contacts at 
$\bar{t} = 0$ are in red, $\bar{t} = 10$ are in yellow, and $\bar{t} = 100$ are in 
cyan. Existing contacts grow, some coalesce, and some new contacts are formed.
Panel (b) is reproduced with permission from \cite{dieterich:3}.
(c) Distribution of contact forces at the initial and final states. Initially,
the area of contact is lower but the average force on a contact is higher.
With time, as the forces relax, the number of contacts increases but the
average force per contact decreases. (d) Evolution of the average contact
size with time (the mean and standard deviation are computed across $15$
realizations of the rough surface). Even though the total contact area
increases, the average contact size does not change significantly.}
\label{fig:initialFinalContactDistributionDieterichComparison}
\end{figure}

After the initial compression, the forces at the contacts start relaxing
because of the viscoelastic behavior. To keep the global normal force constant, more
contacts are formed and the contact area increases.  As in experiments
\cite{dieterich:3}, existing contacts grow with time, some contacts coalesce,
and some new ones are formed (Figure
\ref{fig:initialFinalContactDistributionDieterichComparison}).

%%%%%%%%%%%%%%%%%%%%%%%%%%%%%%%%%%%%%%%%%%%%%%%%%%%%%%%%%%%%%%%%%%%%%%%%%%%%%%%%
%%%%%%%%%%%%%%%%%%%%%%%%%%%%%%%%%%%%%%%%%%%%%%%%%%%%%%%%%%%%%%%%%%%%%%%%%%%%%%%%

As contact forces relax and contact area increases, the force distribution
spreads and the force per unit contact area decreases (Figure
\ref{fig:forceDistributionInitialFinalStaticContact}).  The first moment of the
force distribution, which is the total normal force, is the same for initial and
final states, by the design of the numerical experiment. The zeroth moment (area under the curve), which is the total
contact area, is larger at the final state.

Even though the total contact area increases, the average contact radius,
calculated as $\sqrt{\text{Contact area}/(\pi \times \text{Number of
contacts})}$, remains nearly constant (Figure
\ref{fig:averageContactRadiusTimeStaticContactNoInteraction}).  This is
because, as the contact area increases, the number of contacts also increases
keeping the average contact radius approximately constant.  A similar
observation was made by Greenwood and Williamson in their statistical model of
elastic contacts \cite{greenwood:1} and in experiments \cite{dieterich:3}.

\subsection{Increase of contact area and friction with time}
\label{subsec:IncreaseContactStatic}

%%%%%%%%%%%%%%%%%%%%%%%%%%%%%%%%%%%%%%%%%%%%%%%%%%%%%%%%%%%%%%%%%%%%%%%%%%%%%%%%
\begin{figure}
\centering
\begin{subfigure}[t]{0.49\textwidth}
\includegraphics[width=\textwidth]{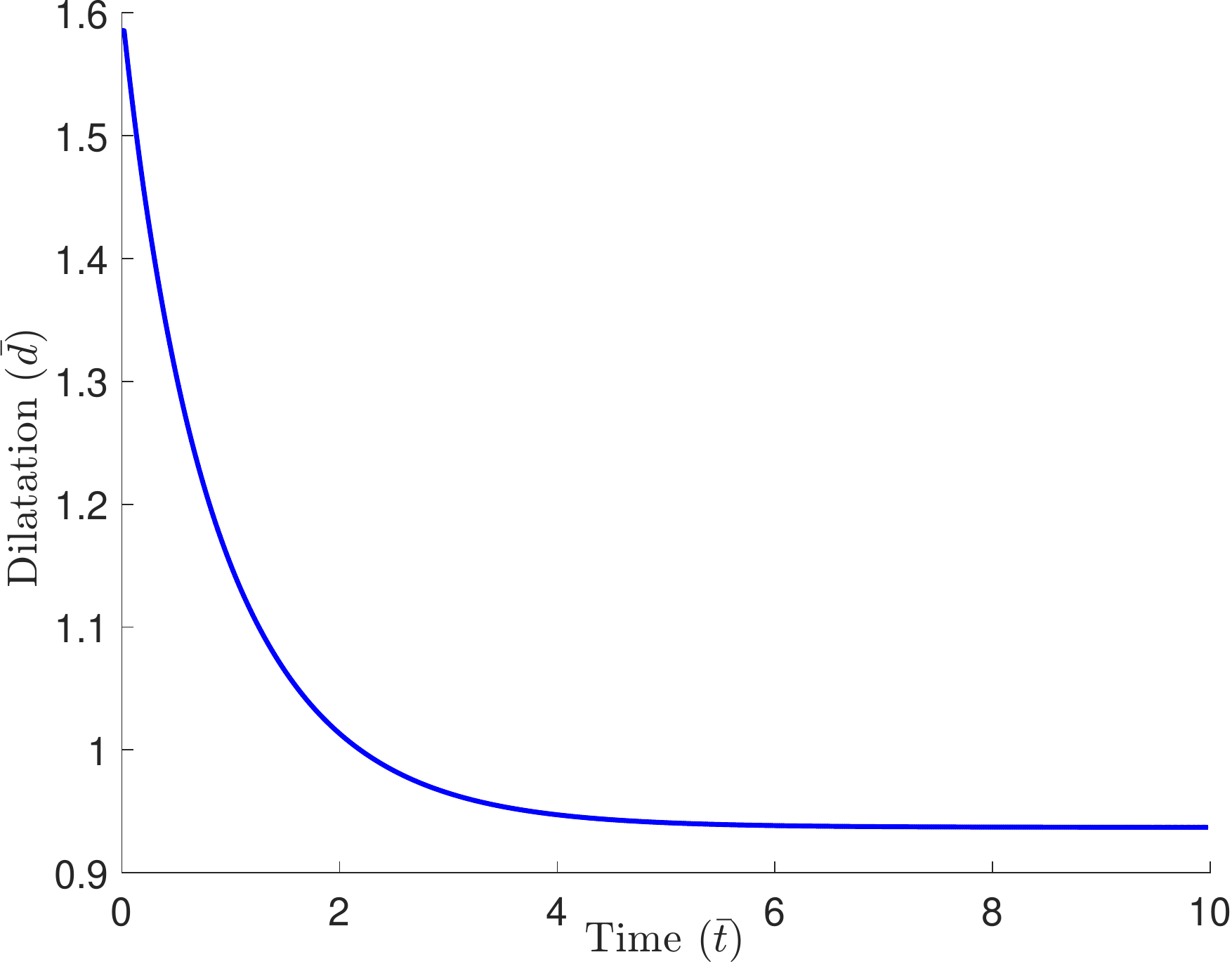}
\subcaption{}
\label{fig:dilatationTimeStaticContact}
\end{subfigure} 
\begin{subfigure}[t]{0.49\textwidth}
\includegraphics[width=\textwidth]{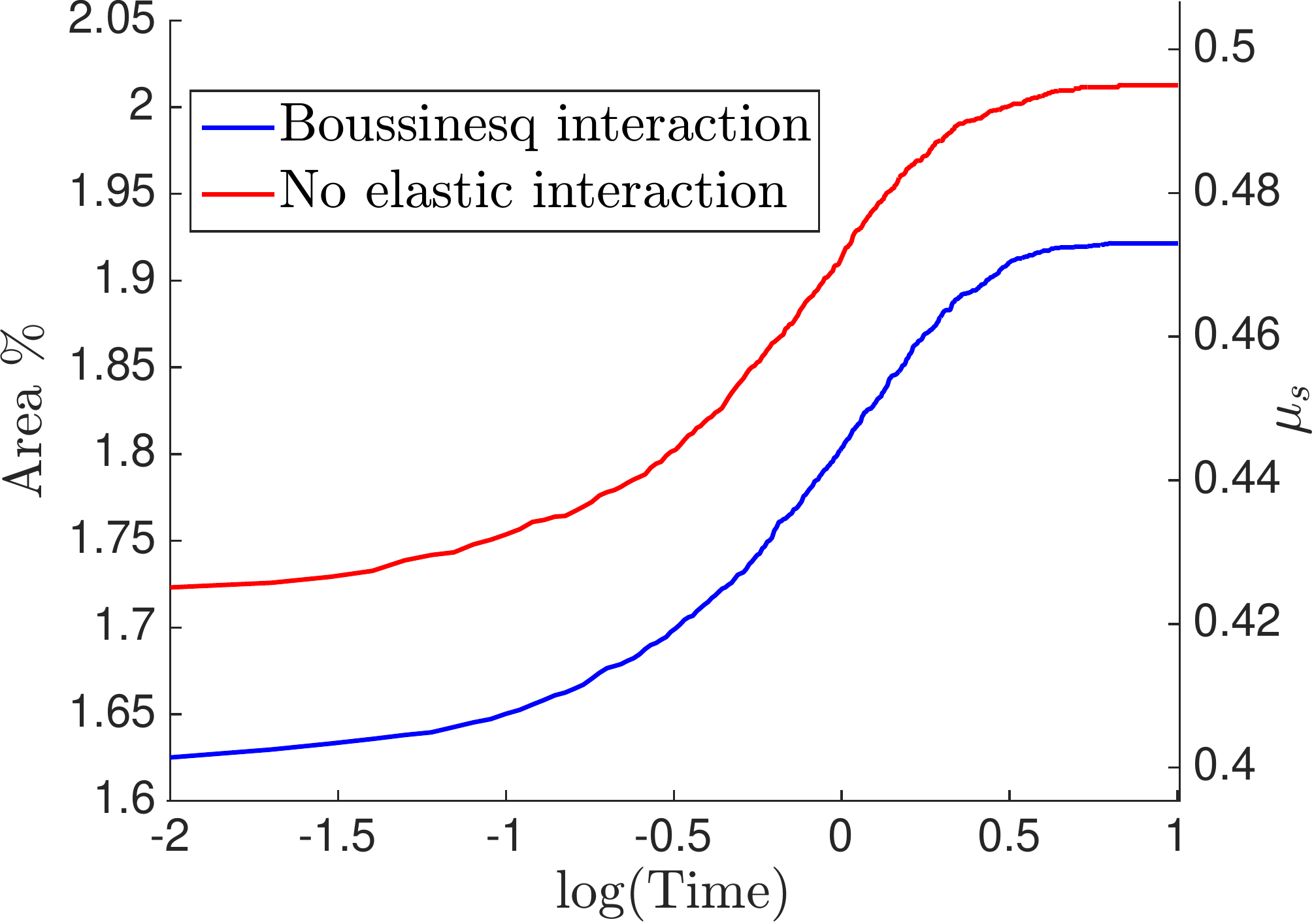}
\subcaption{}
\label{fig:areaFrictionTimeStaticContactBoussinesqNoInteraction}
\end{subfigure} 
\begin{subfigure}[t]{0.49\textwidth}
  \includegraphics[width=\textwidth]{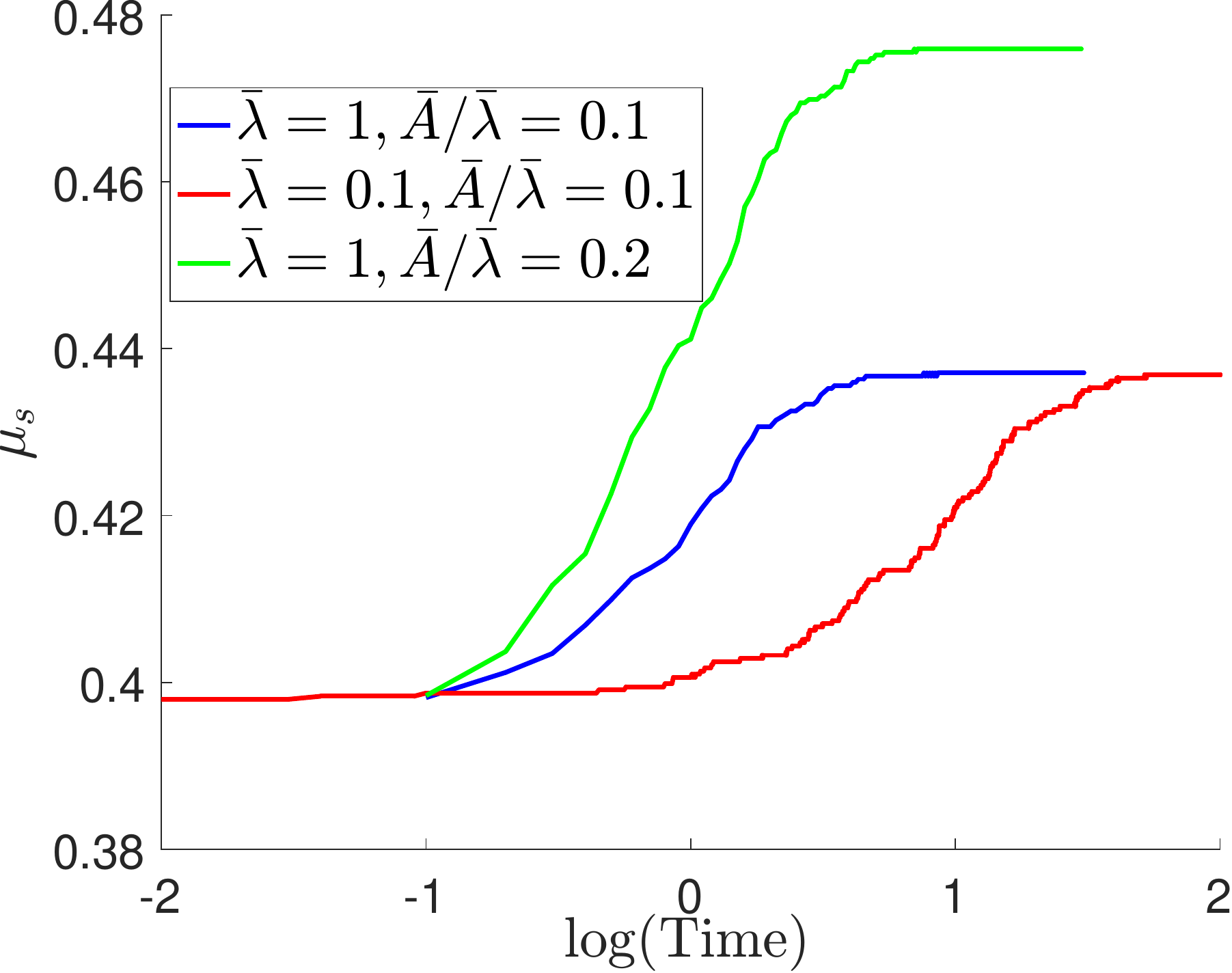}
\subcaption{}
\label{fig:frictionTimeStaticContactViscoelasticityDependence}
\end{subfigure} 
\begin{subfigure}[t]{0.49\textwidth}
\includegraphics[width=\textwidth]{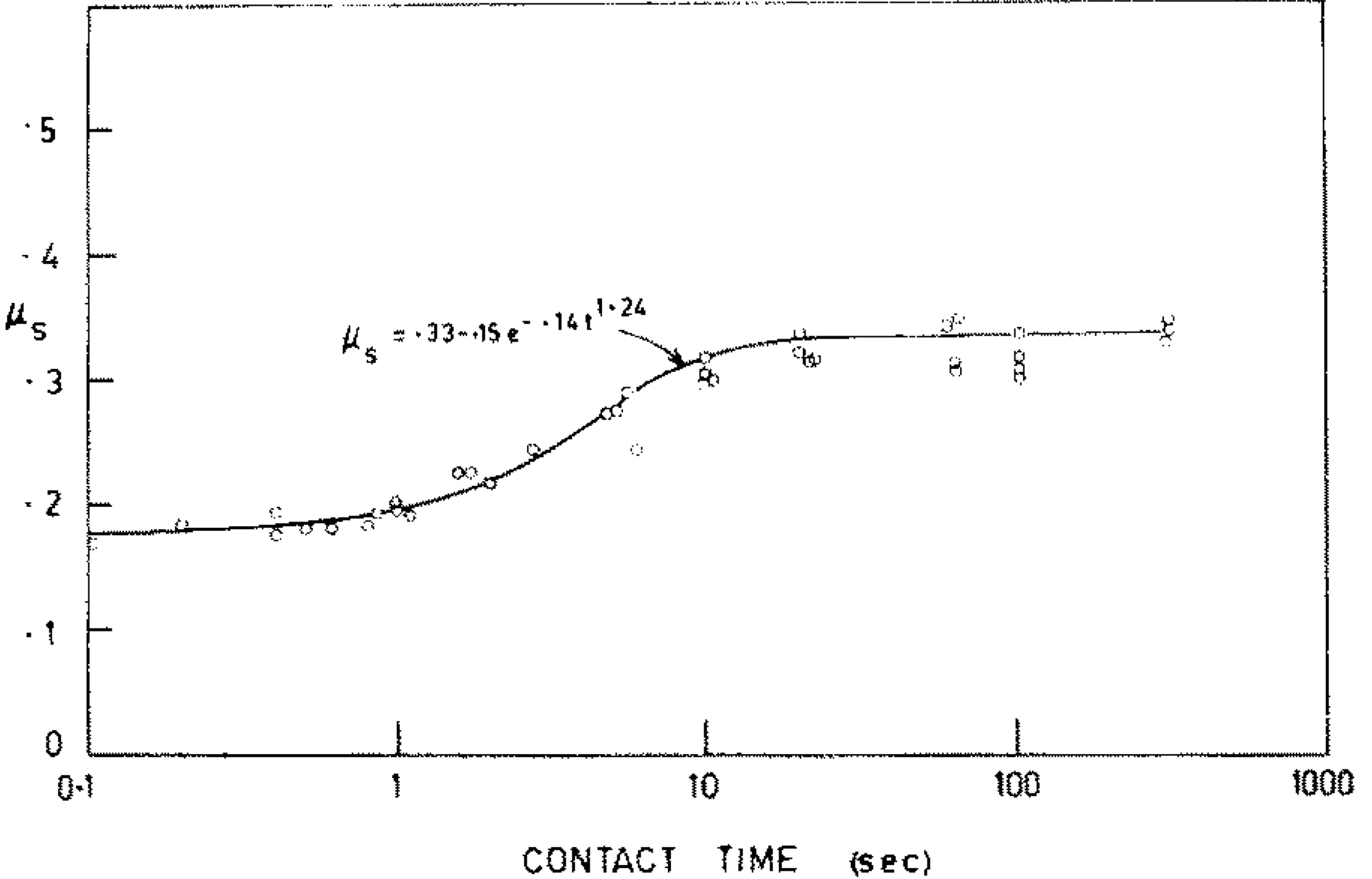}
\subcaption{}
\label{fig:staticFrictionEvolutionRichardsonTolle}
\end{subfigure} 
\caption{Evolution of (a) dilatation and (b) contact-area/friction-coefficient
with the time of contact. Because of viscoelastic relaxation, the surfaces move
closer and the area of contact increases with time. The timescale and the
magnitude of evolution are unchanged by the presence of long-range Boussinesq
interactions. (c) Evolution of friction coefficient for three different
combinations of viscoelastic parameters. The timescale of evolution is
determined by $\bar{\lambda}$ and the difference between initial and final
states is determined by the ratio $\bar{A}/\bar{\lambda}$. For some duration, the growth is logarithmic in time and saturates to a steady
state at long times. (d) Evolution of friction coefficient in experiments of
Richardson et al. (reproduced with permission from \cite{richardsonRSH:1}) showing similar behavior.}
\label{fig:dilatationAreaEvolutionStaticContactBoussinesqNoInteraction}
\end{figure}
%%%%%%%%%%%%%%%%%%%%%%%%%%%%%%%%%%%%%%%%%%%%%%%%%%%%%%%%%%%%%%%%%%%%%%%%%%%%%%%%

As the individual contact forces relax, dilatation decreases (the surfaces move
closer to each other) and the contact area increases (Figure
\ref{fig:dilatationTimeStaticContact}). Consequently, the static friction
coefficient always increases with time (Figure
\ref{fig:areaFrictionTimeStaticContactBoussinesqNoInteraction}) and this has
been universally observed in experiments
\cite{rabinowiczE:1,dieterich:1,richardsonRSH:1}. In Figure
\ref{fig:areaFrictionTimeStaticContactBoussinesqNoInteraction}, there is a
logarithmic growth phase that lasts about $3$ decades in time. If we have only
one relaxation time, then saturation time is inversely proportional to
$\lambda$ as can be seen by from the constitutive equations
(\ref{eq:noInteractionViscoelastic}) and (\ref{eq:boussinesqViscoelastic}). If
$\{u(t,\lambda),F(t,\lambda)\}$ is a solution, then so is $\{u(n t, \lambda/n),
F(n t, \lambda/n)\}$.  In some materials like mild steel, a similar duration of
static friction growth with saturation at small and long times is observed
\cite{richardsonRSH:1}, as shown in Figure
\ref{fig:staticFrictionEvolutionRichardsonTolle}.  Thus, our results are
consistent with these experimental observations.

%%%%%%%%%%%%%%%%%%%%%%%%%%%%%%%%%%%%%%%%%%%%%%%%%%%%%%%%%%%%%%%%%%%%%%%%%%%%%%%%
\begin{figure}
\centering
\includegraphics[scale=0.5]{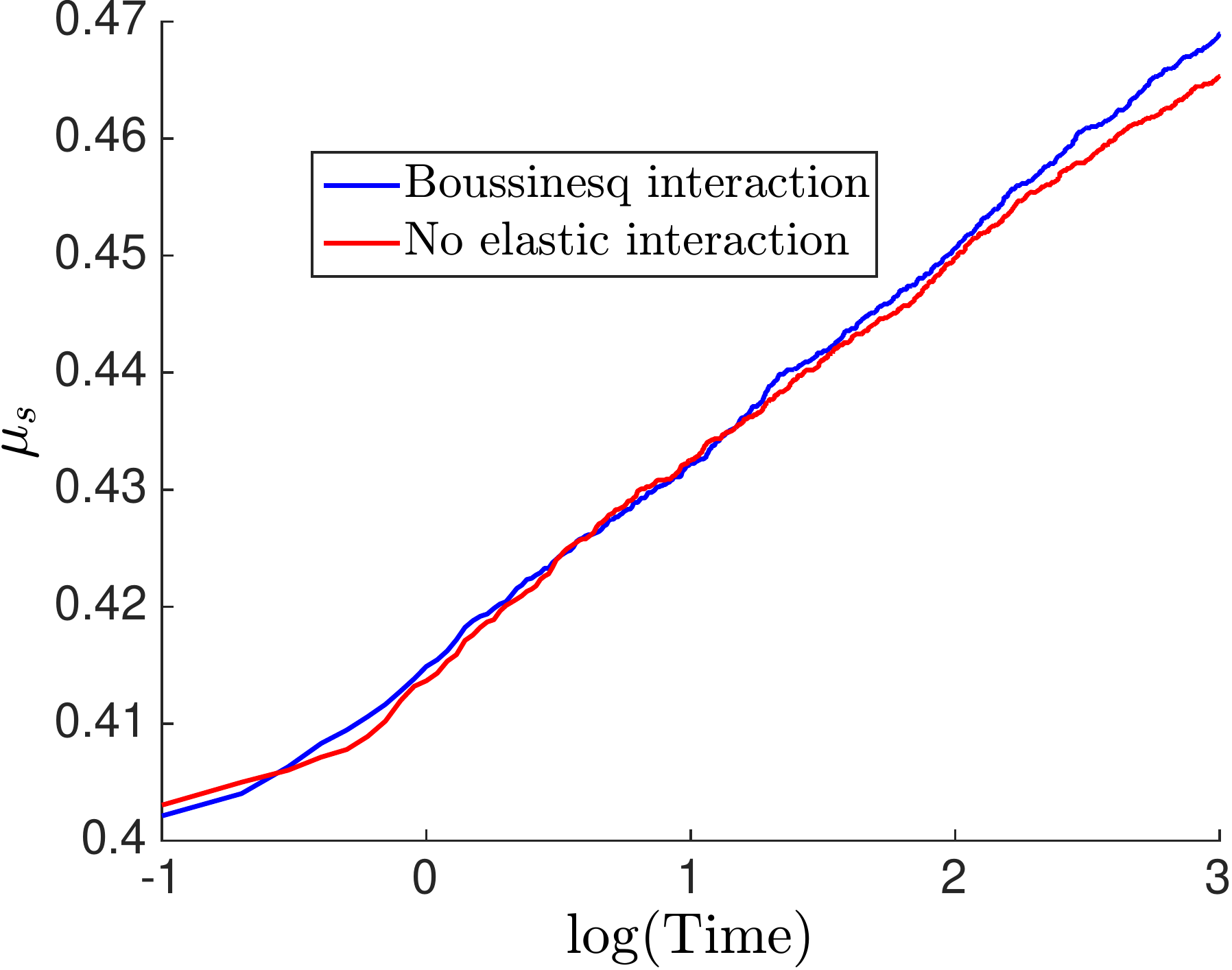}
\caption{Evolution of friction coefficient during static contact for a material
with four viscoelastic timescales. The region of logarithmic growth lasts over 
$4$ decades, as seen in experiments on rocks.}
\label{fig:frictionTimeStaticContactMultipleDecadesBoussinesqNoInteraction}
\end{figure}
%%%%%%%%%%%%%%%%%%%%%%%%%%%%%%%%%%%%%%%%%%%%%%%%%%%%%%%%%%%%%%%%%%%%%%%%%%%%%%%%

In experiments on rocks, the logarithmic growth persists throughout the
duration of the experiments, some of which have lasted up to six decades in time
\cite{dieterich:1,dieterich:2}.   We find that there are two possible reasons.
 
\paragraph{Multiple relaxation times lead to an extended period of increase.}

The simulations above considered a constitutive relation with a single
relaxation time.  However, common materials have multiple relaxation times.
Longer timescales of relaxation lead to longer times of growth in contact area
and hence friction. To study the dependence of friction evolution of the
viscoelastic properties, we perform static contact simulations for three
different combinations of the parameters $\bar{\lambda}$ and $\bar{A}$ (Figure
\ref{fig:frictionTimeStaticContactViscoelasticityDependence}). The blue and
green curves (both have $\bar{\lambda} = 1$) reach steady state at about the
same time but the magnitude of the change in friction is different. The blue
and red curves have different $\bar{\lambda}$ and thus reach steady state at
different times.  However, they have the same $\bar{A}/\bar{\lambda}$ and thus
the steady-state friction coefficient is the same. This tells us that the
timescale of evolution of area and friction is determined by $\bar{\lambda}$
while the steady state stiffness (and the magnitude of the difference between
the initial and final states) is determined by the ratio
$\bar{A}/\bar{\lambda}$. The initial value of friction is determined by the
instantaneous stiffness of the system (stiffness corresponding to fast loading
rates) and is hence independent of the viscoelastic properties.  If we thus
have a material with multiple relaxation times, the growth in friction will
persist over times corresponding to the longest relaxation time. This is
confirmed in Figure
\ref{fig:frictionTimeStaticContactMultipleDecadesBoussinesqNoInteraction} where
the material has four viscoelastic relaxation timescales ($N_T = 4,
\bar{\lambda}_k=1,0.1,0.01,0.001, \bar{A}_k = 0.05,0.005,0.0005,0.00005$). The
linear growth regime of friction now extends over four decades in time.

%
%
%Since $\mu_s$ cannot increase indefinitely, it
%eventually has to reach a steady state. This delayed saturation is not captured
%by our model. We conjecture that the difference between our model response and
%the rock experiments is for the following reason. 

\begin{figure}
\centering
  \includegraphics[scale=0.5]{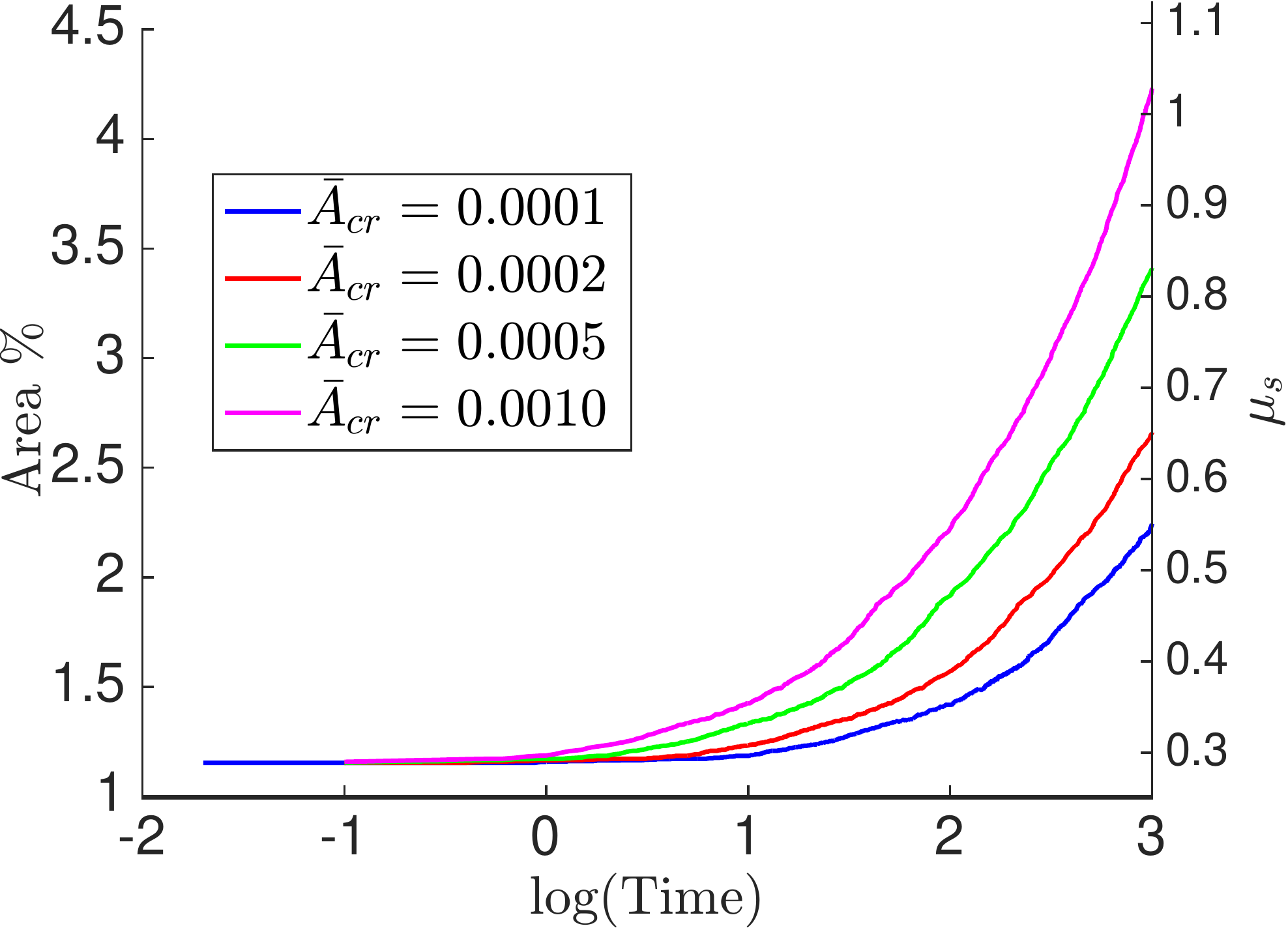}
\label{fig:areaFrictionTimeStaticContactViscoplasticCreepRateDependence}
\caption{ Evolution of contact area and static friction coefficient with time
for a viscoplastic material. $\mu_s$ grows logarithmically with time with no
signs of saturation, as seen in many experiments.}
\label{fig:areaFrictionViscoplasticStaticContact}
\end{figure}

\paragraph{Viscoplasticity also leads to continued increase.} 
We now repeat the simulations assuming 
the viscoplastic constitutive relation (Section \ref{subsubsec:evp}).
As in the viscoelastic case, the area of
contact and thus $\mu_s$ increase with the time of contact and, after an
initial phase, grows logarithmically with time (Figure
\ref{fig:areaFrictionViscoplasticStaticContact}).  In the viscoelastic case,
the growth saturates after about $3$ decades. However, here the growth seems to
continue indefinitely and shows no signs of saturation.  This is expected from
the viscoplastic model without hardening, since the contacts continue to creep under any nonzero
force.  Figure
\ref{fig:areaFrictionViscoplasticStaticContact} also shows the dependence of
the friction evolution on the creep rate $\bar{A}_{cr}$. Larger creep rates
lead to a faster and larger growth in $\mu_s$ (also see \cite{brechetY:1}).

\paragraph{Elastic interactions do not change qualitative behavior.}
We find that that static friction evolution with and without long-range elastic
interactions is qualitatively the same (Figures
\ref{fig:areaFrictionTimeStaticContactBoussinesqNoInteraction} and
\ref{fig:frictionTimeStaticContactMultipleDecadesBoussinesqNoInteraction}). The
elastic interactions do not affect either the duration or the magnitude of
friction growth. 
%%%%%%%%%%%%%%%%%%%%%%%%%%%%%%%%%%%%%%%%%%%%%%%%%%%%%%%%%%%%%%%%%%%%%%%%%%%%%%%%
%%%%%%%%%%%%%%%%%%%%%%%%%%%%%%%%%%%%%%%%%%%%%%%%%%%%%%%%%%%%%%%%%%%%%%%%%%%%%%%%

\subsection{Friction coefficient is nearly independent of the normal force} 
\begin{figure}
\centering
\includegraphics[scale=0.5]{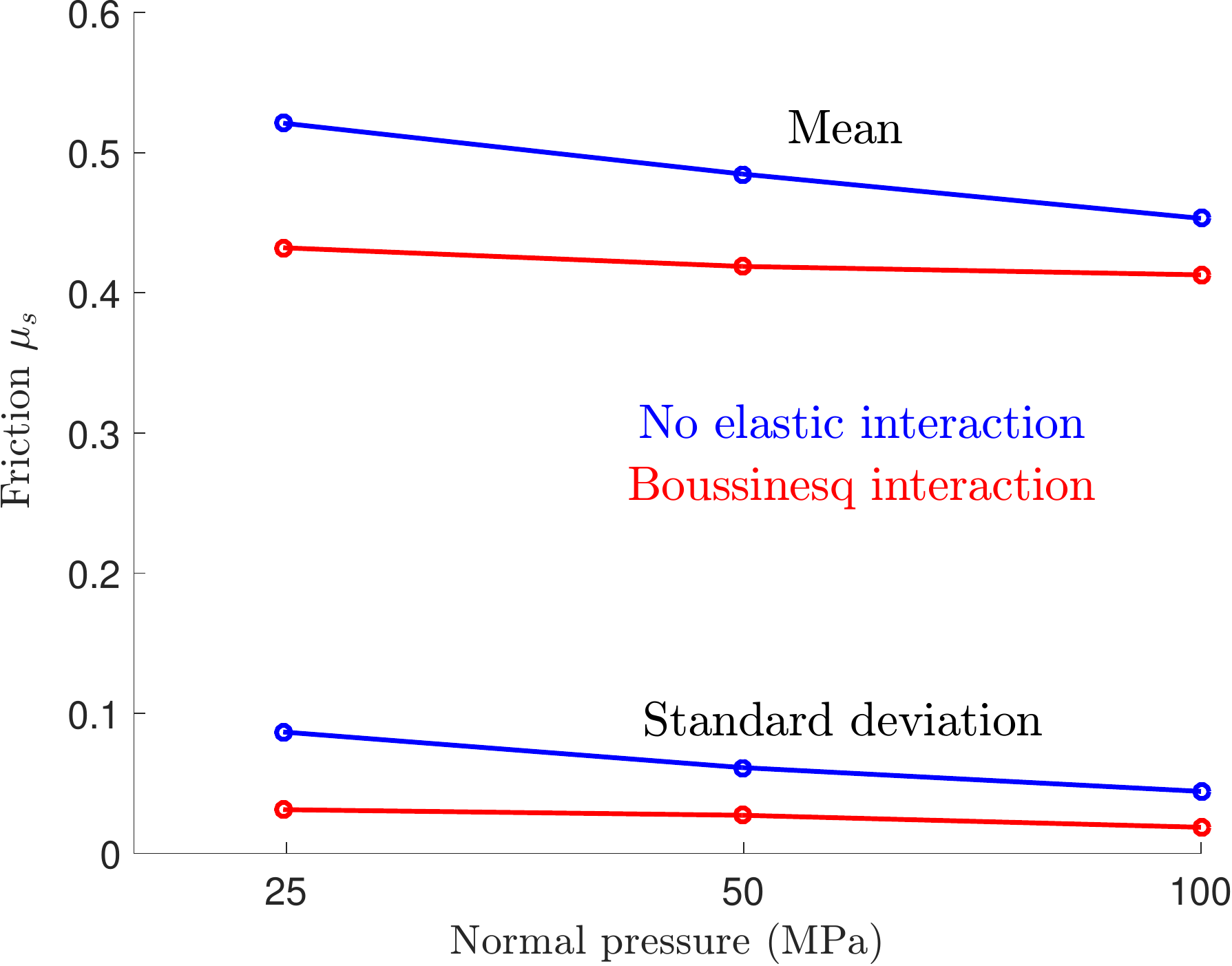}
\caption{Dependence of friction coefficient on normal pressure for three system
sizes. The mean and the standard deviation for $15$ realizations is shown.
Considering the mean and the standard deviation, the friction coefficient is
independent of both the system size and the applied nominal pressure.}
\label{fig:frictionNormalPressureWithAndWithoutInteraction}
\end{figure}

The friction coefficient is known to be independent of the normal force for
macroscopic rough surfaces (Amontons law).  In the Bowden and Tabor model, this
is explained by the plasticity of contacts \cite{bowdenFP:1}, whereas in the
Greenwood-Williamson (GW) model, this is a result of the statistics of the
rough surface \cite{greenwood:1}.  The GW model ignores interactions between
contacts which might be important.  Our simulations suggest that both
with and without elastic interactions, the friction coefficient is nearly
independent of the applied normal pressure (Figure
\ref{fig:frictionNormalPressureWithAndWithoutInteraction}).

%Since the elastic interactions are long-range, the friction coefficient can be
%dependent on the system size. Surprisingly, we find the friction coefficient to
%be nearly independent of the system size  (Figure
%\ref{fig:frictionNormalPressureWithAndWithoutInteraction}).

%%%%%%%%%%%%%%%%%%%%%%%%%%%%%%%%%%%%%%%%%%%%%%%%%%%%%%%%%%%%%%%%%%%%%%%%%%%%%%%%
%%%%%%%%%%%%%%%%%%%%%%%%%%%%%%%%%%%%%%%%%%%%%%%%%%%%%%%%%%%%%%%%%%%%%%%%%%%%%%%%

\section{Sliding contact}\label{sec:slidingContact}

Let us move on to the sliding contact of rough surfaces. The results presented
here are for a $512\times512$ deformable rough surface sliding on a larger $512
\times 1024$ rigid flat surface. We conduct two kinds of numerical
experiments.  First, to study sliding at a given sustained velocity $\bar{v}$,
we compress two rough surfaces to a nominal pressure of $100$ MPa and, starting
at the same initial state, slide at $\bar{v}$ until the contact area and
macroscopic friction reach steady state (sections
\ref{subsec:evolutionOfContacts}-\ref{subsec:AreaVelocity}). 
Note that the correlation length for our rough surface is of the order of 10, and
thus we achieve steady state when we slide over distances of a few times
the correlation length.  Then, we conduct
velocity jump experiments (sections
\ref{subsec:velocityJump}-\ref{subsec:elastoviscoplastic}).

\subsection{Evolution of contacts}\label{subsec:evolutionOfContacts}

\begin{figure}
\centering
\includegraphics[scale=0.49]{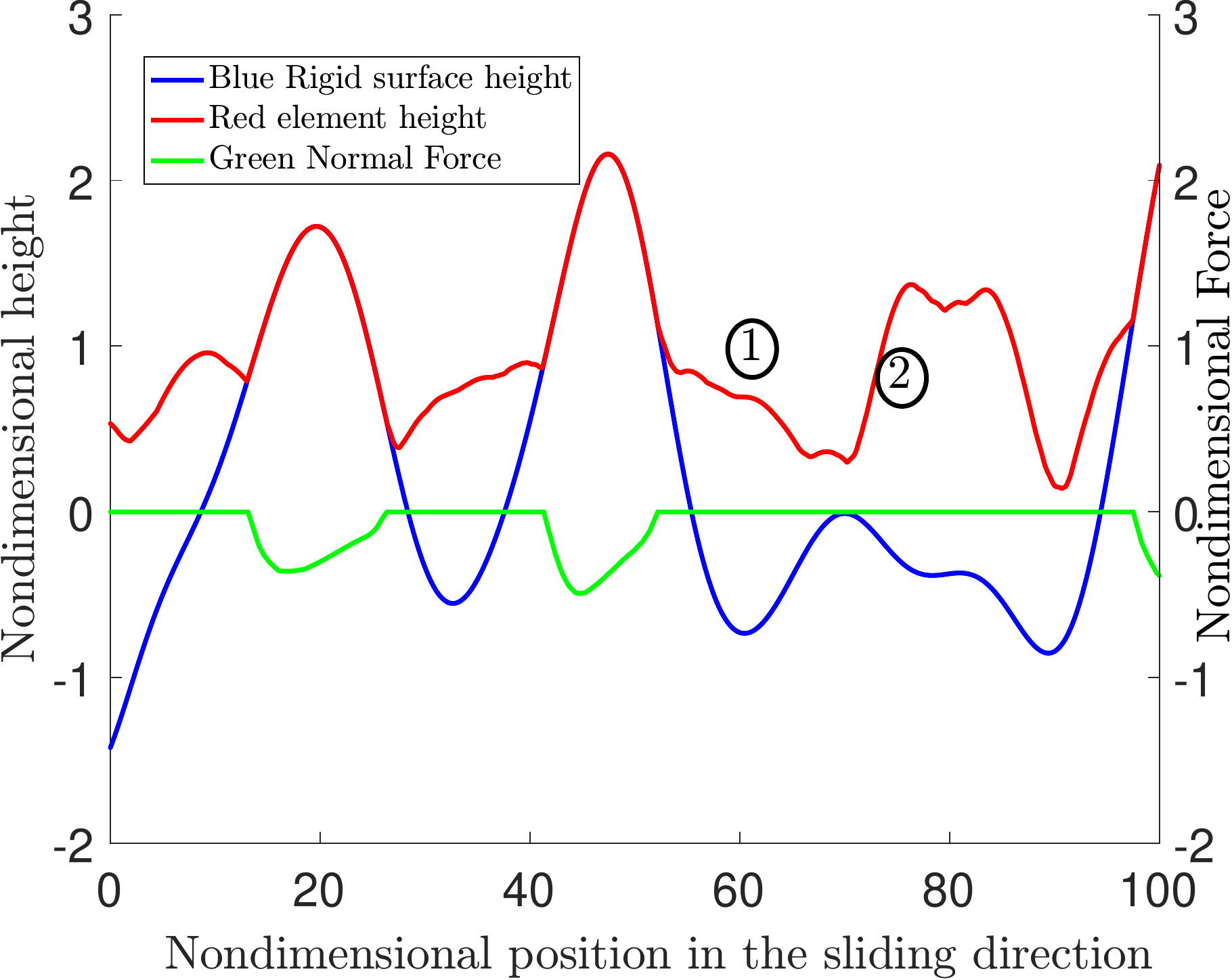}
\caption{Evolution of a single element as it slides along a rough surface in a
simulation with viscoelastic properties and long-range interactions.  The
element is initially out of contact and the force is zero. As it slides, it
repeatedly comes into and goes out of contact with the rigid surface and
correspondingly, the force on it also evolves.  The element also gets perturbed
by the overall dilatation of the sliding surface.}
\label{fig:singleSliderEvolution}
\end{figure}

As a deformable element slides on the rigid rough surface, its length and
normal force evolve depending on the surface profile it encounters, as well on
the long-range interactions with other elements. The evolution of a typical
element during sliding is shown in Figure \ref{fig:singleSliderEvolution}. The
element repeatedly comes into and goes out of contact with the surface. When
not in contact, its normal force is zero, and the change in its length depends
on its viscoelastic relaxation as well as on the evolution of the global
dilatation of the surface, to keep the global normal force constant.  For
example, the wiggles of the element height in the region marked as $1$ are from
the variations in the global dilatation (to keep the normal force constant). In
the region marked $2$, the element height rapidly increases, because of a rapid
change in the global dilatation. 

%%%%%%%%%%%%%%%%%%%%%%%%%%%%%%%%%%%%%%%%%%%%%%%%%%%%%%%%%%%%%%%%%%%%%%%%%%%%%%%%
%%%%%%%%%%%%%%%%%%%%%%%%%%%%%%%%%%%%%%%%%%%%%%%%%%%%%%%%%%%%%%%%%%%%%%%%%%%%%%%%

\begin{figure}
\begin{minipage}{0.5\textwidth}
\begin{subfigure}[t]{1.0\textwidth}
\includegraphics[width=\textwidth]{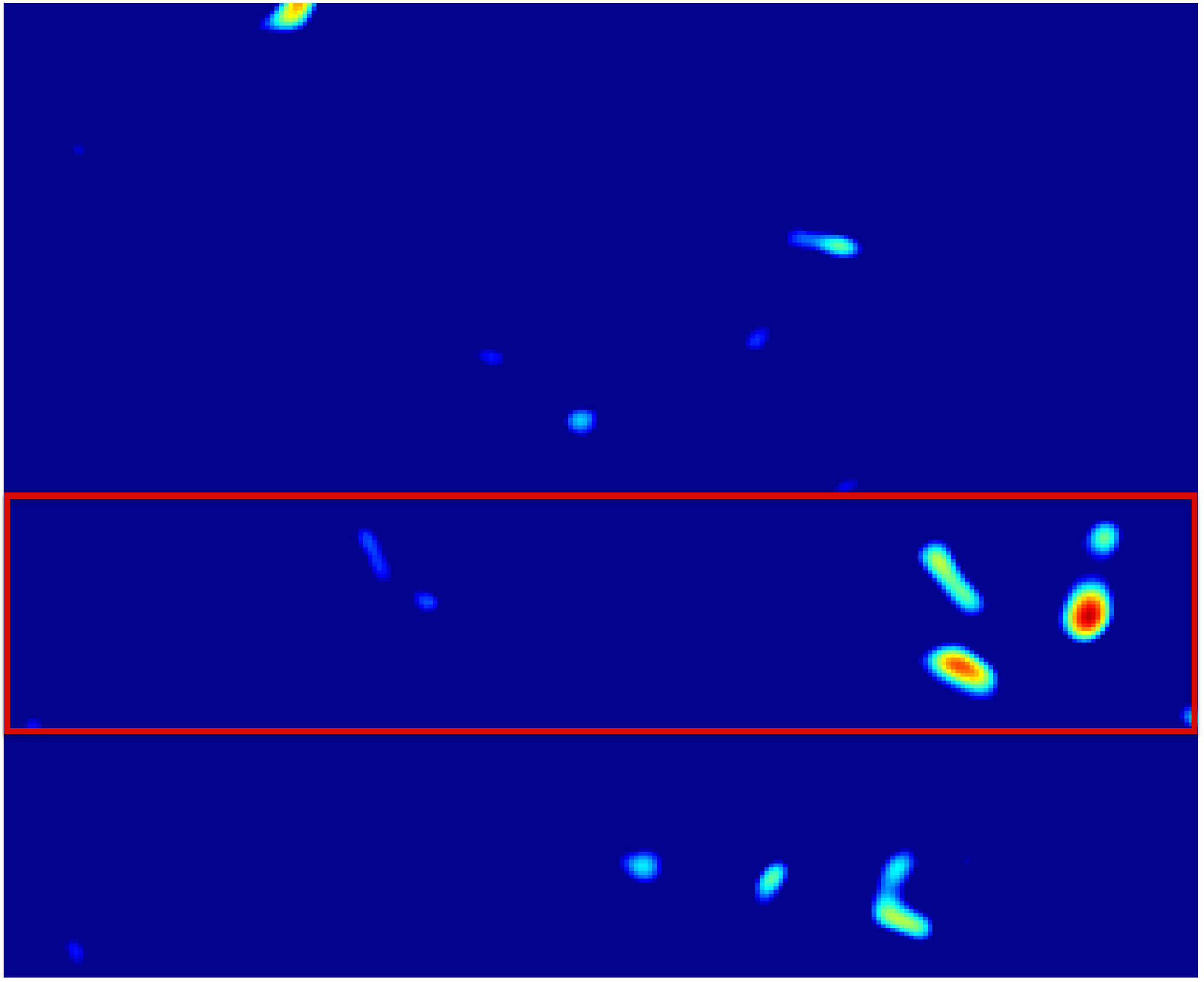}
\subcaption{}
\label{fig:t_001_fullFrame}
\end{subfigure} 
\begin{subfigure}[t]{1.0\textwidth}
\includegraphics[width=\textwidth]{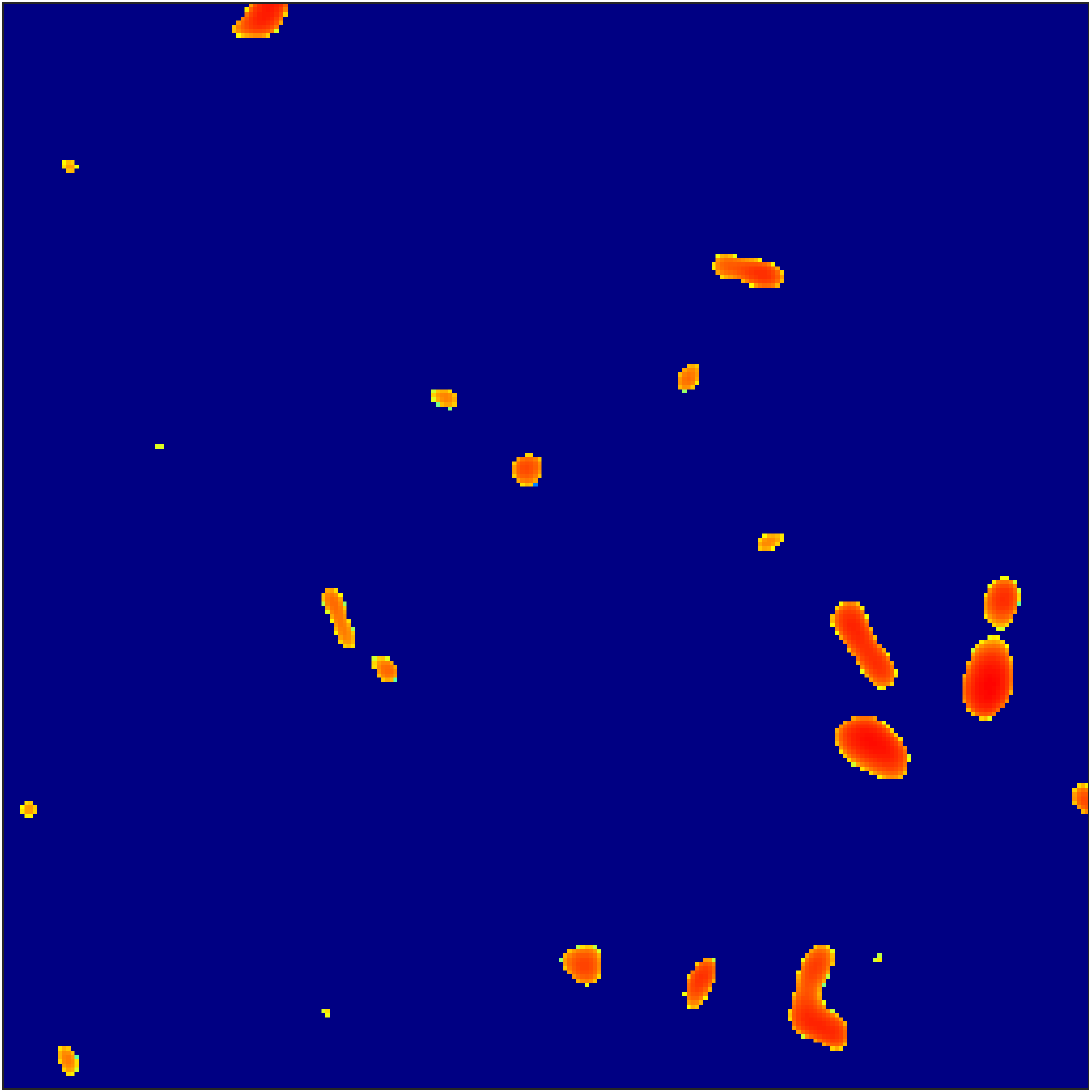}
\subcaption{}
\label{fig:t_001_fullFrame_logarithmic}
\end{subfigure} 
\end{minipage}
\begin{minipage}{0.5\textwidth}
\centering
\begin{subfigure}[t]{1.0\textwidth}
\includegraphics[width=\textwidth]{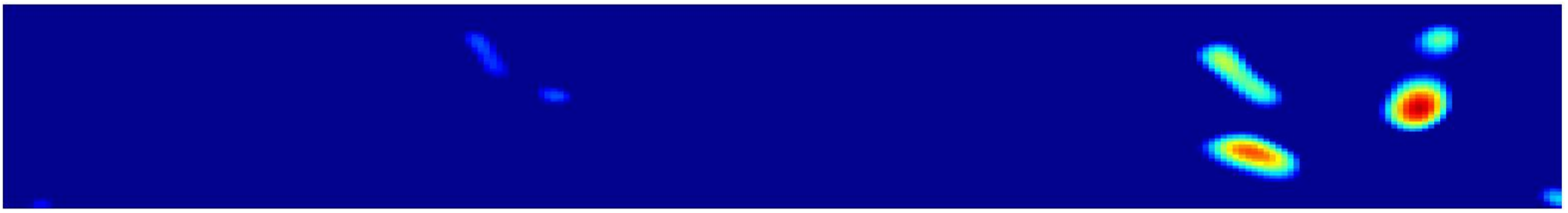}
%\subcaption{}
\label{fig:t_001}
\vspace{-0.3in}
\end{subfigure} 
\vspace{-0.2in}
\begin{subfigure}[t]{1.0\textwidth}
\includegraphics[width=\textwidth]{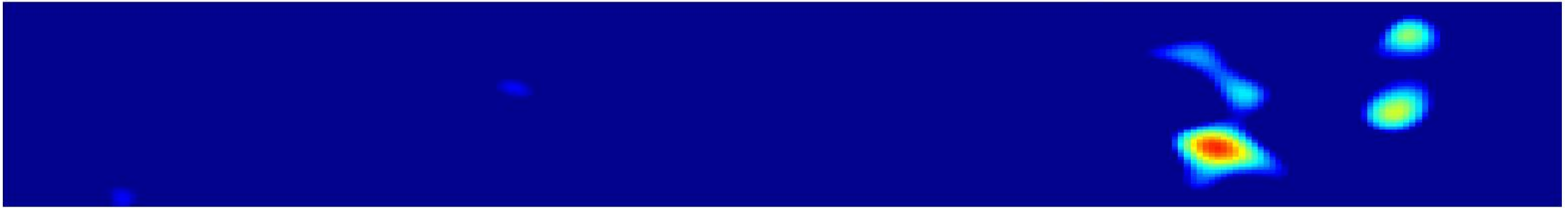}
% \subcaption{}
\label{fig:t_021}
\end{subfigure} 
\vspace{-0.2in}
\begin{subfigure}[t]{1.0\textwidth}
\includegraphics[width=\textwidth]{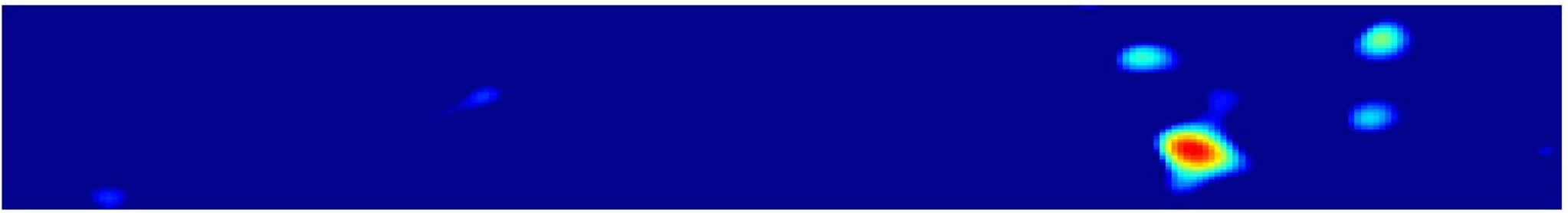}
% \subcaption{}
\label{fig:t_041}
\end{subfigure} 
\vspace{-0.2in}
\begin{subfigure}[t]{1.0\textwidth}
\includegraphics[width=\textwidth]{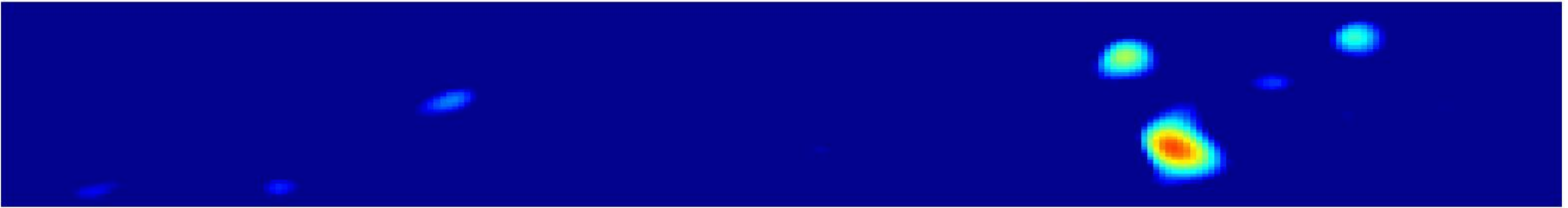}
% \subcaption{}
\label{fig:t_061}
\end{subfigure} 
\vspace{-0.2in}
\begin{subfigure}[t]{1.0\textwidth}
\includegraphics[width=\textwidth]{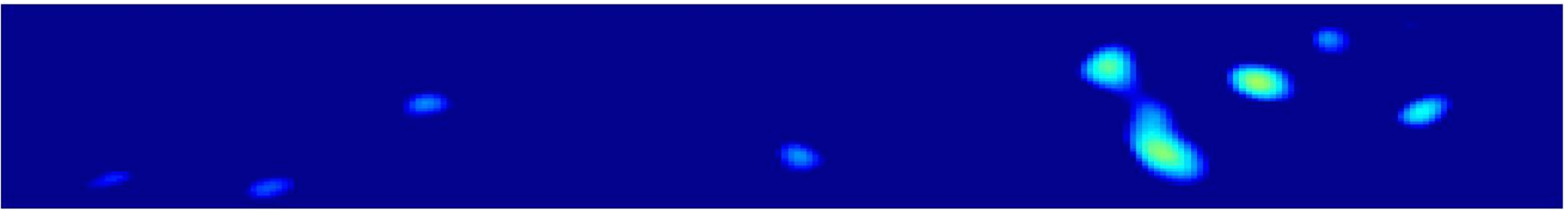}
% \subcaption{}
\label{fig:t_081}
\end{subfigure} 
\vspace{-0.2in}
\begin{subfigure}[t]{1.0\textwidth}
\includegraphics[width=\textwidth]{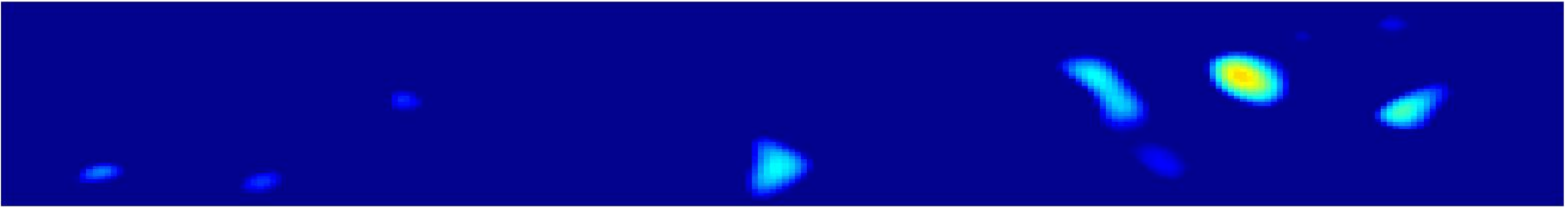}
% \subcaption{}
\label{fig:t_101}
\end{subfigure} 
\vspace{-0.2in}
\begin{subfigure}[t]{1.0\textwidth}
\includegraphics[width=\textwidth]{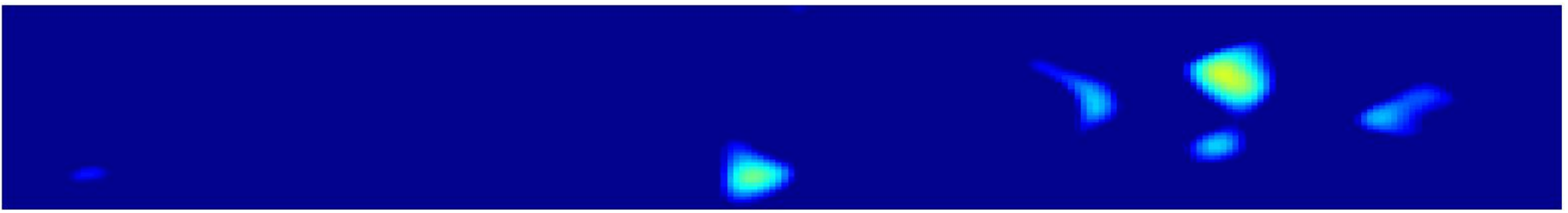}
% \subcaption{}
\label{fig:t_121}
\end{subfigure} 
\vspace{-0.2in}
\begin{subfigure}[t]{1.0\textwidth}
\includegraphics[width=\textwidth]{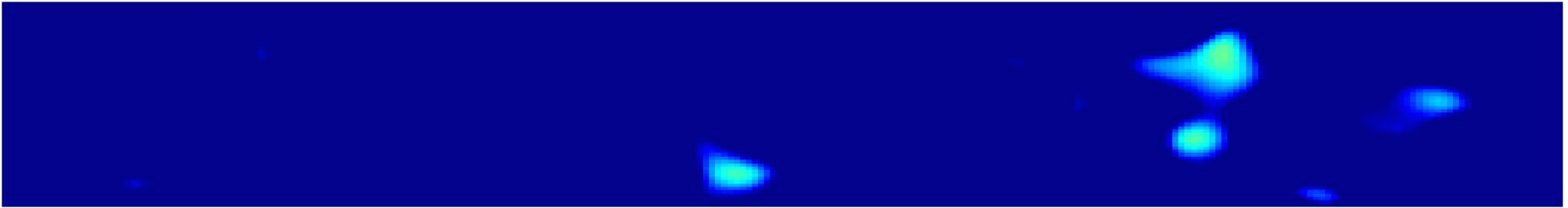}
% \subcaption{}
\label{fig:t_141}
\end{subfigure} 
\vspace{-0.2in}
\begin{subfigure}[t]{1.0\textwidth}
\includegraphics[width=\textwidth]{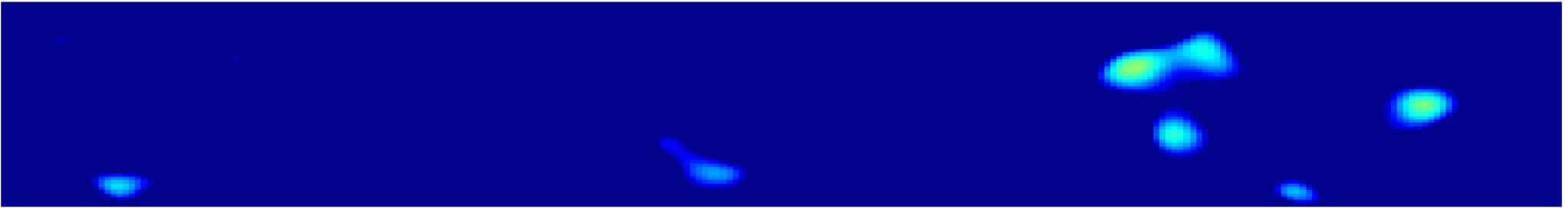}
% \subcaption{}
\label{fig:t_161}
\end{subfigure} 
\vspace{-0.2in}
\begin{subfigure}[t]{1.0\textwidth}
\includegraphics[width=\textwidth]{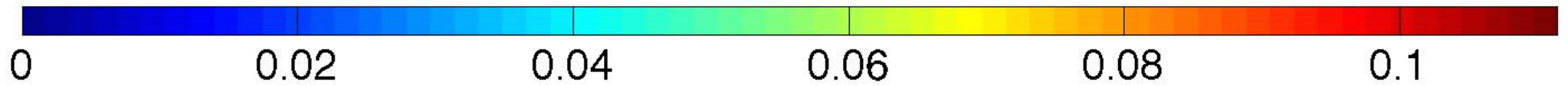}
\subcaption{}
\label{fig:slidingSubset}
\end{subfigure} 
\end{minipage}
\caption{Evolution of forces on a subset of the elements of the surface during
sliding. The subset of elements considered in (c) is marked by a red box in
(a). (b) shows the forces in (a) on a logarithmic scale (to better display the
smaller forces not evident in (a)). The color in (a) and (c) indicates the
nondimensionalized normal force on each element. The time interval between the
snapshots in (c) is $\Delta \bar{t} = 1$ and the sliding speed is $\bar{v} = 5$
(the surfaces slide a distance of $5$ between snapshots, the rms-roughness
$\bar{\sigma} = 1$ and the correlation length $\bar{\beta} = 10$.}
\label{fig:slidingSnapshots}
\end{figure}

As the deformable surface slides on the rigid rough one, the elements and their
forces evolve as an ensemble (Figure \ref{fig:slidingSnapshots}). Contacts
form, some of them grow, while others dwindle.  With continuing slip,
eventually all of them go out of existence, replaced by others. 

%%%%%%%%%%%%%%%%%%%%%%%%%%%%%%%%%%%%%%%%%%%%%%%%%%%%%%%%%%%%%%%%%%%%%%%%%%%%%%%%
%%%%%%%%%%%%%%%%%%%%%%%%%%%%%%%%%%%%%%%%%%%%%%%%%%%%%%%%%%%%%%%%%%%%%%%%%%%%%%%%

\subsection{Distribution of forces}

\begin{figure}
\centering
\begin{subfigure}[t]{0.45\textwidth}
\includegraphics[width=\textwidth]{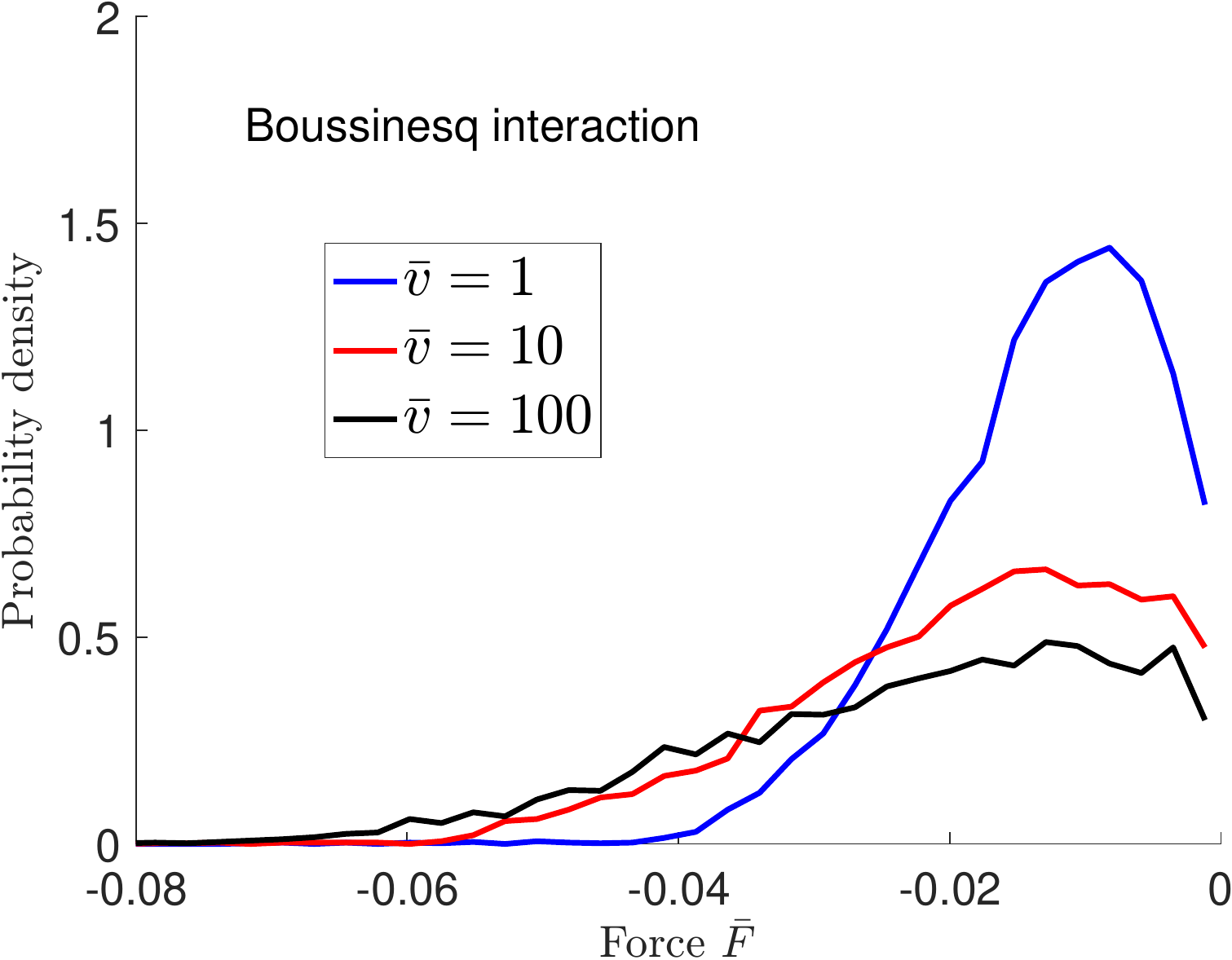}
\subcaption{}
\end{subfigure} 
\begin{subfigure}[t]{0.45\textwidth}
\includegraphics[width=\textwidth]{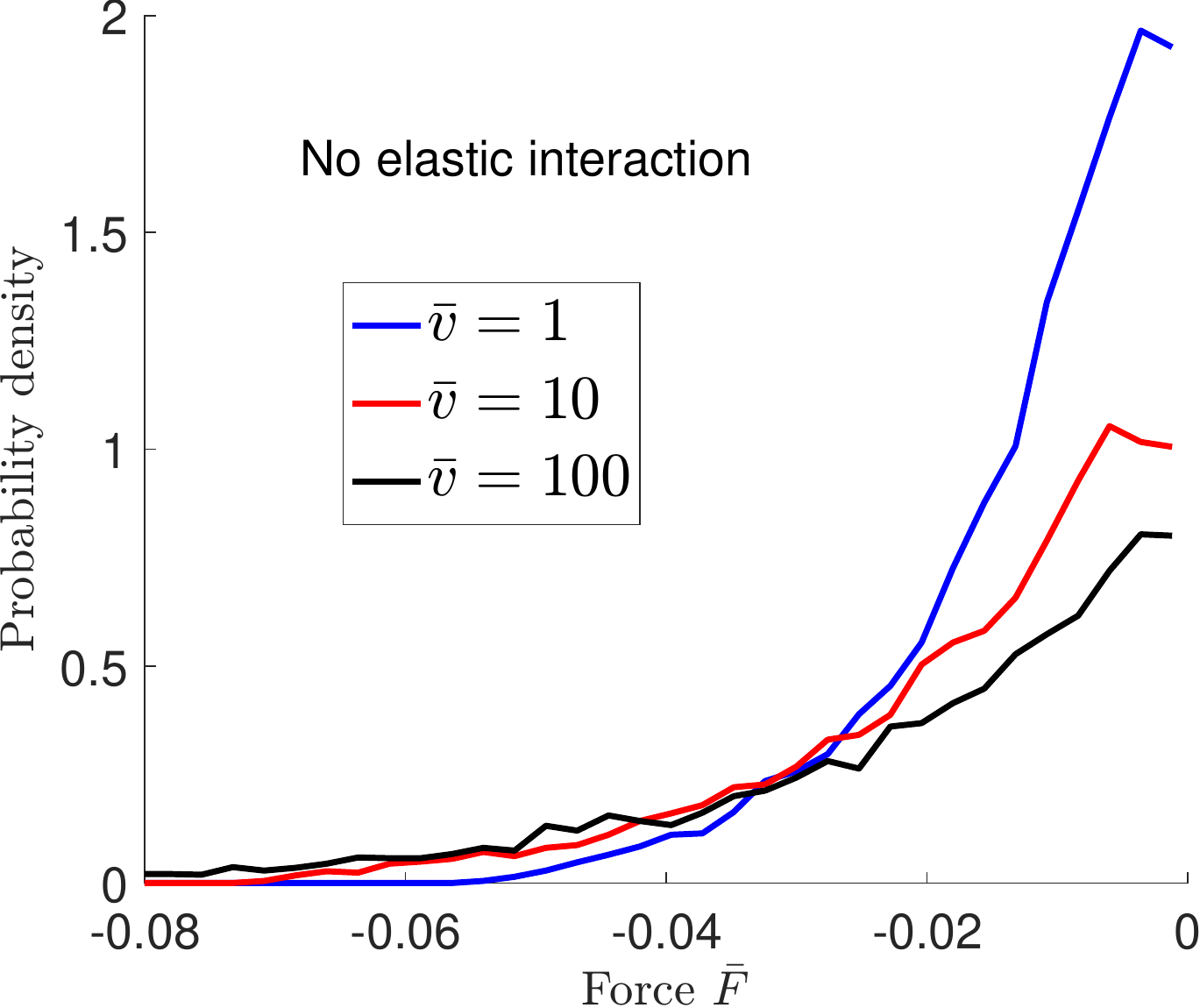}
\subcaption{}
\end{subfigure} 
\caption{Force distribution at contacts at steady state at different sliding 
speeds. The area under the curve is less than $1$ since only a small percentage
of the nominal area is in contact.}
\label{fig:forceDistributionSteadyStateVelocityDependence}
\end{figure}

For a given total normal force $F_N$, the distribution of the normal forces on
the elements depends on the sliding speed.  At higher sliding speeds, for the
same global normal force, fewer sliders need to be in contact since the normal
force on each of the ones in contact is higher on average (because of the
rate-dependent viscoelastic behavior).  This can be seen in Figure
\ref{fig:forceDistributionSteadyStateVelocityDependence} which shows force
distribution at steady state at different sliding speeds. The area under the
curve, which represents the fraction of sliders in contact, is smaller at
higher speeds. However, the first moment of the force distribution, which is
the total normal force is the same for the different velocities, by the design
of the numerical experiments. 
%Thus, on jumping from a low speed to a higher speed, the contact area
%gradually decreases and reaches a steady state corresponding to the new
%sliding speed leading to the characteristic evolution of Figure
%\ref{fig:areaFrictionDistanceVelocityStrengtheningWeakening}.

%%%%%%%%%%%%%%%%%%%%%%%%%%%%%%%%%%%%%%%%%%%%%%%%%%%%%%%%%%%%%%%%%%%%%%%%%%%%%%%%
%%%%%%%%%%%%%%%%%%%%%%%%%%%%%%%%%%%%%%%%%%%%%%%%%%%%%%%%%%%%%%%%%%%%%%%%%%%%%%%%

\subsection{Decreasing contact area with increasing velocity}\label{subsec:AreaVelocity}
\begin{figure}
\centering
\begin{subfigure}[t]{0.49\textwidth}
  \includegraphics[width=\textwidth]{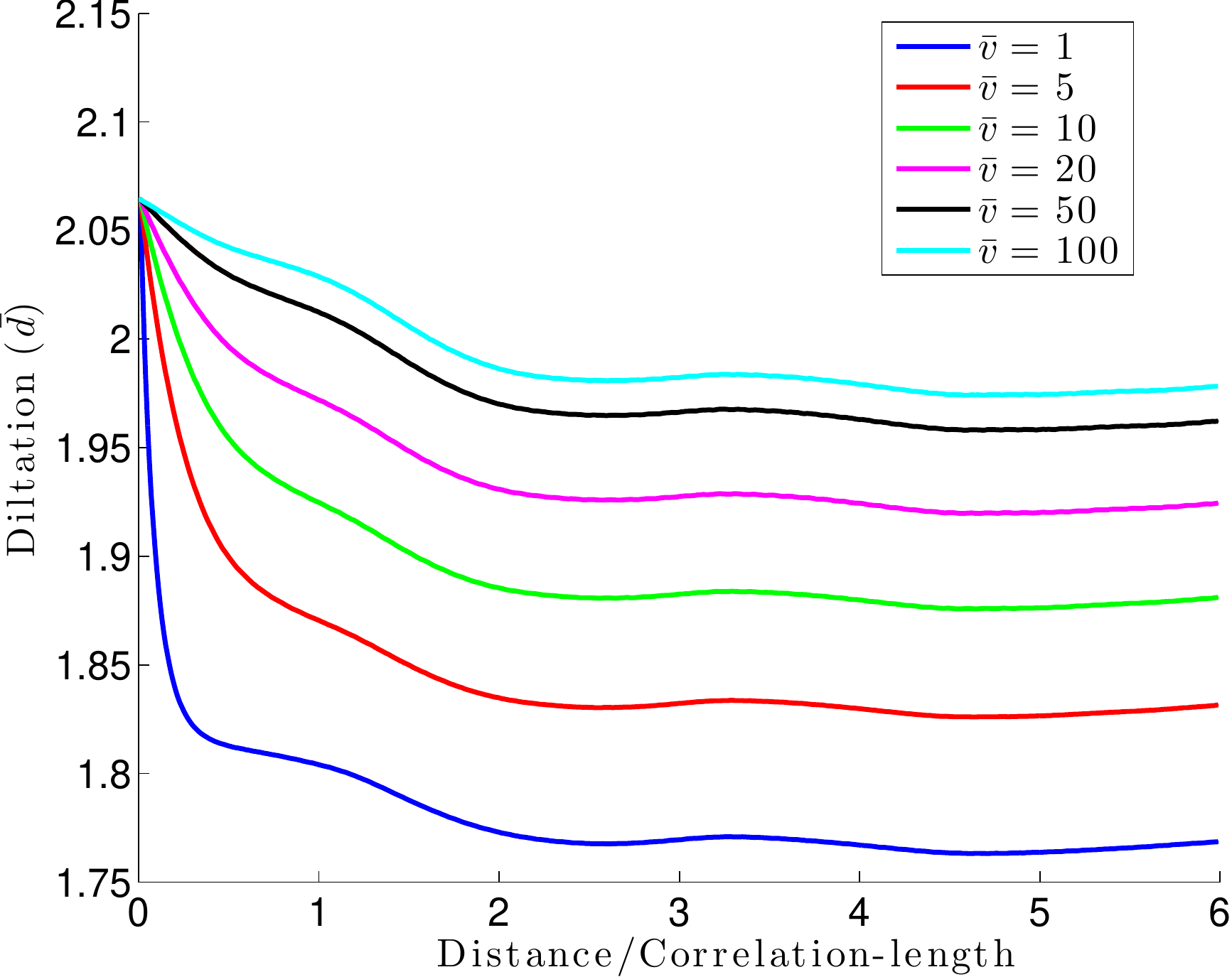}
\subcaption{}
\label{fig:dilatationDistanceVelocityDependence}
\end{subfigure} 
\begin{subfigure}[t]{0.49\textwidth}
  \includegraphics[width=\textwidth]{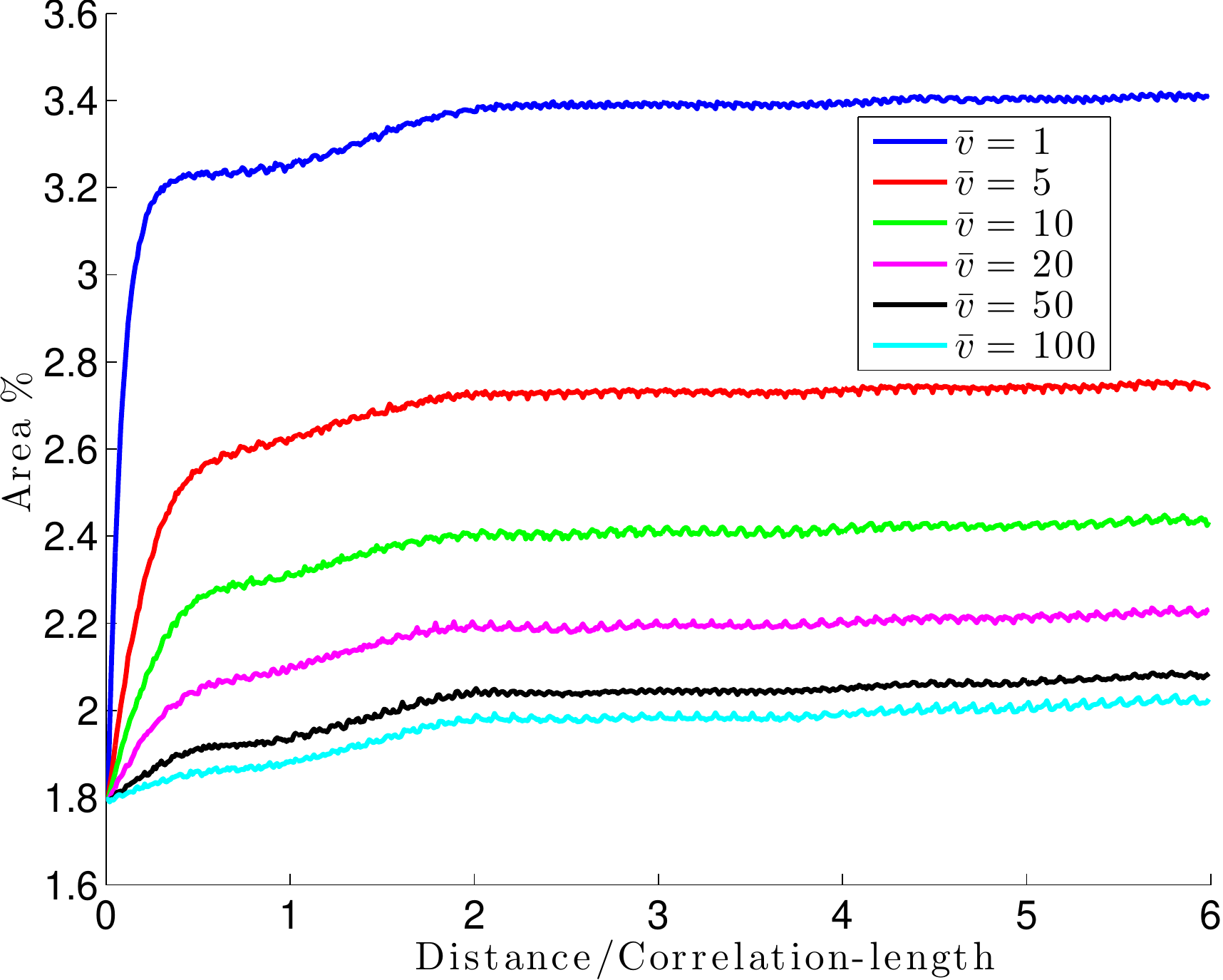}
\subcaption{}
\label{fig:areaDistanceVelocityDependence}
\end{subfigure} 
\caption{Evolution of (a) dilatation and (b) contact area starting from the
same initial state and sliding at different speeds.  At higher sliding speeds,
the average force on a contact is higher and thus, to sustain the same global
normal force, the dilatation is higher and the total area of contact is
smaller.}
\label{fig:velocityDependence}
\end{figure}

With increasing slip velocity, the steady-state dilatation always increases
and the total contact area always decreases (Figure
\ref{fig:velocityDependence}).  The two timescales relevant to sliding contact
are the viscoelastic relaxation timescale (determined by $\bar{\lambda}$) and
the ratio of the correlation length to the sliding speed,
$\bar{\beta}/\bar{v}$.

\begin{figure}
\centering
\begin{subfigure}[t]{0.49\textwidth}
  \includegraphics[width=\textwidth]{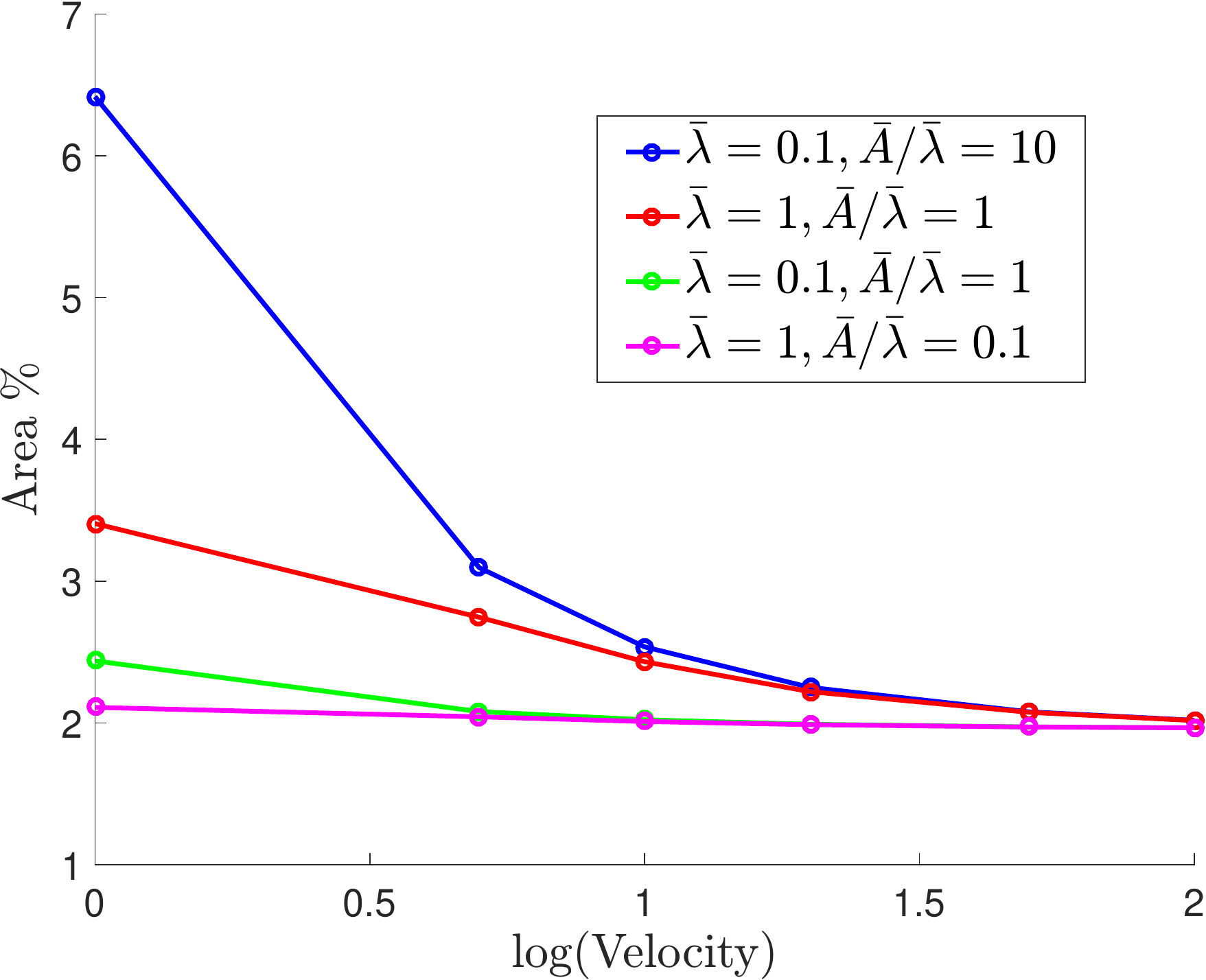}
\subcaption{}
\label{fig:areaVelocitySteadyStateViscoelasticityDependenceNoInteraction}
\end{subfigure} 
\begin{subfigure}[t]{0.49\textwidth}
  \includegraphics[width=\textwidth]{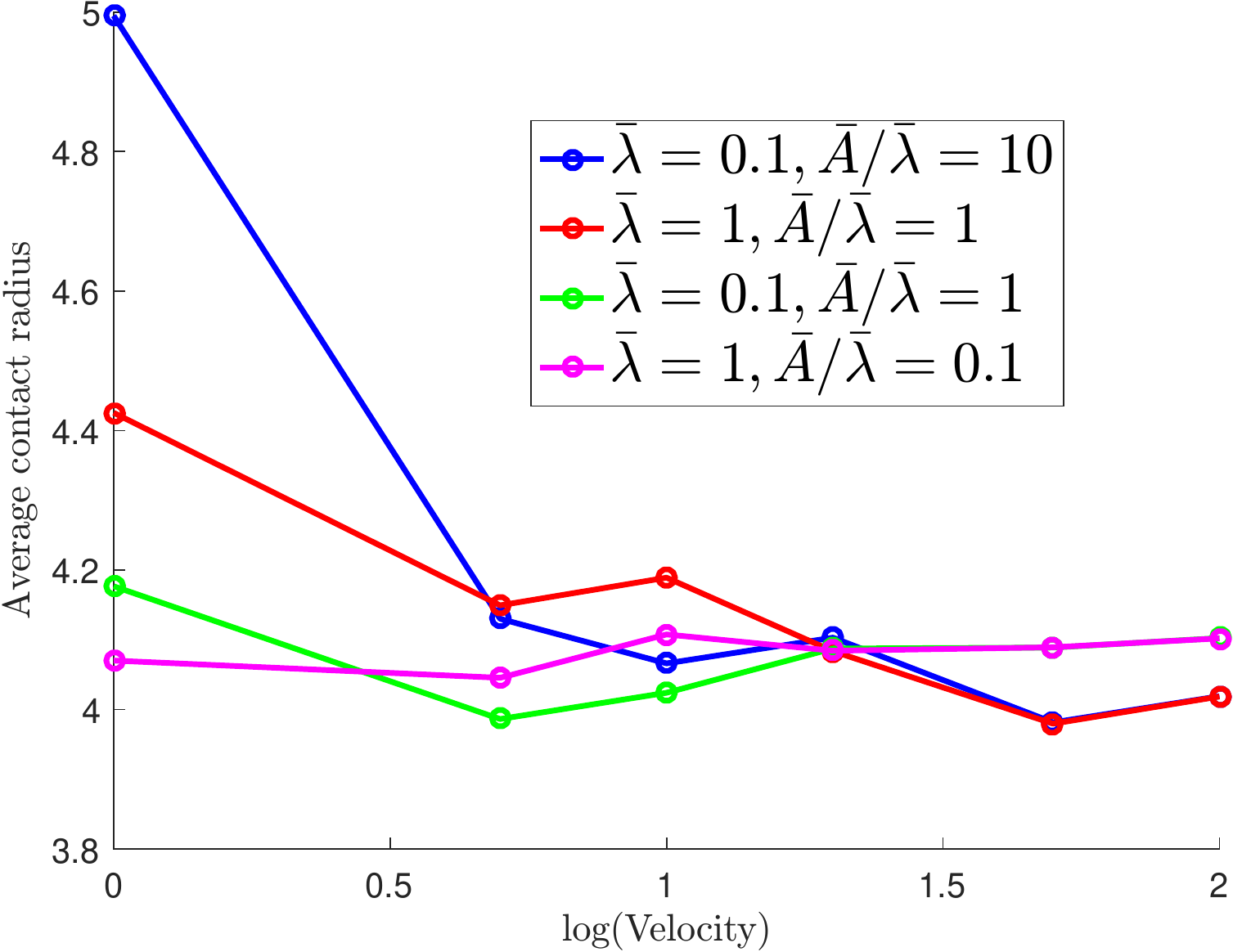}
\subcaption{}
\label{fig:averageContactRadiusSteadySlidingStaticContactNoInteractionVelocityDependence}
\end{subfigure} 
\begin{subfigure}[t]{0.49\textwidth}
  \includegraphics[width=\textwidth]{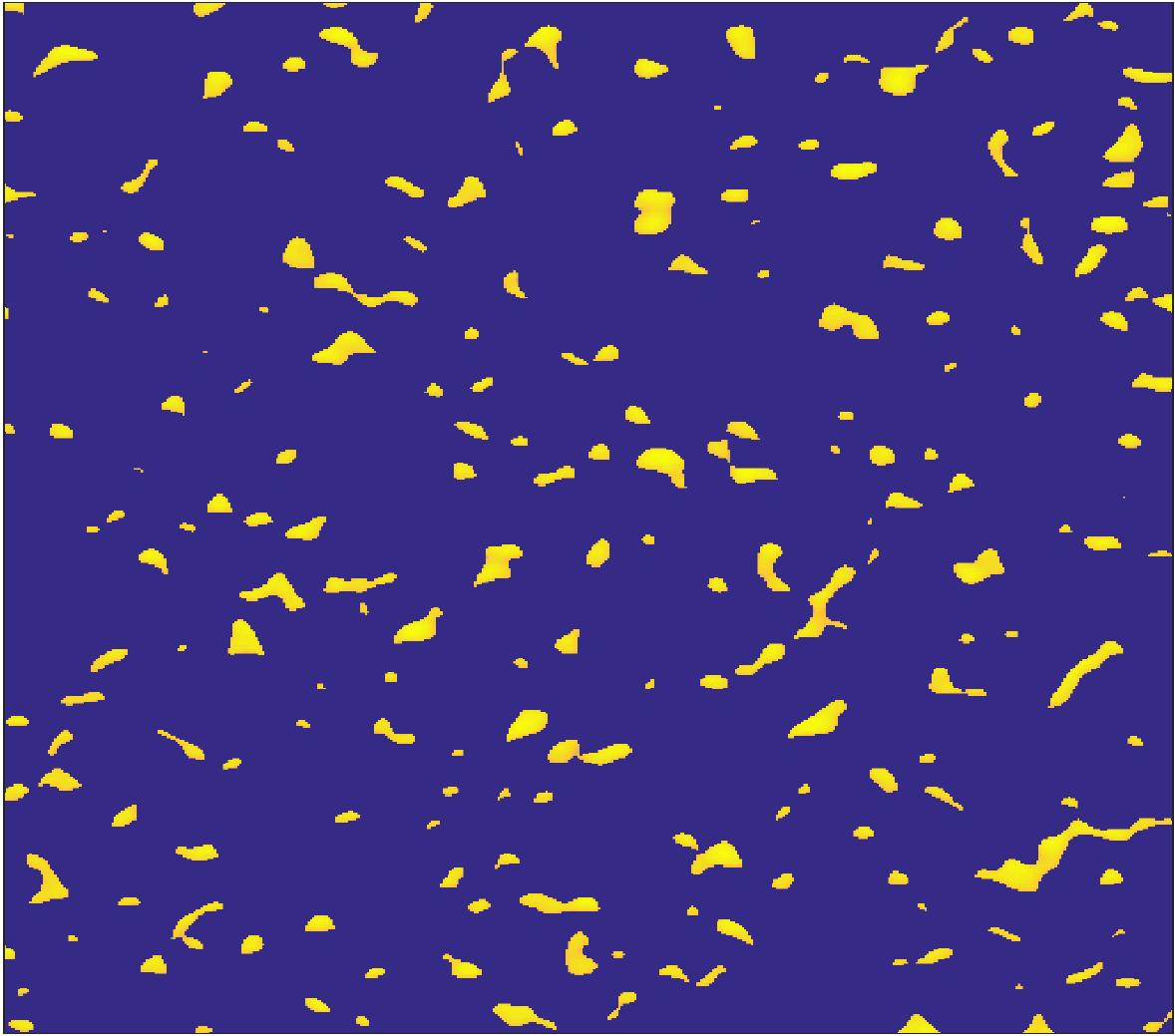}
\subcaption{}
\label{fig:contactsFinal512by512GaussianCorrelationDiscretization10VelocityDependence_Velocity1}
\end{subfigure} 
\begin{subfigure}[t]{0.49\textwidth}
  \includegraphics[width=\textwidth]{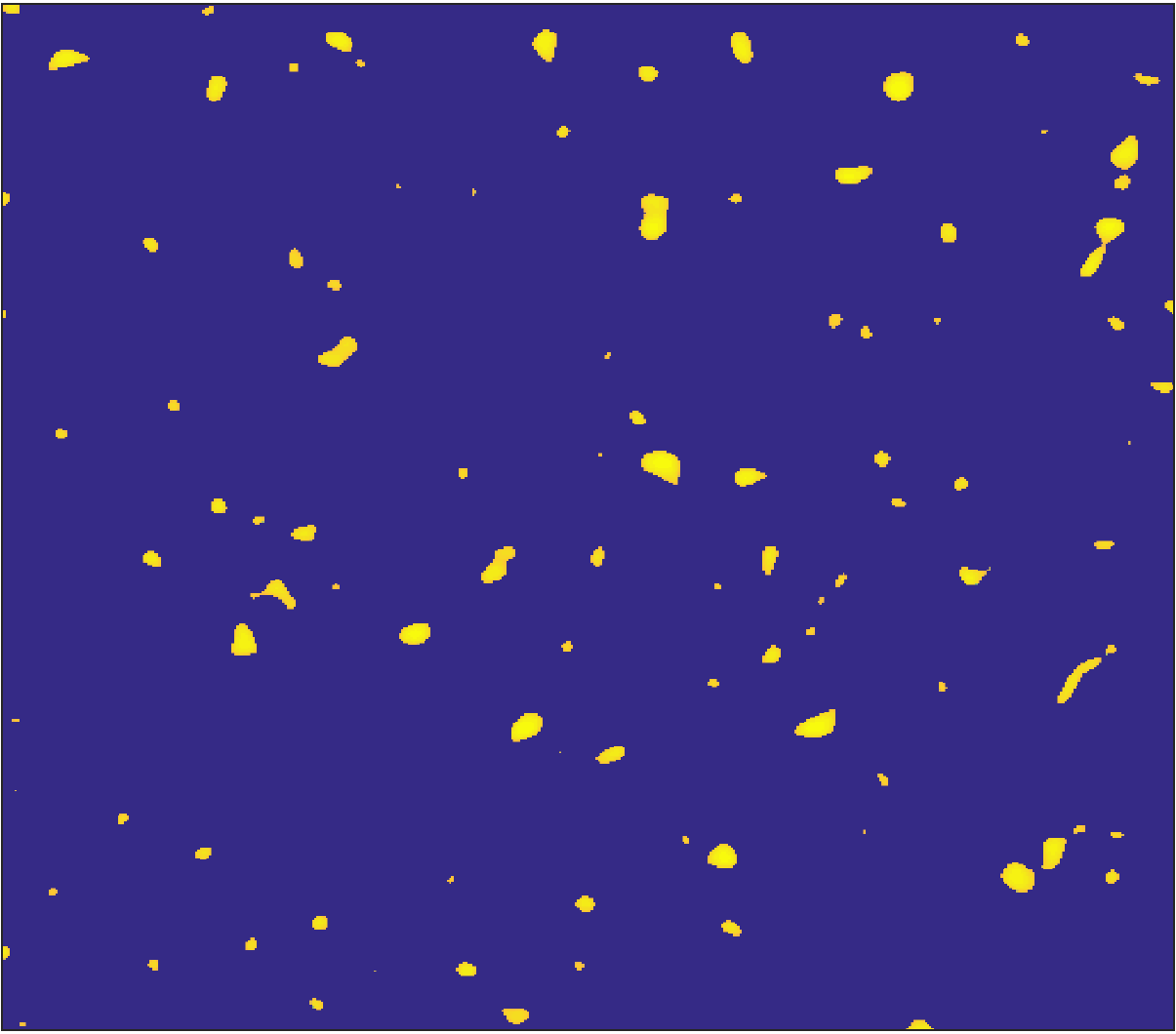}
\subcaption{}
\label{fig:contactsFinal512by512GaussianCorrelationDiscretization10VelocityDependence_Velocity100}
\end{subfigure} 
\caption{(a) Steady state contact area  at different sliding speeds for four
combinations of the viscoelastic parameters. The contact area always decreases
with increasing speed. The magnitude of change and the velocities where
$\mathrm{d}$Area/$\mathrm{d}$Velocity is large depends on the viscoelastic
properties (see Section \ref{subsec:AreaVelocity}). (b) The average contact
remains approximately constant with velocity. This is similar to the static
contact case where the average contact size remained constant with time (Figure
\ref{fig:averageContactRadiusTimeStaticContactNoInteraction}).  (c) and (d)
show the contacts at steady state after sliding the same distance for
velocities $\bar{v} = 1$ and $\bar{v} = 100$ (for the case $\bar{\lambda} =
0.1, \bar{A}/\bar{\lambda} = 10$). At $\bar{v} = 1$, the total contact area is
larger than that at $\bar{v} = 100$ but so are the number of contacts and the
average contact size remains approximately the same.}
\label{fig:AreaViscoelasticityAndVelocityDependence}
\end{figure}

Apart from the surface roughness, the steady-state contact area at a given
velocity depends on the viscoelastic properties. In static contact (section
\ref{sec:staticContact}), the duration of  friction evolution is determined by
$\bar{\lambda}$ and the magnitude of the friction increase by
$\bar{A}/\bar{\lambda}$ (Section \ref{subsec:IncreaseContactStatic}).
Instantaneous and steady-state behavior of the static contact is somewhat
analogous to sliding at very high and very low velocity. Thus, $\bar{\lambda}$
is related to the range of slip velocities at which the system is
velocity-dependent and $\bar{A}/\bar{\lambda}$ is related to the magnitude of
the sensitivity. This is reflected in Figure
\ref{fig:areaVelocitySteadyStateViscoelasticityDependenceNoInteraction} which
shows the steady-state area as a function of the sliding speed for four
combinations of $\bar{A},\bar{\lambda}$ and three values of
$\bar{A}/\bar{\lambda}$ of 10, 1, and 0.1. Cases with the same value of
$\bar{\lambda}$ but larger value of $\bar{A}/\bar{\lambda}$ have larger area
changes over the velocity range shown. The two cases with the same
$\bar{A}/\bar{\lambda} = 1$, but different relaxation times of $\bar{\lambda} =
1$ (blue line) and $\bar{\lambda} = 0.1$ (purple line), have similar area
changes for a range of slip velocities shifted by an order of magnitude;
compare velocity ranges from 10 to 100 for $\bar{\lambda} = 1$ (blue line) and
from 1 to 10 for $\bar{\lambda} = 0.1$ (purple line).

The average contact size remains approximately independent slip velocity
(Figure
\ref{fig:averageContactRadiusSteadySlidingStaticContactNoInteractionVelocityDependence}).
With increasing velocity, the total contact area decreases but so do the number
of contacts and the average contact size remains approximately the same. This
is similar as in static contact where the average contact radius remains nearly
constant with time under a constant total normal force (Figure
\ref{fig:averageContactRadiusTimeStaticContactNoInteraction}). 

%%%%%%%%%%%%%%%%%%%%%%%%%%%%%%%%%%%%%%%%%%%%%%%%%%%%%%%%%%%%%%%%%%%%%%%%%%%%%%%%
%%%%%%%%%%%%%%%%%%%%%%%%%%%%%%%%%%%%%%%%%%%%%%%%%%%%%%%%%%%%%%%%%%%%%%%%%%%%%%%%

%%%%%%%%%%%%%%%%%%%%%%%%%%%%%%%%%%%%%%%%%%%%%%%%%%%%%%%%%%%%%%%%%%%%%%%%%%%%%%%%
%%%%%%%%%%%%%%%%%%%%%%%%%%%%%%%%%%%%%%%%%%%%%%%%%%%%%%%%%%%%%%%%%%%%%%%%%%%%%%%%

\subsection{Comparison with experiments - velocity jump test}
\label{subsec:velocityJump}

As in experiments, we perform velocity jump simulations. Two rough surfaces are
brought into contact to a total force equivalent to a pressure of $100$ MPa.
The surfaces are slid at this constant normal force at velocity $\bar{v} = 10$
for the slip distance several times larger than the correlation length, to
establish steady-state sliding. The sliding velocity is then instantaneously
changed to $\bar{v} = 1$. Jumps to $\bar{v} = 10$ and $\bar{v} = 1$ are
repeated. Since the nondimensionalizing length and time scales are $1\mu$m and
$1$ s, respectively, $\bar{v} = 10$ corresponds to a sliding velocity of
$10\mu$m/s, a typical value in experiments \cite{dieterich:3}.

The resulting evolution of the macroscopic friction coefficient contains the
direct and transient effects, as observed in experiments (Figure 14). The
friction coefficient changes instantaneously with the velocity jumps because of
the velocity-strengthening term in local shear resistance (\ref{eq:shear}), and
the change has the same sense as that of the velocity jump, i.e., the friction
coefficient increases (decreases) when the velocity increases (decreases).
Following the standard rate and state terminology, we call this jump the {\it
direct effect}. The jump is followed by an evolution towards a steady state,
because of the evolution of the real contact area.  We call this evolution the
{\it transient}.  The transient changes the friction coefficient in the
direction opposite to the direct effect, since the contact area decreases for
higher velocities, as already discussed.   

\begin{figure}
\centering
\begin{subfigure}[t]{0.4\textwidth}
\includegraphics[width=\textwidth]{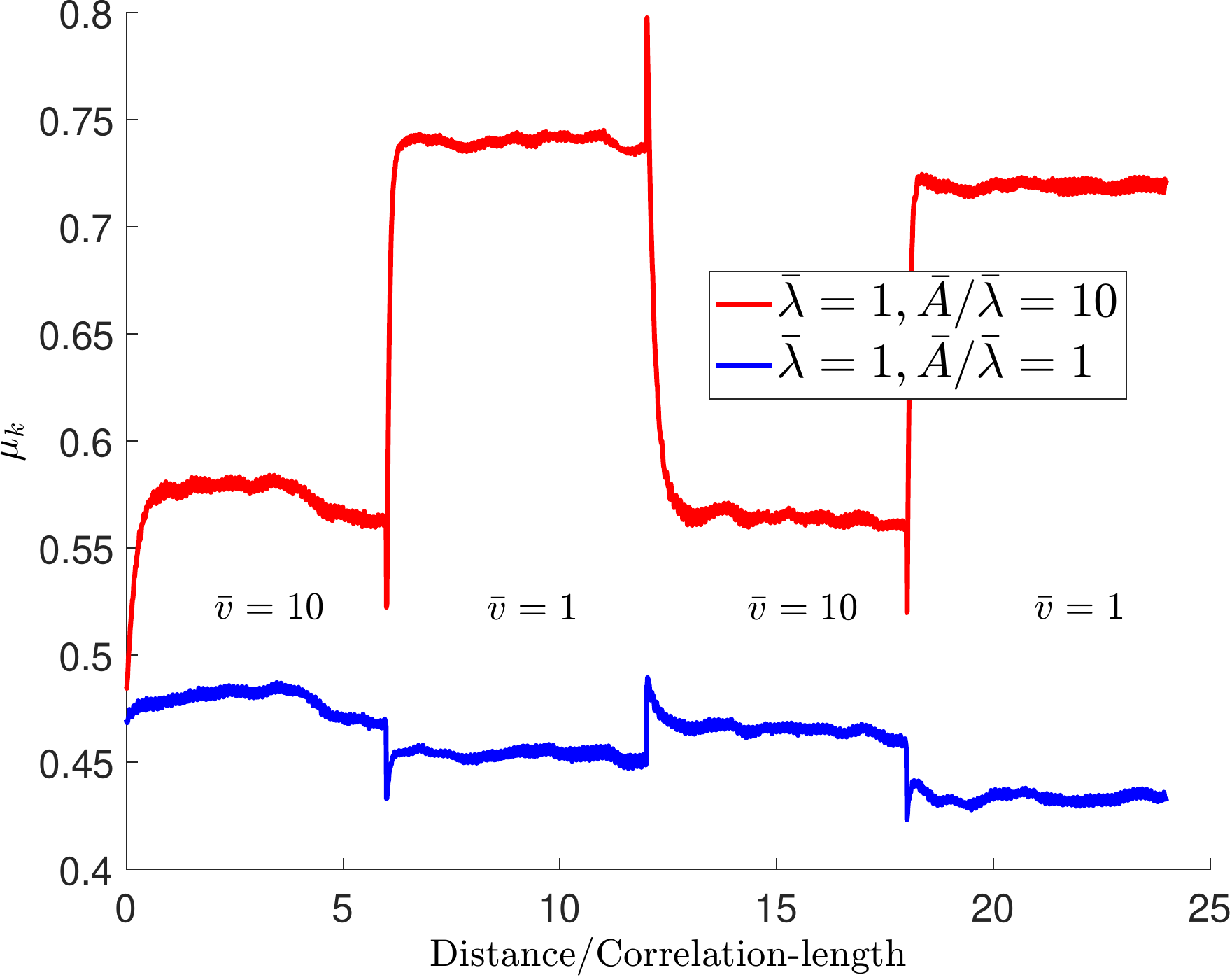}
\subcaption{}
\label{fig:frictionDistanceVelocityStrengtheningWeakening}
\end{subfigure} 
\begin{subfigure}[t]{0.4\textwidth}
\includegraphics[width=\textwidth]{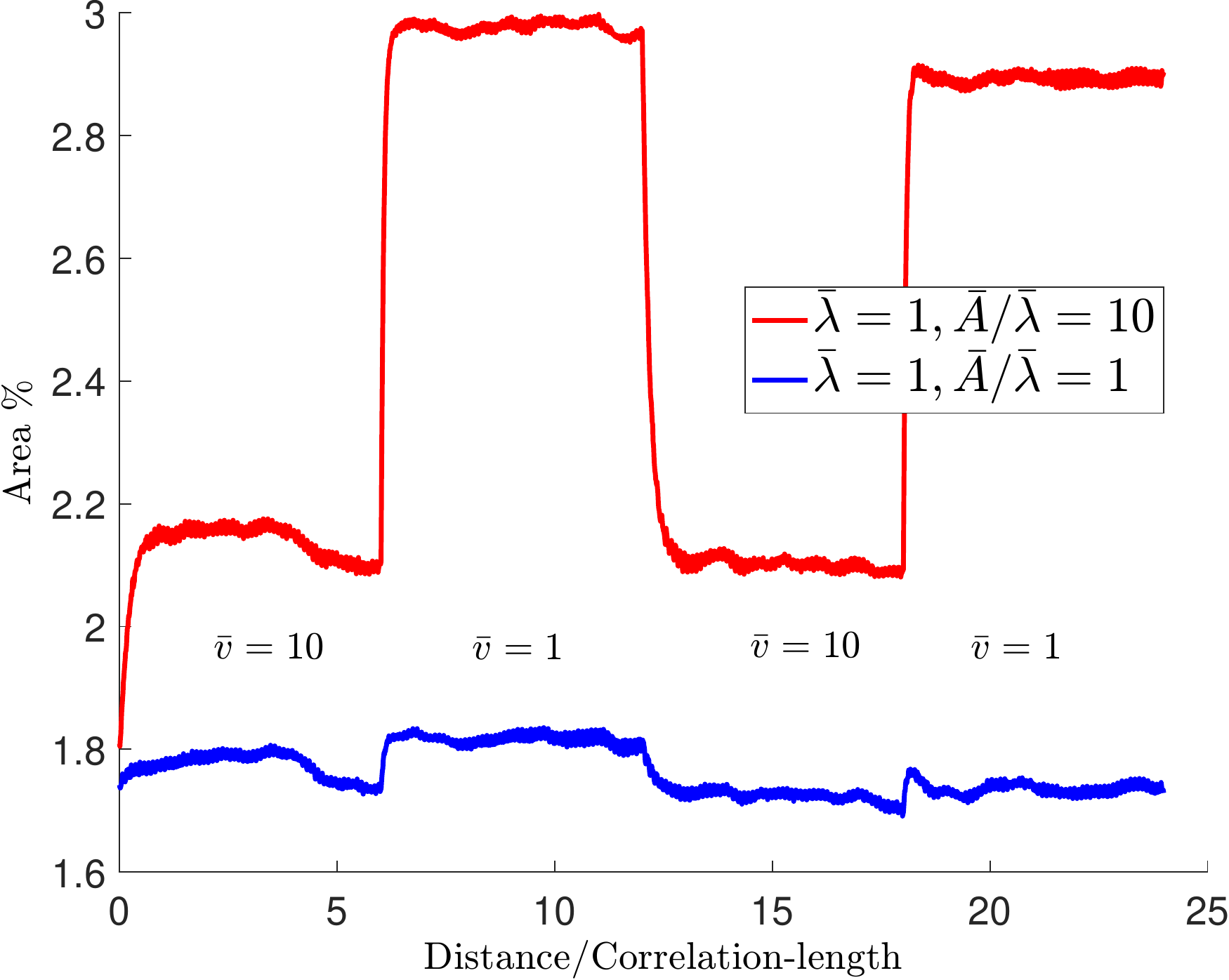}
\subcaption{}
\label{fig:areaDistanceVelocityStrengtheningWeakening}
\end{subfigure} 
\begin{subfigure}[t]{0.45\textwidth}
\includegraphics[width=\textwidth]{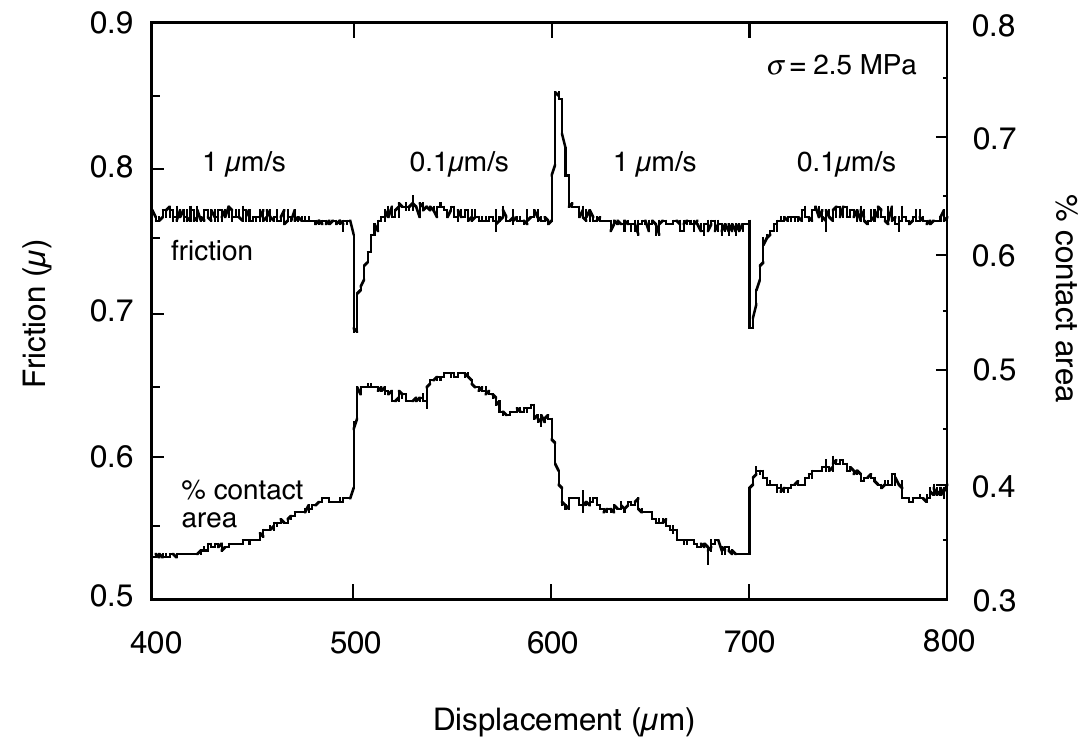}
\caption{}
\label{fig:frictionAreaEvolutionVelocityJumpDieterich_1994}
\end{subfigure} 
\begin{subfigure}[t]{0.4\textwidth}
  \includegraphics[width=\textwidth]{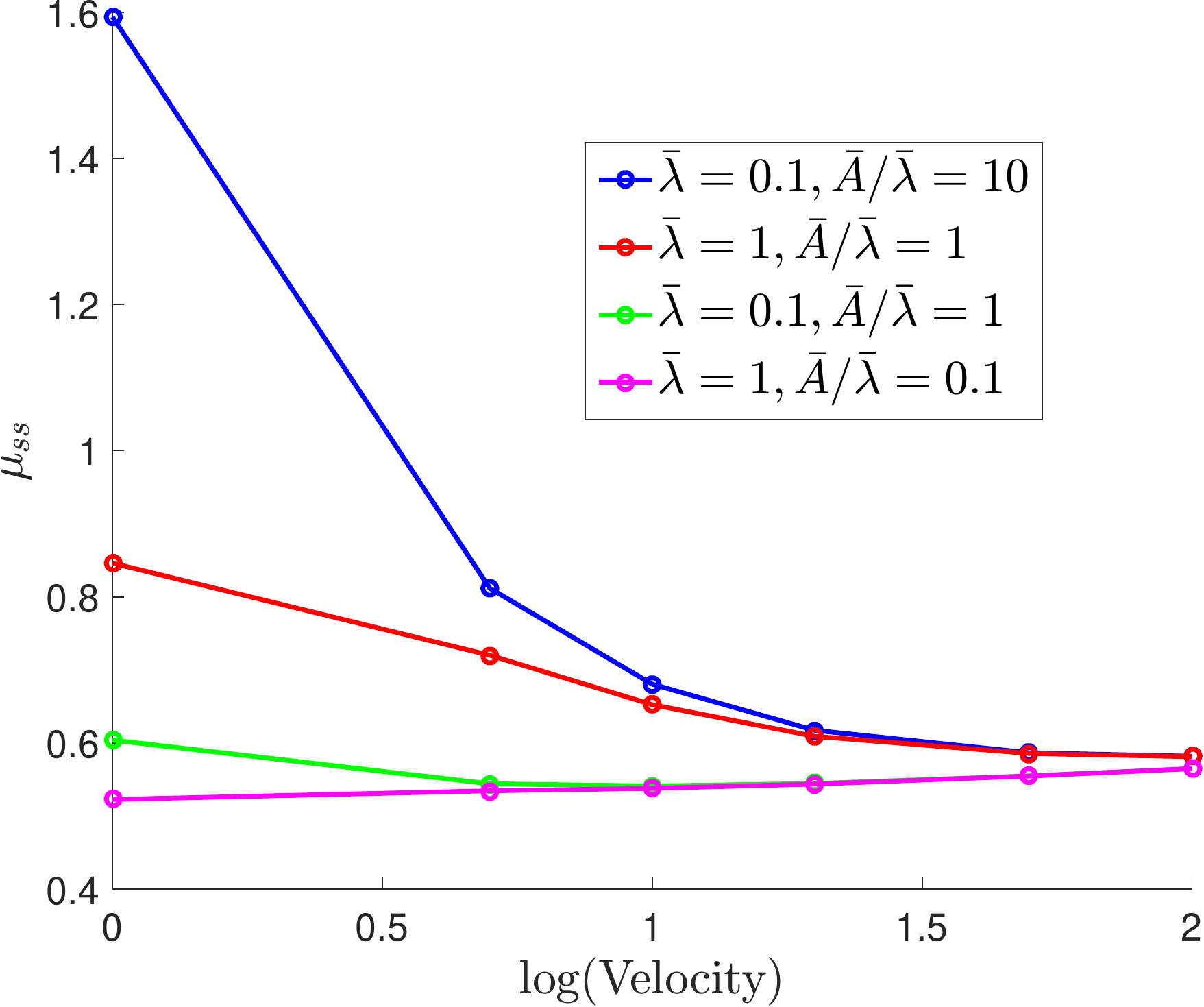}
\subcaption{}
\label{fig:frictionVelocitySteadyStateViscoelasticityDependenceNoInteraction}
\end{subfigure} 
\begin{subfigure}[t]{0.4\textwidth}
\includegraphics[width=\textwidth]{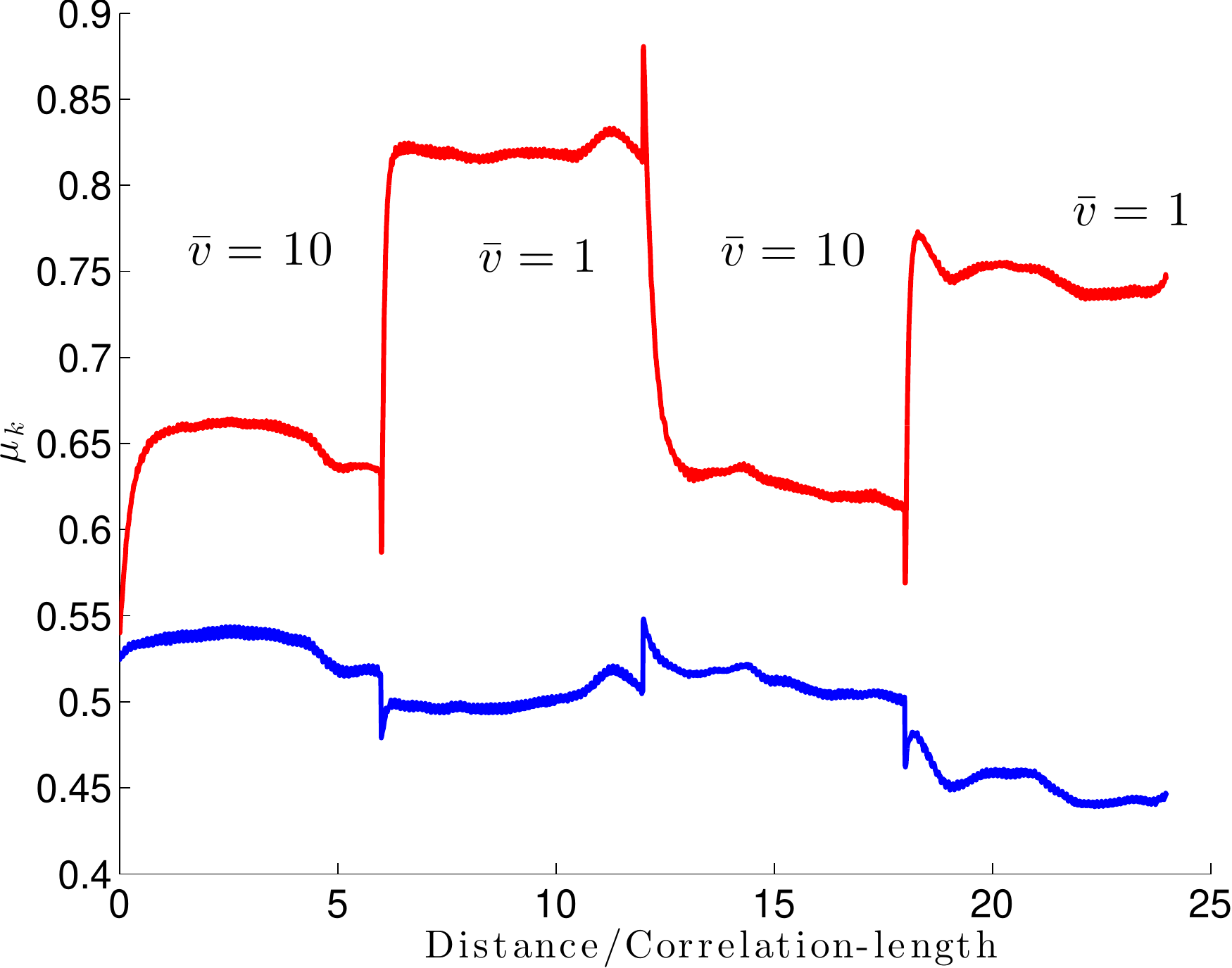}
\subcaption{}
\label{fig:frictionDistanceVelocityStrengtheningWeakeningNoInteraction}
\end{subfigure} 
\begin{subfigure}[t]{0.4\textwidth}
\includegraphics[width=\textwidth]{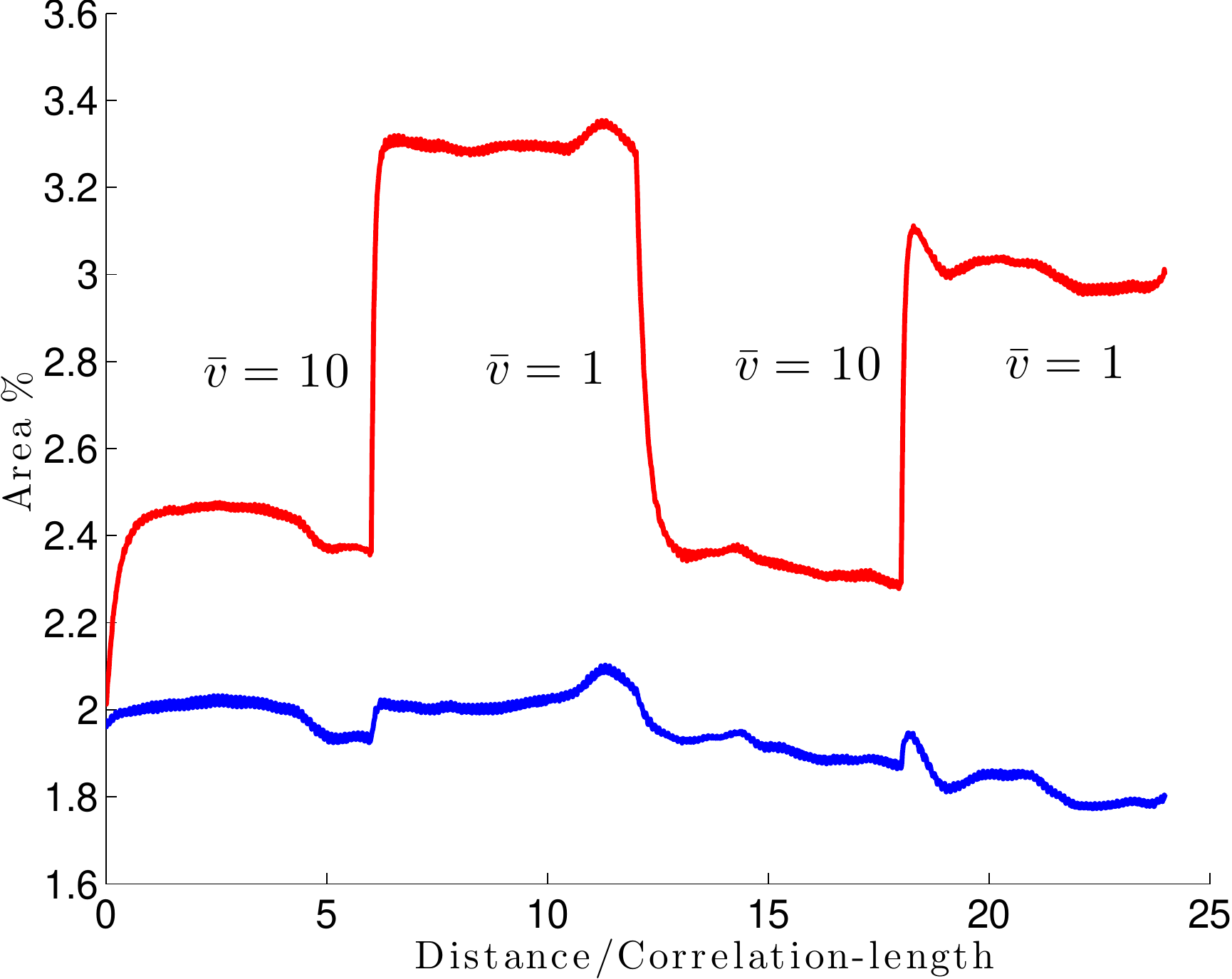}
\subcaption{}
\label{fig:areaDistanceVelocityStrengtheningWeakeningNoInteraction}
\end{subfigure} 
\caption{Evolution of (a) friction coefficient and (b) contact area during
velocity jump tests with Boussinesq elastic interactions.  Because of
viscoelasticity, the average force on a contact is higher at higher speeds, and
thus for the same global normal force, contact area is smaller at higher
speeds. With a jump in sliding speed, $\mu_k$ changes instantaneously and this
is followed by an evolution to a steady state. Depending on the material
parameters, the steady state value can increase or decrease with increasing
sliding speed.  The two cases correspond to velocity strengthening (blue -
$\bar{\lambda} = 1, \bar{A} = 1$ ) and velocity weakening (red - $\bar{\lambda}
= 1, \bar{A} = 10$ ) respectively. (c) Evolution of contact area and friction
during a velocity jump test (reproduced with permission from Dieterich
\cite{dieterich:3}). (d) Steady state friction coefficient at different sliding
speeds for four combinations of the viscoelastic parameters.  The contact area
always decreases with increasing speed but the friction coefficient might
increase or decrease depending on whether the direct effect or the transient
effect dominates. Evolution with no elastic interactions (e) and (f) is
qualitatively similar to (a) and (b).}
\label{fig:areaFrictionDistanceVelocityStrengtheningWeakening}
\end{figure}

Even though the contact area always decreases with increasing sliding velocity,
the steady-state friction coefficient can either increase or decrease,
depending on whether the direct effect or the transient change dominates. If
the transient is smaller than the direct effect, the steady-state friction is
higher for higher velocity, resulting in {\it velocity strengthening} behavior.
If the transient is larger than the direct effect, the steady-state value is
lower for higher velocity, producing {\it velocity weakening} behavior. The
area evolution depends on both the sliding velocity and viscoelastic
properties. For given viscoelastic properties and a single relaxation
timescale, the area evolution is largest for a certain range of velocities  In
part, the velocity dependence can transition between strengthening and
weakening for different sliding velocities (Figure
\ref{fig:frictionVelocitySteadyStateViscoelasticityDependenceNoInteraction}).
Such transitions have been observed in experiments as well \cite{shimamoto:1}.
However, some materials have been shown to be consistently velocity-weakening
or velocity-strengthening for a wide range of sliding velocities.  Such
sustained behavior likely results from multiple relaxation timescales in the
viscoelastic response.

The qualitative aspects of the sliding friction remain unchanged when the
elastic interactions between elements are turned off.  During sliding, the
evolution of area and friction, velocity strengthening and velocity weakening,
are determined largely by the viscoelastic properties and only quantitatively
changed by the long-range elastic interactions.

%%%%%%%%%%%%%%%%%%%%%%%%%%%%%%%%%%%%%%%%%%%%%%%%%%%%%%%%%%%%%%%%%%%%%%%%%%%%%%%%

%%%%%%%%%%%%%%%%%%%%%%%%%%%%%%%%%%%%%%%%%%%%%%%%%%%%%%%%%%%%%%%%%%%%%%%%%%%%%%%%

\subsection{Macroscopic direct effect vs. microscopic hardening assumption} 

According to the assumed friction law (\ref{eq:shear}) and the microscopic
friction assumption (\ref{eq:shear}), the magnitude of the direct effect for a
velocity jump from $\bar{v}_{\text{old}}$ to $\bar{v}_{\text{new}}$ is given by
$$\Delta \mu_{\text{direct}} = \frac{\alpha
\log(\bar{v}_{\text{new}}/\bar{v}_{\text{old}})\bar{A}(\bar{v}_{\text{old}})}{\bar{F}_N}.$$
Since $\bar{A}(\bar{v})$ decreases monotonically with increasing $\bar{v}$, the
magnitude of the direct effect also decreases monotonically, for given
$\bar{v}_{\text{new}}/\bar{v}_{\text{old}}$, if $\alpha$ is a constant.
However, in experiments, it is observed that the direct effect remains constant
over a large range of velocities \cite{linker:1}.  This is possible only if the
velocity-dependent term in the microscopic friction (\ref{eq:shear}) increases
faster than $\log(\bar{v})$. Moreover, the faster increase should exactly
compensate for the evolution in the area, which is related to the viscoelastic
properties. Note that this is a general observation based only on the
assumption that the total shear force is proportional to the real area of
contact, and it does not depend on any other particular features of our model.
 
%%%%%%%%%%%%%%%%%%%%%%%%%%%%%%%%%%%%%%%%%%%%%%%%%%%%%%%%%%%%%%%%%%%%%%%%%%%%%%%%

\subsection{Elasto/viscoplastic contacts}\label{subsec:elastoviscoplastic}

%%%%%%%%%%%%%%%%%%%%%%%%%%%%%%%%%%%%%%%%%%%%%%%%%%%%%%%%%%%%%%%%%%%%%%%%%%%%%%%%
\begin{figure}
\centering
\begin{subfigure}[t]{0.5\textwidth}
  \includegraphics[width=\textwidth]{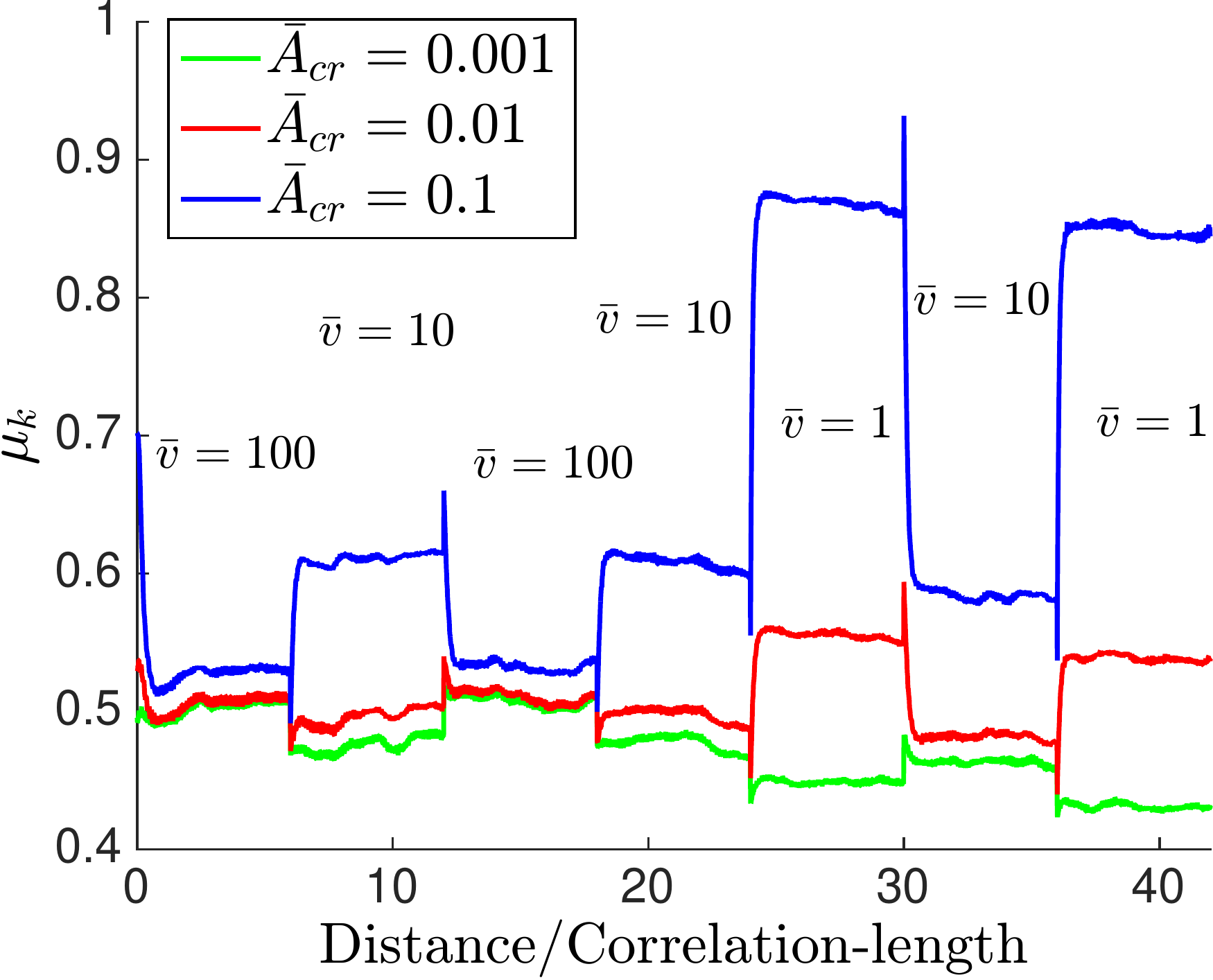}
\end{subfigure} 
\caption{Evolution of $\mu_k$ during velocity jump tests for elasto/viscoplastic contacts. Larger creep rates
lead to steady-state velocity-weakening friction while smaller creep rates result in
steady-state velocity-strengthening behavior.}
\label{fig:slidingContactViscoplastic}
\end{figure}
%%%%%%%%%%%%%%%%%%%%%%%%%%%%%%%%%%%%%%%%%%%%%%%%%%%%%%%%%%%%%%%%%%%%%%%%%%%%%%%%

%%%%%%%%%%%%%%%%%%%%%%%%%%%%%%%%%%%%%%%%%%%%%%%%%%%%%%%%%%%%%%%%%%%%%%%%%%%%%%%%
\begin{figure}
\centering
\begin{subfigure}[t]{0.4\textwidth}
  \includegraphics[width=\textwidth]{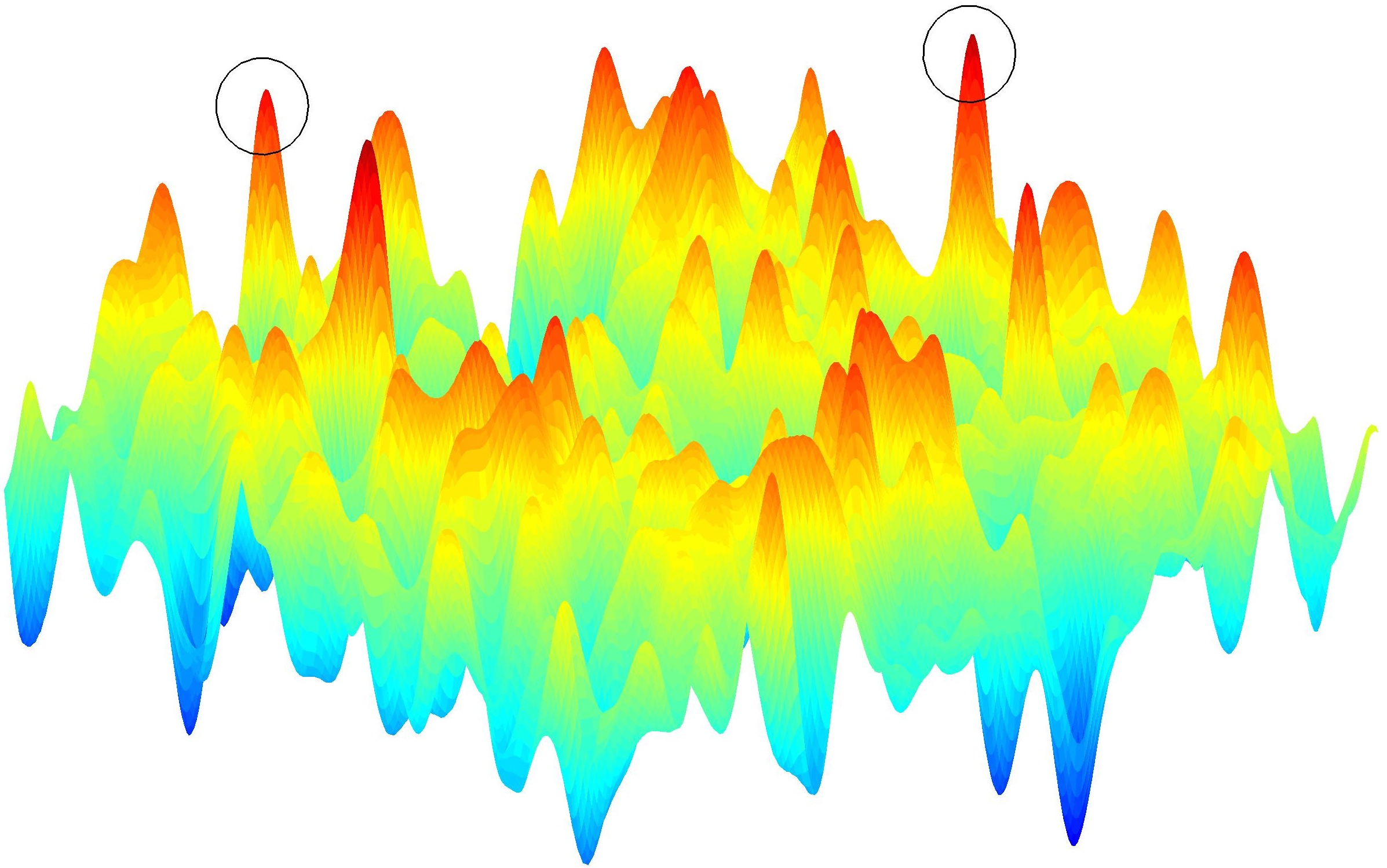}
\subcaption{}
\label{fig:initialSurface}
\end{subfigure} 
\begin{subfigure}[t]{0.1\textwidth}
  \includegraphics[scale=0.06]{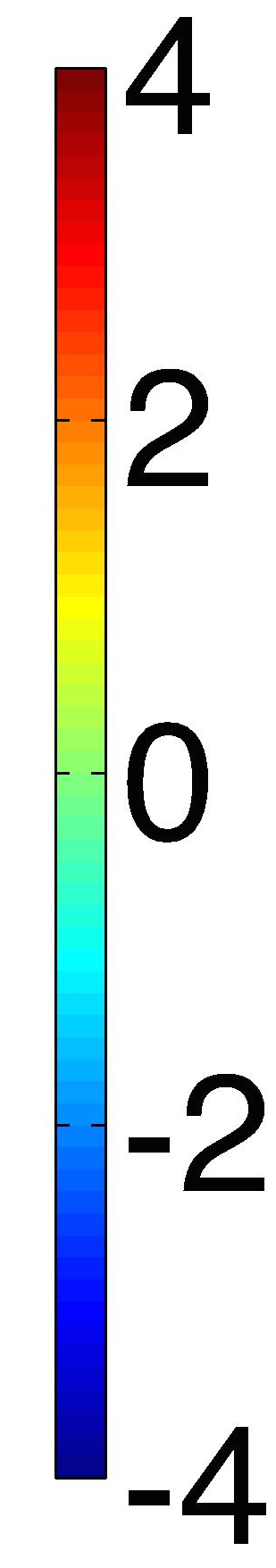}
%\subcaption{}
%\label{fig:scale}
\end{subfigure} 
\begin{subfigure}[t]{0.4\textwidth}
  \includegraphics[width=\textwidth]{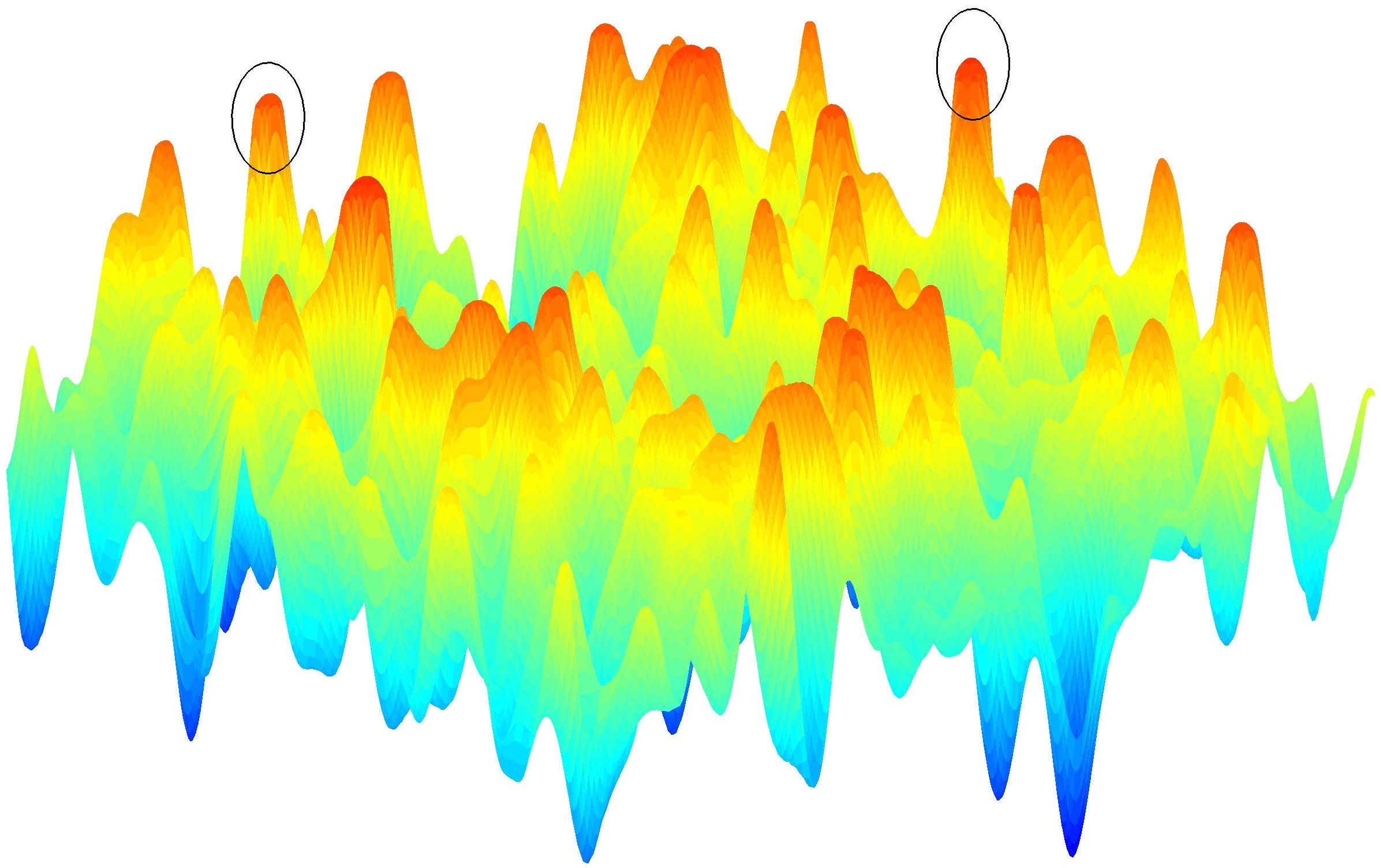}
\subcaption{}
\label{fig:finalSurface}
\end{subfigure} 
\begin{subfigure}[t]{0.45\textwidth}
  \includegraphics[width=\textwidth]{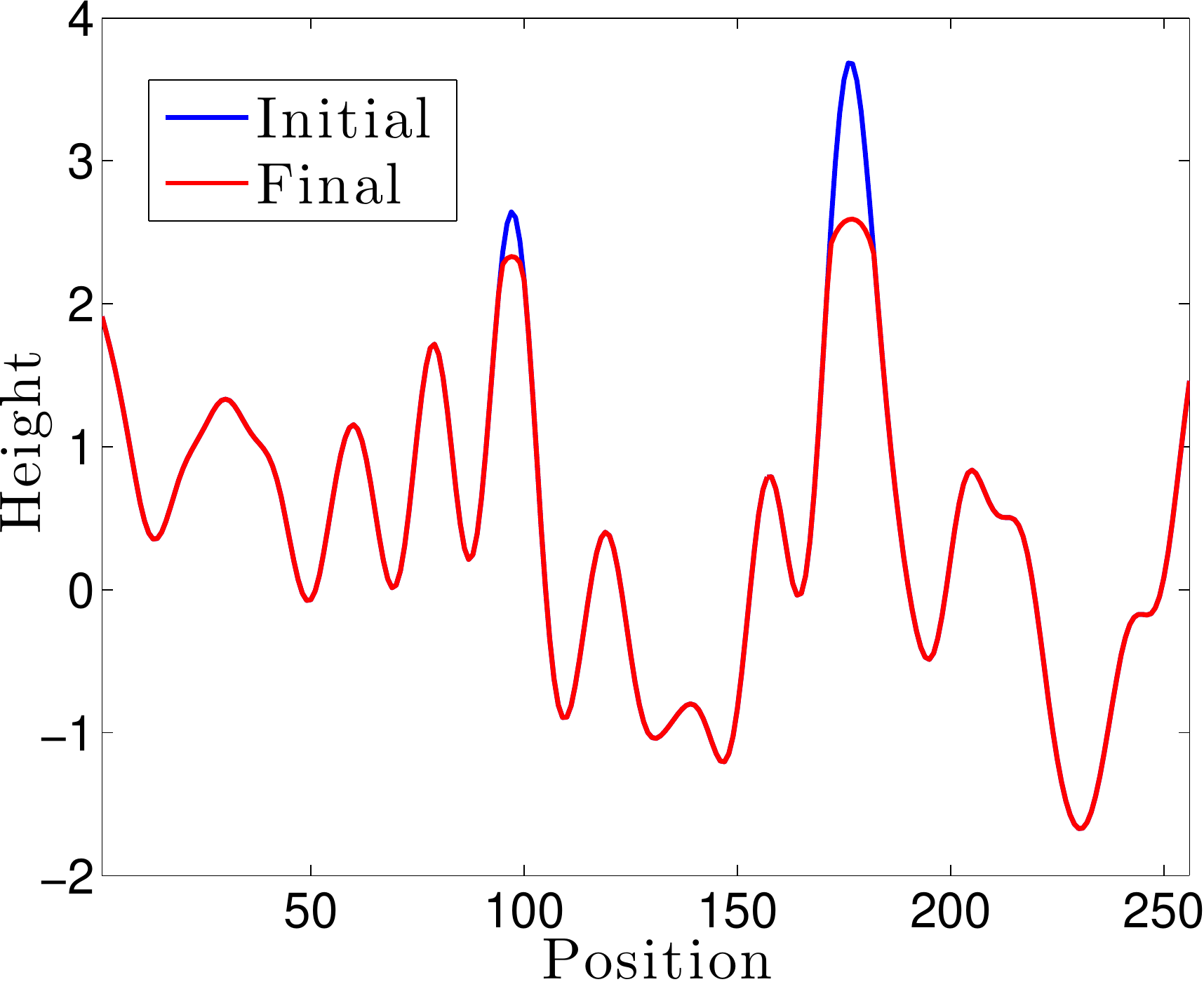}
\subcaption{}
\label{fig:initialFinalSurfaceProfiles}
\end{subfigure} 
\begin{subfigure}[t]{0.45\textwidth}
  \includegraphics[width=\textwidth]{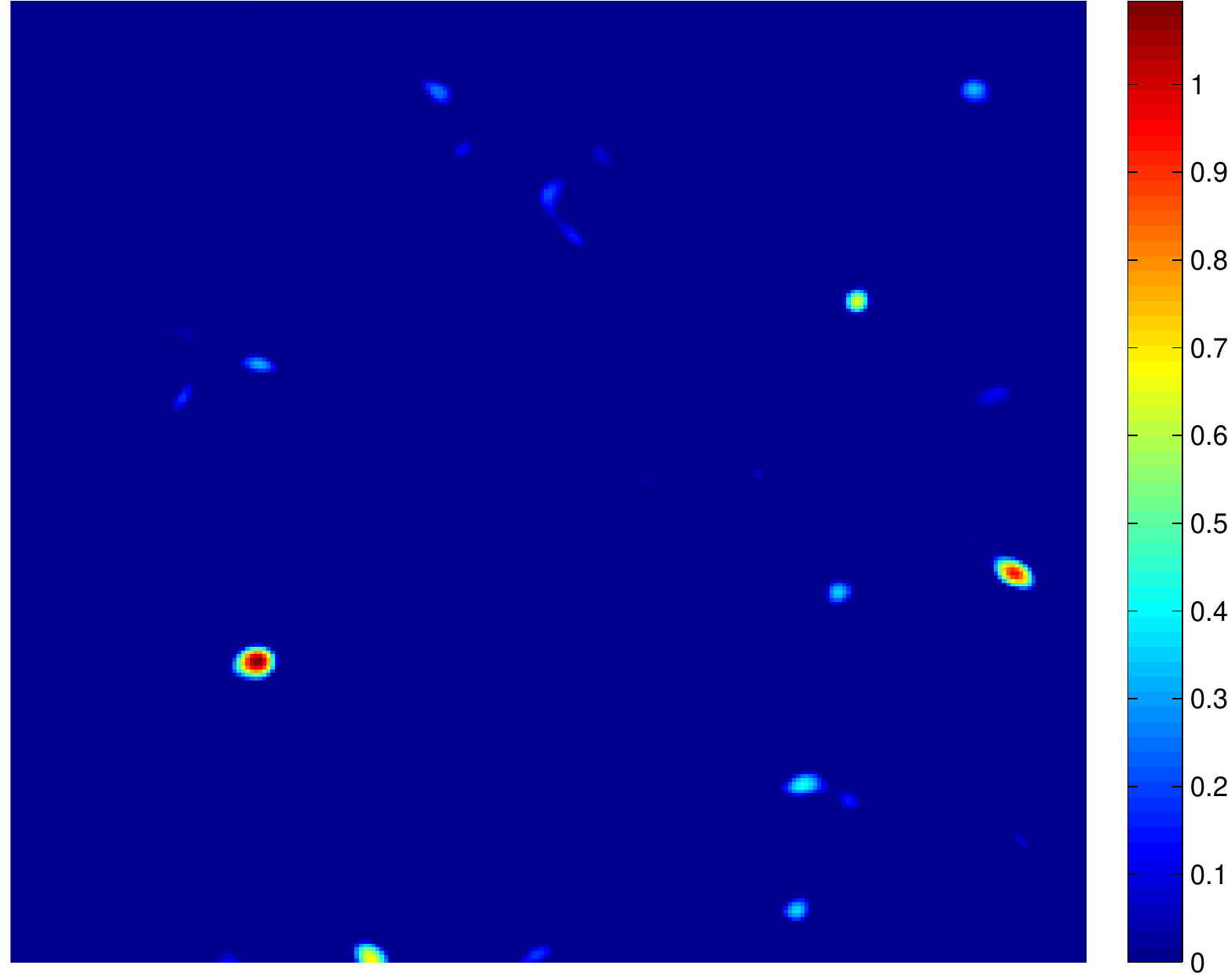}
\subcaption{}
\label{fig:plasticDeformation}
\end{subfigure} 
\caption{Evolution of the surface (i.e., $\bar{u}_i^p + \bar{h}(\bar{x}_i^0+\bar{x}, \bar{y}_i^0)$)  during sliding: (a) initially undeformed and
(b) final deformed surfaces. The blunting of the peaks due to permanent deformation
is evident (black circles). (c) a profile showing a section across the
surface and (d) plastic length changes at the contacts.}
\label{fig:initialAndFinalSurfaces}
\end{figure}
%%%%%%%%%%%%%%%%%%%%%%%%%%%%%%%%%%%%%%%%%%%%%%%%%%%%%%%%%%%%%%%%%%%%%%%%%%%%%%%%

The qualitative features of the contact area and friction evolution described
for the viscoelastic contacts in sections  hold true for an
elastic/viscoplastic constitutive assumption for the elements (Figure
\ref{fig:slidingContactViscoplastic}). As expected, the friction coefficient
$\mu_k$ strongly depends on the creep rate $A_{cr}$. Larger creep rates lead to
larger area changes, larger transients, and hence velocity weakening, whereas
smaller creep rates result in velocity-strengthening behavior.  Note that
higher creep rates can be caused by higher temperatures
\cite{frost1982deformation}; however, higher temperatures also result in higher
values for the microscopic rate-hardening of the contact friction (larger
$\alpha$ in (14)), and the latter effect dominates at high enough temperatures
in some materials (\cite{blanpied1991fault,blanpied1995frictional,scholz:2}).

One key difference between the viscoelastic and elasto/viscoplastic
formulations is that, in the elasto/viscoplastic case, there is permanent
deformation. So, the surfaces permanently change due to static or sliding
contact.  Figure \ref{fig:initialAndFinalSurfaces} shows an example of the
permanent surface change due to slip, with several peaks flattened due to slip,
in particular, the two peaks marked with circles. The distribution of the
plastic length change over the surface and a profile through the surface
further illustrate the point. 

%An interesting open
%question is if the plastic deformation carries a signature of the sliding
%direction and surface-roughness properties.

\begin{figure}
\centering
\begin{subfigure}[t]{0.45\textwidth}
  \includegraphics[width=\textwidth]{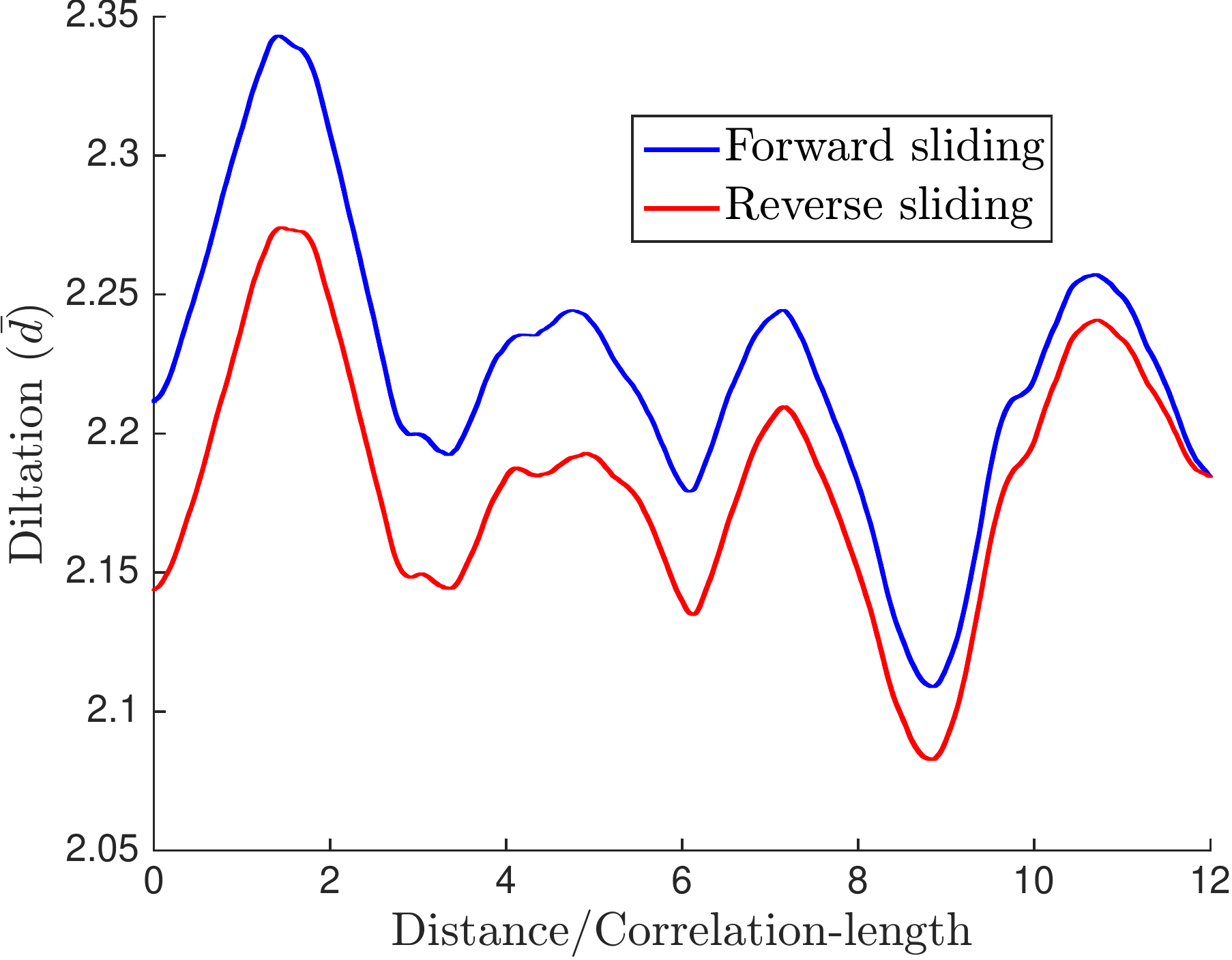}
\subcaption{}
\label{fig:dilatationEvolutionReverseSliding_512_1}
\end{subfigure} 
\begin{subfigure}[t]{0.45\textwidth}
  \includegraphics[width=\textwidth]{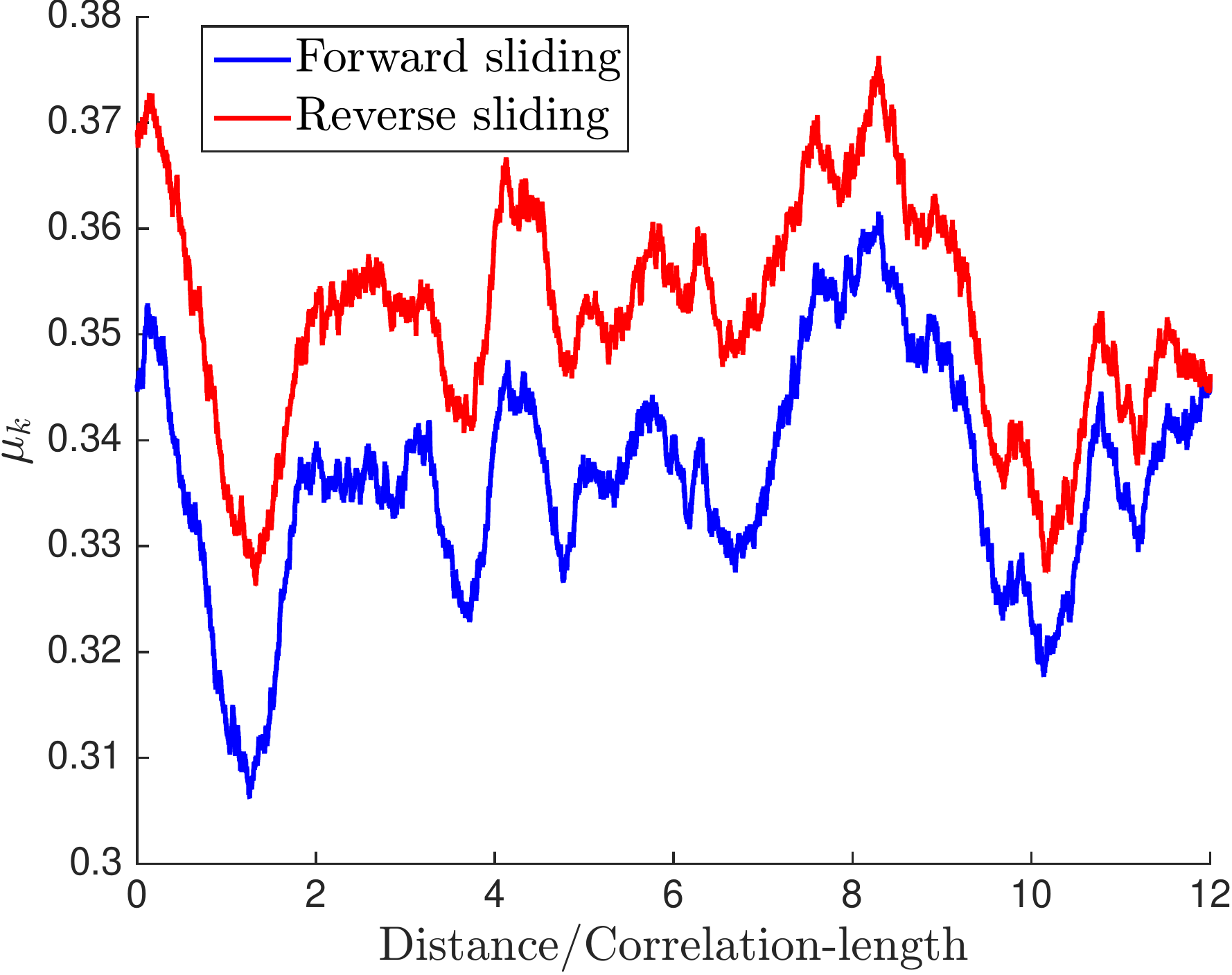}
\subcaption{}
\label{fig:frictionEvolutionReverseSliding_512_1}
\end{subfigure} 
\caption{Evolution of dilatation and friction coefficient in forward and reverse sliding.}
\label{fig:frictionDilatationEvolutionReverseSliding_512_1}
\end{figure}

This permanent change of the surface leads to history dependence and
potentially carries a signature of the sliding direction.   This is shown in
Figure \ref{fig:frictionDilatationEvolutionReverseSliding_512_1}.  One surface
is slid over another for a certain distance and then the direction of sliding
is reversed.  We see that the dilatation decreases and the coefficient of
friction increases on reversal.  The deformable elements that pass over the
highest peaks of the rigid surface get permanently deformed during forward
motion and thus move closer and have higher contact area with the peaks on
reverse motion.

%%%%%%%%%%%%%%%%%%%%%%%%%%%%%%%%%%%%%%%%%%%%%%%%%%%%%%%%%%%%%%%%%%%%%%%%%%%%%%%%

\section{Conclusion}\label{sec:conclusion} 

We have developed a numerical framework to study the time- and
velocity-dependent behavior of viscoelastic and elasto/viscoplastic rough
surfaces in static and sliding contact, including the long-range elastic
interactions between contacts.  We prescribe the material and surface
properties at the microscale and infer the macroscopic friction behavior.  We
find that this framework reproduces main qualitative features of experimental
observations. Further, our framework is able to identify which factors
conceptually influence the macroscopic behavior and which ones are lost in the
averaging process. Surprisingly, we find that, in both static and sliding
contact in our models, long-range elastic interactions do not change the main
qualitative aspects of macroscopic friction, although they are important from a
quantitative perspective.  

In static contact, both viscoelastic and elasto/viscoplastic surfaces exhibit
an increase in static friction that is widely observed in experiments. In our
models, this is achieved through creep-induced contact area increases. For
viscoelastic surfaces, the duration of friction growth is determined by the
viscoelastic relaxation times; multiple relaxation times scales are needed to
reproduce the sustained increase in static friction coefficient with the
logarithm of time observed in the laboratory experiments on some materials such
as rocks (\cite{dieterich:1}). For viscoplastic surfaces, the growth
persists without saturation, in the absence of hardening. In both cases, the
rate of growth can be related to the viscoelastic and viscoplastic properties.
Surface roughness plays an important role in determining the absolute contact
area and thus friction, but it does not change either the duration or the
amount of the friction increase during static contact.

During sliding with velocity jumps, both viscoelastic and elasto/viscoplastic
surfaces show the direct and transient effects as observed in the laboratory.
The direct effect in our models arises from the assumed velocity-strengthening
shear strength of contacts. For an abrupt velocity increase/decrease, the total
contact area remains constant, but the shear resistance abruptly
increases/decreases. However, larger sliding velocities correspond to smaller
steady-state contact areas, because the average normal force on a contact
increases with the sliding velocity. Hence, after a velocity jump, the contact
area evolves, leading to the transient effect. Depending on the amplitude of
the direct effect and how the contact area evolves for different material
properties and sliding velocities, the steady-state friction response can be
velocity strengthening or velocity weakening. In our model, as in the previous
theoretical studies \cite{baumberger:2,putelat:1}, the evolution of the contact
area provides the physical basis for the state variable used in empirical
rate-and-state models.  An important difference between viscoelastic and
elasto/viscoplastic surfaces is that plasticity leads to permanent changes in
the surface profile. 

To reproduce the more specific features encapsulated in the empirical
rate-and-state formulations (Equations \ref{eq:RS}-\ref{eq:RS_steady}), our
model would require more involved constitutive models for both the deforming
elements and for the microscopic shear strength of contacts. In the
viscoelastic models with a single relaxation time scale explored in this study,
the area change is substantial for a certain range of sliding velocities.  As
higher velocities are imposed, the area change decreases and eventually
disappears, effectively corresponding to parameter $b$ from the empirical
rate-and-state laws (Equations \ref{eq:RS}-\ref{eq:RS_steady}) being a function
of sliding velocity $v$. This leads to velocity-strengthening behavior at high
enough sliding velocities, due to the presence of the direct effect. Such
transition to velocity-strengthening behavior at higher velocities have been
observed in some experiments (\cite{shimamoto:1}), but many materials display
velocity-weakening friction over several orders of magnitude in sliding
velocities, with the transient effect being proportional to the logarithm of
the velocity jump (\cite{marone:1}). We hypothesize that such sustained
velocity-weakening behavior can be achieved in our model by including multiple
shorter relaxation time scales. 

Furthemore, the macroscopic direct effect in our model is not proportional to
the logarithm of the velocity jump, as observed in many experiments, but it is
modulated by the area changes with sliding velocity. In other words, our model
results in parameter $a$ of (2-3) that depends on the contact area and hence
$a$ is also a function of $v$. Since the evolution of the total contact area
with sliding velocity is a robust feature of our model, this discrepancy cannot
be fixed by simply changing the constitutive properties of the elements, e.g,
adding additional relaxation time scales.  Rather, this discrepancy signifies
the need to a more sophisticated shear strength assumption on the microscale,
one that reflects the assumed constitutive properties of the bulk material.

Our model assumes local shear resistance that does not depend on the individual
conditions of the element in contact, and all elements are assumed to move with
the same macroscopic sliding velocity. Hence, the evolving asperity population
affects the macroscopic friction only through the evolution of the total
contact area. The developed numerical framework can be used to study the
consequences of relaxing this assumption. One idea that can be explored in our
framework is that the local shear resistance increases with the local time in
contact (or contact maturity), due to local processes allowing for better
atomic scale matching, desorption of trapped impurities, etc \cite{riceJR:1}.
This would make the local shear resistance depend not only on the sliding
velocity but also on the (evolving) individual asperity size, with larger
asperities being stronger per unit area. Another possibility to explore is that
the local shear resistance depends on the local state of the element, e.g., its
normal force.  In both cases, the evolution of macroscopic friction, and hence
the effective state variable, may no longer be primarily dependent on the
evolution of the total contact area, but additionally reflect the variation in
the distribution of the contact sizes and forces. Such a modification would
enhance the importance of the surface roughness in controlling the macroscopic
friction behavior. In such models, the long-range elastic interactions may
become more conceptually important, since our simulations show that they
significantly affect the distribution of the contact forces and asperity sizes.  

Finally, we use a Gaussian autocorrelation function, while natural surfaces
are thought to be fractal.  We have done some preliminary studies on
surfaces with an exponential autocorrelation,
$$R_{zz}(\delta_x,\delta_y) = \mathrm{E}\{ z(x,y) z(x+\delta_x,y+\delta_y)\} = \sigma^2 e^{-(\delta_x^2+\delta_y^2)^{1/2}/\beta},$$ 
where $\beta$ is the correlation
length.  The power spectral density for this surface is given by,
\begin{equation}\label{eq:psdExponential} S(\omega_x,\omega_y) =
\frac{2\pi\sigma^2\beta^2}{\left[1+\left(\omega_x^2+\omega_y^2\right)\beta^2\right]^{3/2}}.
\end{equation}
So, if  $\left(\omega_x^2+\omega_y^2\right)^{1/2}\beta \gg 1$, the power spectrum
decays as $1/\omega^3$ (where $\omega = (\omega_x^2+\omega_y^2)^{1/2}$) which
corresponds to a fractal dimension of $2.5$ (or a Hurst exponent of $0.5$).
We have verified that the qualitative friction behavior during both
static/sliding contact remains the same as in the Gaussian autocorrelation
case. However, a detailed parameter study, as well as the study of other 
fractal surfaces remains a
topic of future work.
\vspace{\baselineskip}

To summarize, various experimentally observed features of friction -- growth of
the static friction coefficient with time, velocity and history dependence of
the dynamic friction coefficient, velocity-strengthening and velocity-weakening
behavior of steady-state friction --  can be captured by a bare minimum of
ingredients, a time-dependent and velocity-dependent contact at the microscale
with statistical averaging due to rough surfaces. The main qualitative aspects
of the evolution are the same in the viscoelastic and elasto/viscoplastic
cases, showing that the macroscopic frictional response is robust with respect
to a wide range of microscopic material behavior. The developed framework can
be used to study which assumption on the microscale produce the specific
features of the macroscopic behavior captured in empirical rate-and-state laws.
%
%\section{Long and reverse sliding}
%
%\subsection{Long sliding}
%
%Evolution of $\mu_k$ with viscoplasticity, I do not see any systematic
%variation in the friction coefficient with sliding distance.
%
%%%%%%%%%%%%%%%%%%%%%%%%%%%%%%%%%%%%%%%%%%%%%%%%%%%%%%%%%%%%%%%%%%%%%%%%%%%%%%%%%
%\begin{figure} \centering \begin{subfigure}[t]{0.5\textwidth}
%\includegraphics[width=\textwidth]{./figuresViscoplastic/slidingContact/frictionEvolutionLongSliding.pdf}
%\end{subfigure} \caption{} \label{fig:longSlidingContactViscoplastic}
%\end{figure}
%%%%%%%%%%%%%%%%%%%%%%%%%%%%%%%%%%%%%%%%%%%%%%%%%%%%%%%%%%%%%%%%%%%%%%%%%%%%%%%%%
%
%\subsection{Reverse sliding}
%

\section*{Acknowledgment}
We gratefully acknowledge the support for this study from the National Science
Foundation (grant EAR 1142183) and the Terrestrial Hazards Observations and
Reporting center (THOR) at Caltech.

\bibliographystyle{unsrt}
\bibliography{2dViscoelastic}

\end{document}